\documentclass[a4paper,11pt]{article}

\usepackage{stmaryrd}
\usepackage{bm}
\usepackage{cite}
\usepackage{color}
\usepackage{dsfont}
\usepackage{nicefrac}
\usepackage{multirow}
\usepackage{graphics}
\usepackage{graphicx}
\usepackage{epsfig}
\usepackage{subfigure}
\usepackage{pgf}
\usepackage{amsmath,amssymb,amsfonts}
\usepackage{wasysym}
\usepackage{yhmath}
\usepackage{booktabs}

\begin{document}

\title{Topology optimization of low-temperature type-II superconductors with superconductor-dielectric/vacuum interfaces based on Ginzburg-Landau theory under Weyl gauge}

\author{Yongbo Deng\footnote{yongbo.deng@kit.edu (Y.~Deng)}, 
Jan G. Korvink\footnote{jan.korvink@kit.edu (J.~G.~Korvink)} \\
Institute of Microstructure Technology (IMT), \\
Karlsruhe Institute of Technology (KIT), \\
Hermann-von-Helmholtzplatz 1, \\
Eggenstein-Leopoldshafen 76344, Germany.}


\maketitle

\abstract{Geometrical design is a crucial and challenging strategy for improving the performance of type-II superconductors, because the proper placement of intended defects in the current path contribute to flux pinning, a reduction in dissipation, and an increase in achievable current density. Topology optimization is currently one of the most powerful approaches used to determine consistent structural geometries. Therefore, a topology optimization approach is presented to inversely design structural geometries of low-temperature type-II superconductors with superconductor-dielectric/vacuum interfaces. In the presented approach, the magnetic response of low-temperature type-II superconductors is modeled using the Ginzburg-Landau theory, where the temporal evolution of the order parameter and vector potential is described by the time-dependent Ginzburg-Landau equations under the Weyl gauge. The presented approach is developed using a material distribution method, where material interpolations are applied to the Ginzburg-Landau parameter, Landau free energy, and magnetic energy. The material density used in the material interpolations is derived through a sequence of steps, including a PDE filter, piecewise homogenization, and threshold projection of a design variable defined over the design domain, where the piecewise homogenization step is employed to remove gradients of the filtered design variable during adjoint analysis and ensure smoothness of the adjoint sensitivity. The topology optimization model is constructed to enhance the robustness of the mixed state in a low-temperature type-II superconductor. It is regularized by an essential constraint of the volume fraction of the superconducting material in the design domain. To circumvent complexity of the adjoint analysis arising from numerical discretization of time derivatives in the time-dependent Ginzburg-Landau equations, continuous adjoint analysis is employed. It is implemented by expressing the complex order parameter in real and imaginary parts, and further splitting the gauged time-dependent Ginzburg-Landau equations. This splitting preserves the consistency between the real-valued properties of the design variable, and the adjoint sensitivity of the optimization objective. Using the derived adjoint sensitivities of the optimization objective and volume fraction, the model is solved numerically via an iterative procedure. Numerical studies explore the influence of the volume fraction, applied magnetic field, Ginzburg-Landau parameter, and functional mechanisms of optimized topologies. By recasting the optimization objective to minimize the least-square difference between the order parameters of the mixed and Meissner states, the topology optimization model can be adopted to delay the appearance of regions in the intermediate and normal states before the occurrence of second-order phase transitions at the upper critical magnetic fields. The presented topology optimization approach holds potential applications in nuclear magnetic resonance and quantum computing.
}

\section{Introduction} \label{sec:Introduction}

Superconductivity, a state in which certain materials known as superconductors conduct electricity with zero resistance and expel magnetic fields below a critical temperature, was first discovered in 1911 by the Dutch physicists H. K. Onnes et al during their investigation of solid mercury's electrical conductivity using liquid helium cooling  \cite{OnnesProceedings1911}. In contrast to ordinary conductors, superconductors can sustain electrical currents indefinitely without energy loss, as they produce no resistive heating. This property underpins a range of transformative applications, including microelectronics \cite{BairamkulovCSM2024}, nuclear magnetic resonance \cite{MaedaJMR2019}, lossless electricity transmission \cite{ThomasRSER2016}, highly efficient motors \cite{ChowER2023}, fusion reactors \cite{WolfFED2021}, and quantum computers \cite{DiCarloNature2009}, to name the most prominent.

\begin{figure}[!htbp]
  \centering
  \subfigure[]
  {\includegraphics[width=0.4\textwidth]{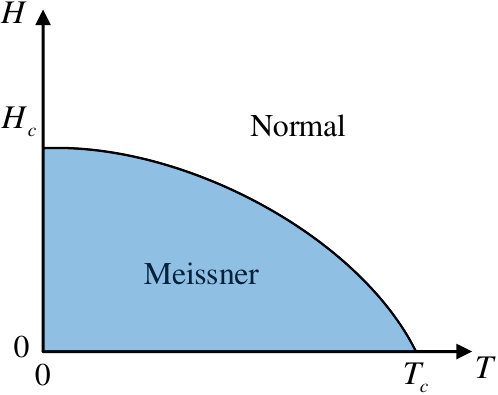}} \hspace{4ex}
  \subfigure[]
  {\includegraphics[width=0.4\textwidth]{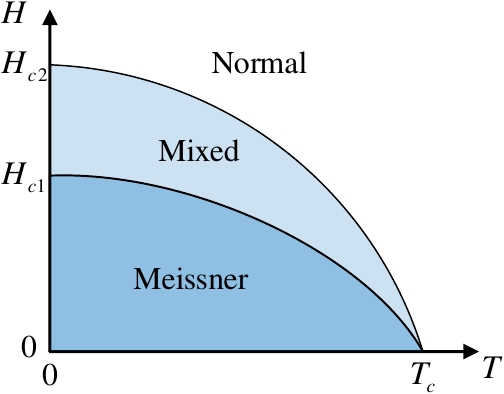}} \\
  \subfigure[]
  {\includegraphics[width=0.87\textwidth]{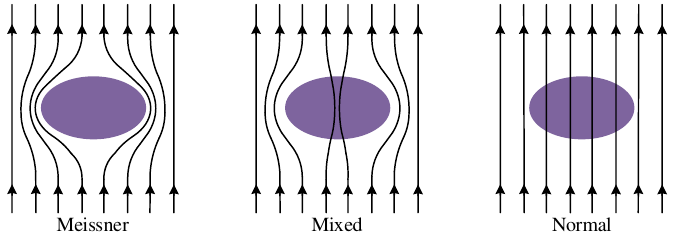}}
  \caption{Schematic phase diagrams of superconductors, and sketches for the possible states of a type-II superconductor. (a) Phase diagram of a type-I superconductor, where $H$ is the externally applied magnetic field, $H_c$ is the thermodynamic critical magnetic field, $T$ is the temperature and $T_c$ is the critical value of the temperature for the phase transition, respectively. (b) Phase diagram of a type-II superconductor, where $H_{c1}$ and $H_{c2}$ are the lower and upper critical values of the magnetic field for the first and second second-order phase transitions, respectively. (c) Sketches for the distribution of magnetic induction lines in the Meissner, mixed and normal states, where the lines with arrows are magnetic induction lines and the superconductors are marked in purple color. In (c), the magnetic field cannot penetrate into the interior of the superconductor in the Meissner state; it can penetrate completely into the interior in the normal state. The intermediate state between the Meissner and normal states is called a mixed state, where the magnetic field can penetrate normal cores of the flux lines sketched in Fig.~\ref{fig:SketchFluxLineAndMixedState}.}
  \label{fig:PhaseDiagramSupercond}
\end{figure}

The interior of a superconductor cannot be penetrated by an externally applied magnetic field. This is a phenomenon known as the Meissner effect or Meissner-Ochsenfeld effect, discovered by W. Meissner and R. Ochsenfeld in 1933 \cite{MeissnerOchsenfeldNatur1933}. When the applied magnetic field becomes too large, superconductivity breaks down. According to how the breakdown occurs, superconductors can be divided into two types, commonly known as type-I and type-II. 
In type-I superconductors, superconductivity can be abruptly destroyed via a first-order phase transition, when the strength of the applied magnetic field rises above the thermodynamic critical magnetic field as sketched in Fig.~\ref{fig:PhaseDiagramSupercond}a. This behavior is different from type-II superconductors discovered by J. N. Rjabinin and L. W. Shubnikov in 1935 \cite{RjabininSchubnikowNature1935}.
A type-II superconductor exhibits two critical magnetic fields for second-order phase transitions as sketched in Fig.~\ref{fig:PhaseDiagramSupercond}b. The first second-order phase transition occurs at the lower critical magnetic field, which begins penetrating into the material as quantized vortices. But the material remains superconducting outside of those microscopic vortices. The entire material becomes non-superconducting, when the vortex density becomes too large. This corresponds to the second-order phase transition occuring at the upper critical magnetic field. 
Three states defined by the two critical magnetic fields can exist in a type-II superconductor as sketched in Fig.~\ref{fig:PhaseDiagramSupercond}c. 
Type-I and type-II superconductors can be quantitatively defined by the Ginzburg-Landau parameter. They are those with $0 < \kappa < 1/\sqrt{2}$ and $\kappa > 1/\sqrt{2}$, respectively, where $\kappa$ is the Ginzburg-Landau parameter \cite{GinzburgLandau1950,AbrikosovPASUSSR1952}. Type-II superconductors can be further categorized into two types, known as low- and high-temperature superconductors. 
The first high-temperature type-II superconductor with a critical temperature around $35.1$ K was discovered by J. G. Bednorz and K. A. M\"{u}ller in 1986 \cite{BednorzMueller1986}. It was then modified by C. W. Chu et al in 1987, where the critical temperature was increased to $93$ K \cite{WuChuPRL1987}. 
Owing to existing debates on the fundamental mechanism of high-temperature type-II superconductors \cite{SinghJSNM2025,LiShenPRL2025,LeeMTP2025}, this paper concentrates on low-temperature type-II superconductors.

\begin{figure}[!htbp]
  \centering
  \includegraphics[width=0.57\textwidth]{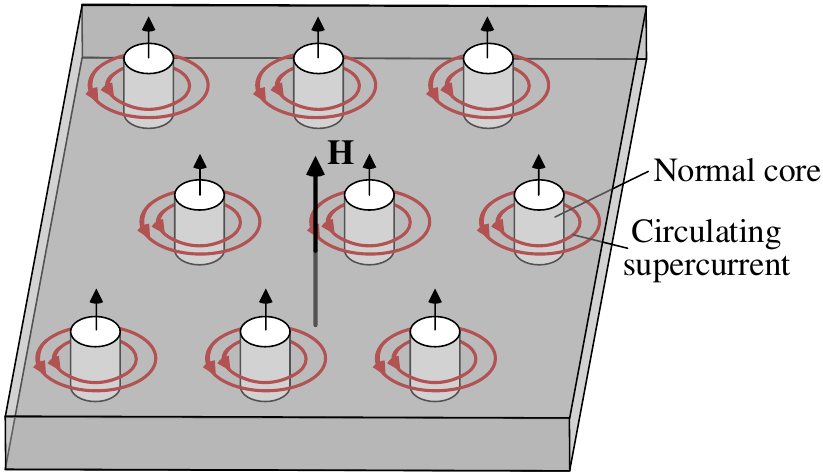}
  \caption{Sketch for flux lines/supercurrent vortices in the mixed state of a type-II superconductor, where $\mathbf{H}$ is the applied magnetic field. In the sketch, every flux line has a normal core which can be approximated by a thin cylinder with its axis parallel to the applied magnetic field, and this normal core is surrounded by circulating supercurrents.}\label{fig:SketchFluxLineAndMixedState}
\end{figure}

In type-II superconductors, mutual repulsion between vortices can arise from superconducting currents named supercurrents, which circulate in the same direction around non-superconducting cores named normal cores. The vortices are arranged into regular lattices, also referred to as Abrikosov lattices \cite{AbrikosovRMP2004}, in an ideal type-II superconductor. The circulating supercurrents and the normal cores compose the flux lines sketched in Fig.~\ref{fig:SketchFluxLineAndMixedState}. The flux lines achieve a mixed state in a type-II superconductor. Each vortex in the mixed state carries one magnetic flux quantum, i.e., the presence of vortices allows the magnetic flux in integral multiples of the flux quantum $h/2e$, where $h$ is the Planck constant and $e$ is the electronic charge. This is the mechanism of an imperfect and incomplete Meissner effect found in type-II superconductors. At the upper critical magnetic field sketched in Fig.~\ref{fig:PhaseDiagramSupercond}, the normal cores of the flux lines overlap so much that the entire superconductor becomes normal; thereby, superconductivity is destroyed. The mixed state can be stable up to a large critical magnetic field. Thus, type-II superconductors have the advantage of operating in higher magnetic fields and carrying higher currents compared to type-I superconductors, making them suitable for practical and large-scale applications \cite{ZhouNationalScienceReview2023}. 

When a current flows in a type-II superconductor, it exerts Lorentz forces on the flux lines. The Lorentz forces tend to move the flux lines in a direction perpendicular to both the electric and magnetic fields. If the superconductor does not contain defects, the flux lines move freely and lead to energy dissipation and finite resistance within the material. This can make the superconductor useless for technical applications. The flux lines can be trapped or pinned by defects to ensure their stability, and further, to conduct the current without loss. Therefore, the superconducting performance can be enhanced by adding defects to type-II superconductors in order to pin the flux lines \cite{MatsushitaSpringer2007}. A real material usually contains defects, such as dislocations, twins, microcavities, and grain boundaries. Those defects interact with the flux lines, which tend to position themselves in the defect zones acting as pinning centers. A pinned flux line is subjected to two forces, the pinning force and the Lorentz force. It does not move until the Lorentz force exceeds the pinning force. A critical Lorentz force density is then defined corresponding to a critical current density \cite{RuizHaenisch2026}. 

The geometrical design of type-II superconductors aims to introduce artificial defects, including microcavities, edges, and corners, to effectively pin or confine the flux lines. It can strongly influence how flux lines enter, move through, and destabilize the mixed state, by determining where they enter (e.g., edges, corners, or thin regions), how densely they pack, and how easily they move. Poorly designed geometries concentrate magnetic fields locally and accelerate vortex motion, which leads to energy dissipation and loss of superconductivity. Optimized geometries can spread the field more uniformly and delay vortex entry. Geometric design is thereby a crucial strategy with the primary goal to prevent premature quenching and enhance the overall performance of type-II superconductors. Currently, several micro/nano-fabrication technologies have been developed \cite{HeJingAFM2025,LangEncyclopedia2024,ZhakinaDonnellyAFM2025}. This promises the manufacturing of type-II superconductors with complex geometries. However, geometric design of type-II superconductors remains challenging, because existing approaches rely on heuristic/surrogate optimization methods \cite{TomassettiSST2025} and shape/topology optimization methods without considering the quantum effects \cite{TamuraJRCS2020,YangITAS2023,ElhautITAS2012,PozziITAS2013,ByunITAS2004}, and cannot yield an optimal configuration. Therefore, the following questions naturally arise: what is the optimal geometric configuration of a type-II superconductor, and how can it be found?

To address these questions, a topology optimization approach for low-temperature type-II superconductors is developed in this paper. Topology optimization is a robust computational method, which can be used to determine an optimal geometric configuration corresponding to material distribution within a structure \cite{BendsoeAndSigmund2003}. Unlike size and shape optimization, which relies on adjusting a limited number of geometric parameters, topology optimization explores the full design space to generate structures that meet user-defined performance objectives. Because it is less dependent on initial design guesses, topology optimization offers greater flexibility and robustness. Consequently, it currently serves as one of the most powerful tools to optimize a structural topology.

Optimization of structural topology was investigated as early as 1904 for truss structures \cite{Michell1904}. Modern topology optimization originated from structural optimization problems in elasticity and compliant mechanisms \cite{Bendsoe1988,ChengOlhoff1981,Sigmund2001,Sigmund1997,Saxena2005}. It was later extended to a variety of fields, including acoustics, electromagnetics, fluidics, optics, thermodynamics, etc \cite{Borrvall2003,Gerborg-Hansen2006,Nomura2007,Sigmund2008,Duhring2008,Akl2008,XueHaoCMAME2024}. As one of the most widely used methods in topology optimization, the material distribution method, also known as the density method, performs topology optimization by assigning a density variable that represents the amount of material present at each point within the design domain \cite{BendsoeAndSigmund2003}. It offers advantages such as rapid convergence, low sensitivity to the initial design guess, and the capability to handle multiple constraints. Therefore, it is adopted in this paper.

By using the material distribution method, topology optimization of low-temperature type-II superconductors is implemented based on the Ginzburg-Landau theory, which is a phenomenological and macroscopic theory proposed by V. L. Ginzburg and L. D. Landau in 1950 \cite{GinzburgLandau1950}. Because the solutions of the stationary Ginzburg-Landau equations are not unique in general, the time-dependent Ginzburg-Landau equations are used to describe superconductivity \cite{DuAppAna1994}. 
They were derived by Gor'kov and \'{E}liashberg in 1968 \cite{GorkovEliashbergSPJETP1968}, based on an averaging of the BCS theory proposed by J. Bardeen, L. N. Cooper and J. R. Schrieffer in 1957 \cite{BardeenCooperSchrieffer1957}.
The Weyl gauge, also known as the temporal gauge or zero electric potential gauge, is chosen for the time-dependent Ginzburg-Landau equations, because it is suitable for studying the dynamic response of a superconductor under an applied magnetic field and in the absence of any external currents and charges \cite{FleckingerPelleHamburg1995}. This is the kernel of the code used at Argonne National Laboratory for the numerical simulation of vortex dynamics in type-II superconductors \cite{GroppLeafSIAMConf1993}. 

The goal of designing geometries of low-temperature type-II superconductors is to enhance the robustness of the mixed states. Therefore, the optimization objective is set to minimize the supercurrent density under an applied magnetic field. This is equivalent to minimizing the Lorentz force density to avoid the supercurrent density exceeding the critical current density. The volume fraction of the superconducting material in the design domain is constrained to a specified value, to avoid the unreasonable zero minima of the supercurrent density caused by a zero volume of the superconducting material. Because of the temporal property of the time-dependent Ginzburg-Landau equations, continuous adjoint analysis is employed to derive the adjoint sensitivities used in the iterative solution of the topology optimization model and to avoid the complexity of discrete adjoint analysis, where the complexity is caused by the numerical discretization of the time derivatives. In the numerical implementation, the finite element method is used to solve the relevant partial differential equations.

The sections remained are organized as follows.
In Section~\ref{sec:TOOPLowTempSupercond}, the methodology for topology optimization of low-temperature type-II superconductors is presented, including the time-dependent Ginzburg-Landau equations, material interpolation, material density, handling of objectives and constraints, topology optimization model, adjoint analysis, and numerical implementation.
In Section~\ref{sec:ResultsDiscussionTOOPSuperCond}, results are presented, and discussion is provided, to demonstrate the achieved topology optimization, including an investigation of the volume fraction, applied magnetic field and Ginzburg-Landau parameter. 
In Sections \ref{sec:ConclusionsTOOPSuperCond} and \ref{sec:AcknowledgementsTOOPSuperCond}, the paper is concluded.
Section~\ref{sec:AppendixTOOPSuperCond} contains additional materials. 
Throughout the paper, all mathematical formulations are expressed in Cartesian coordinates; vectors are represented in column form by default; and the gradient of a vector function is defined such that the gradients of its components appear as column vectors.


\section{Topology optimization of low-temperature type-II superconductors} \label{sec:TOOPLowTempSupercond}

In this section, the material distribution method is applied to develop the topology optimization approach for low-temperature type-II superconductors.

\subsection{Ginzburg-Landau model for low-temperature type-II superconductors} \label{sec:GinzburgLandauModelLowTempTypeIISupercond}
 
The temporal evolution of the superconducting properties of a low-temperature type-II superconductor, under an applied magnetic field, can be described by using the time-dependent Ginzburg-Landau equations.

\subsubsection{Time-dependent Ginzburg-Landau equations}

\begin{figure}[!htbp]
  \centering
  \includegraphics[width=1.0\textwidth]{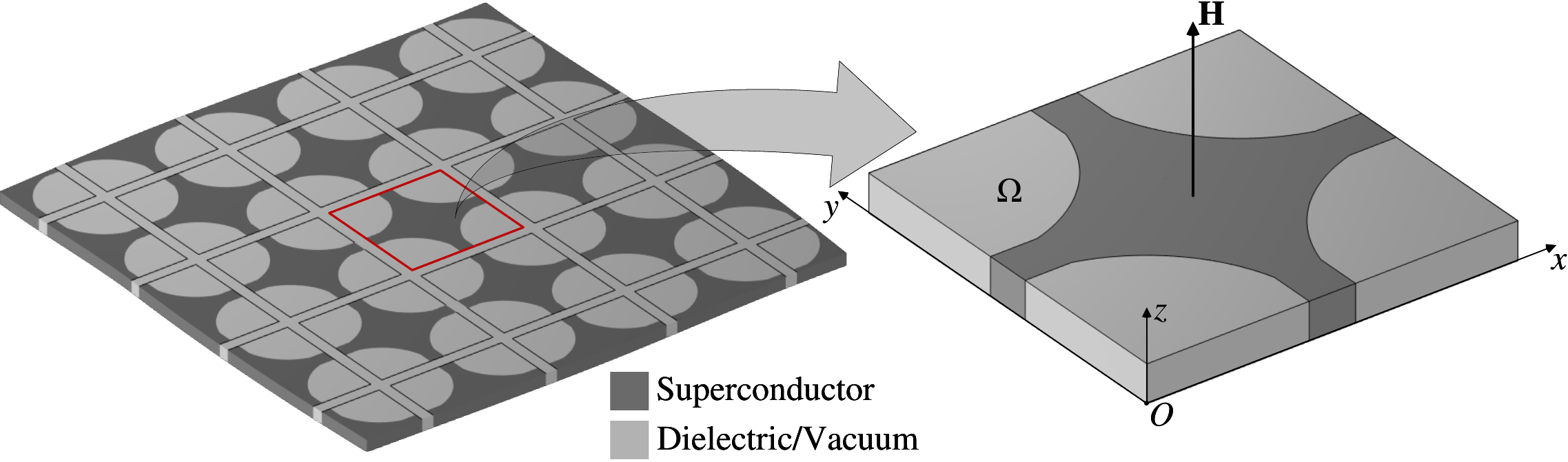}
  \caption{Sketch for a low-temperature type-II superconductor under an externally applied magnetic field, where the structural unit in a dielectric matrix is optimized. $\Omega$ denotes the cuboid-shaped design domain of the structural unit, $\mathbf{H}$ the applied magnetic field, and $O$-$xyz$ the Cartesian system with.}\label{fig:SketchDom1}
\end{figure}

In the BCS theory of superconductivity, the electrons in the vicinity of the Fermi level exist in Cooper pairs. The condensation of Cooper pairs yields the superconducting states. In the Ginzburg-Landau theory, an order parameter is used to describe the macroscopic wave function of the Cooper pairs \cite{BardeenCooperSchrieffer1957}. In a low-temperature type-II superconductor sketched in Fig.~\ref{fig:SketchDom1}, the dynamics of the order parameter, the vector potential of the magnetic induction, and the scalar potential of the electric field, can be described by the time-dependent Ginzburg-Landau equations, where the modulus of the order parameter is the density of the Cooper pairs. 

For low-temperature type-II superconductors, the non-dimensionalized Gibbs free energy $\mathcal{G}$ is \cite{DuSIAM1992}
\begin{equation}\label{eq:GibbsFreeEnergy}
  \mathcal{G}\left( \psi, \mathbf{A} \right) = \int_\Omega f_0 + {1\over2} \left( \left|\psi\right|^2 - 1 \right)^2 + \left| \left( {i\over\kappa} \nabla + \mathbf{A} \right) \psi \right|^2 + \left| \nabla \times \mathbf{A} - \mathbf{H} \right|^2 \,\mathrm{d}\Omega,
\end{equation}
where $f_0$ is the free energy density of the material in the non-superconducting state; the remaining three terms of the integrand correspond to the Landau free energy, kinetic energy and magnetic energy in the superconducting state, respectively; the Ginzburg-Landau parameter can be derived as $\kappa = \lambda_s/\xi_s$ with $\lambda_s$ and $\xi_s$ representing the penetration depth and coherence length of the superconducting material, respectively; $\psi$ is the complex-valued order parameter in the unit of $\sqrt{\left|\alpha\right|/\beta}$, the equilibrium state of the order parameter, with $\alpha$ and $\beta$ representing the phase transition parameters, respectively; $\mathbf{A}$ is the real-valued vector potential in the unit of $\sqrt{2}H_c\lambda_s$, with $H_c = \sqrt{4\pi\left|\alpha\right|^2/\beta}$ representing the thermodynamic critical field; $\Omega$ is the computational domain, its sizes are in the unit of $\lambda_s$ and it coincides with the design domain of topology optimization; $\mathbf{H}$ is the applied magnetic field in the unit of $\sqrt{2}H_c$, and it is divergence free, i.e. $\nabla \cdot \mathbf{H}= 0$ at all times; and $i=\sqrt{-1}$ is the imaginary unit. The time-dependent Ginzburg-Landau equations are related to the Gibbs free energy through the identities as \cite{DuJMP2005}
\begin{equation}\label{eq:TDGLIdentities}
\left\{\begin{split}
  & \eta \left( {\partial \over \partial t} + i \kappa \phi \right) \psi = - {1\over2} {\partial \mathcal{G}\left( \psi, \mathbf{A} \right) \over \partial \psi}, ~ \forall \left(\mathbf{x},t\right) \in \Omega \times \left(0, +\infty\right), \\
  & \sigma\left({\partial \mathbf{A} \over \partial t} + \nabla \phi \right) = - {1\over2} {\partial \mathcal{G}\left( \psi, \mathbf{A} \right) \over \partial \mathbf{A}}, ~ \forall \left(\mathbf{x},t\right) \in \Omega \times \left(0, +\infty\right), \\
\end{split}\right.
\end{equation}
where $\eta$ and $\sigma$ are the dimensionless friction coefficient and conductivity in the units of $\kappa^{-2}$ and $4\pi\sigma_sD/c^2$, respectively, with $\sigma_s$, $D$ and $c$ representing the conductivity of the normal phase, diffusion coefficient and light speed in vacuum, respectively; 
$t$ is the time in the unit of $\lambda_s^2/D$; $\mathbf{x}$ is the Cartesian coordinate in the unit of $\lambda_s$; and $\phi$ is the real-valued scalar potential in the unit of $\sqrt{2}H_cD/c$. Based on Eqs. \ref{eq:GibbsFreeEnergy} and \ref{eq:TDGLIdentities}, the dimensionless time-dependent Ginzburg-Landau equations can be derived as
\begin{equation}\label{eq:TDGLEqs}
\left\{\begin{split}
& \eta \left( {\partial \over \partial t} + i \kappa \phi \right) \psi = - \left( {i\over\kappa} \nabla + \mathbf{A} \right)^2 \psi + \left( 1 - \left| \psi \right|^2 \right) \psi, ~ \forall \left(\mathbf{x},t\right) \in \Omega \times \left(0, +\infty\right), \\
& \sigma\left({\partial \mathbf{A} \over \partial t} + \nabla \phi \right) = - \nabla \times \nabla \times \mathbf{A} + \mathbf{j}_s + \nabla \times \mathbf{H}, ~ \forall \left(\mathbf{x},t\right) \in \Omega \times \left(0, +\infty\right), \\
\end{split}\right.
\end{equation}
where $\mathbf{j}_s$ is the supercurrent density expressed as
\begin{equation}
  \mathbf{j}_s = {1\over2i\kappa} \left( \psi^* \nabla \psi - \psi \nabla \psi^* \right) - \left| \psi \right|^2 \mathbf{A} 
\end{equation}
with the superscript $*$ denoting the conjugation of a complex.
The penetration depth and coherence length of the superconducting material are
\begin{equation}
  \lambda_s = \sqrt{m_s c^2 \beta \over 4 \pi e_s^2 \left| \alpha \right|}
\end{equation}
and
\begin{equation}
  \xi_s = {\hbar \over \sqrt{2m_s\left|\alpha\right|}},
\end{equation}
where $m_s=2m_e$ and $e_s=2e$ are the effective mass and charge of a Cooper pair, respectively; $m_e$ is the electronic mass; and $\hbar = h / 2\pi$ is the reduced Planck constant. 
From the vector and scalar potentials, the magnetic induction, electric field and current can be derived as
\begin{equation}\label{eq:MagneticInductionOfVectorPotential}
  \mathbf{B} = \nabla \times \mathbf{A}
\end{equation}
and
\begin{equation}
  \mathbf{E} = - {\partial \mathbf{A} \over \partial t} - \nabla \phi,
\end{equation}
where $\mathbf{B}$ and $\mathbf{E}$ are the magnetic induction and electric field, respectively. The time-dependent Ginzburg-Landau equations in Eq.~\ref{eq:TDGLEqs} are satisfied everywhere in the low-temperature type-II superconductor at all times.

Because there are no Cooper pairs penetrating through the superconductor boundary and no surface current and charge existing there, the boundary conditions of the order parameter and vector potential satisfy
\begin{equation}\label{eq:TDGLEqsBndConds}
\left\{\begin{split}
& \mathbf{n} \cdot \left( {i\over\kappa} \nabla + \mathbf{A} \right) \psi + {i\over\kappa} \gamma \psi = 0, ~ \forall \left(\mathbf{x},t\right) \in \partial\Omega \times \left(0, +\infty\right), \\
& \mathbf{n} \times \left( \nabla \times \mathbf{A} - \mathbf{H} \right) = \mathbf{0}, ~ \forall \left(\mathbf{x},t\right) \in \partial\Omega \times \left(0, +\infty\right), \\
\end{split}\right.
\end{equation}
where $\mathbf{n}$ is the unitary outward normal at $\partial\Omega$; $\gamma$ is a non-negative function, and it in analogy with the energy of the interface between two thermodynamic phases can be referred as a surface tension. When the boundary of the design domain is the interface between the superconductor and dielectric/vacuum, $\gamma$ is $0$ at $\partial\Omega$, i.e. $\gamma=0$ at $\forall \left(\mathbf{x},t\right) \in \partial\Omega \times \left(0, +\infty\right)$; when the boundary is the interface between the superconductor and normal metal, $\gamma$ is positive at $\partial\Omega$, i.e. $\gamma>0$ at $\forall \left(\mathbf{x},t\right) \in \partial\Omega \times \left(0, +\infty\right)$.

\subsubsection{Gauge invariance}

The Weyl gauge is used to study the dynamic response of the type-II superconductors under a magnetic field and in the absence of any external currents and charges. 
The time-dependent Ginzburg-Landau equations are invariant under gauge transformation:
\begin{equation}
  \mathcal{G}_\chi: \left(\psi, \mathbf{A}, \phi\right) \mapsto \left( \psi e^{i\kappa\chi}, \mathbf{A} + \nabla\chi, \phi-{\partial\chi \over \partial t} \right),
\end{equation}
where $\chi$ is the gauge, and it can be any sufficiently smooth, real and scalar-valued function of space and time \cite{Alstr0m2011,Fleckinger1998}. The $\omega$-gauge of $\phi = -\omega \nabla\cdot\mathbf{A}$ is adopted with $\omega$ representing a real non-negative parameter. This gauge is determined by taking $\chi$ as a solution of the problem defined as
\begin{equation}
\left\{\begin{split}
  & \left( {\partial \over \partial t} - \omega \Delta \right) \chi = \phi + \omega \nabla \cdot \mathbf{A}, ~ \forall \left(\mathbf{x},t\right) \in \Omega \times \left(0, +\infty\right), \\
  & \omega \left( \mathbf{n} \cdot \nabla \chi \right) = - \omega \mathbf{n} \cdot \mathbf{A}, ~ \forall \left(\mathbf{x},t\right) \in \partial \Omega \times \left(0, +\infty\right).
\end{split}\right.
\end{equation}
In this gauge, $\mathbf{A}$ and $\phi$ satisfy
\begin{equation}\label{eq:GaugeIdentities}
\left\{\begin{split}
  & \phi + \omega \nabla \cdot \mathbf{A} = 0, ~ \forall \left(\mathbf{x},t\right) \in \Omega \times \left(0, +\infty\right), \\
  & \omega \mathbf{n} \cdot \mathbf{A} = 0, ~ \forall \left(\mathbf{x},t\right) \in \partial \Omega \times \left(0, +\infty\right).
\end{split}\right.
\end{equation}
If the triple $\left(\psi, \mathbf{A}, \phi \right)$ satisfies the time-dependent Ginzburg-Landau equations, Eq.~\ref{eq:GaugeIdentities} can be reduced to
\begin{equation}\label{eq:StrengthenedGaugeIdentities}
\left\{\begin{split}
  & \phi + \omega \nabla \cdot \mathbf{A} = 0, ~ \forall \left(\mathbf{x},t\right) \in \Omega \times \left(0, +\infty\right), \\
  & \mathbf{n} \cdot \mathbf{A} = 0, ~ \forall \left(\mathbf{x},t\right) \in \partial \Omega \times \left(0, +\infty\right).
\end{split}\right.
\end{equation}
The gauge choice fixes $\phi$ such that $\int_\Omega \phi \, \mathrm{d}\mathbf{x} = 0$ at all times, because of $\int_\Omega \phi \,\mathrm{d}\Omega = - \int_\Omega \omega \nabla \cdot \mathbf{A} \,\mathrm{d}\Omega = - \int_{\partial\Omega} \omega \mathbf{A} \cdot \mathbf{n} \,\mathrm{d}\Omega = 0$. 
Under the limit of $\omega = 0$, the $\omega$-gauge of $\phi = -\omega \nabla \cdot \mathbf{A}$ can degenerate into the Weyl gauge:
\begin{equation}\label{eq:WeylGauge}
\left\{\begin{split}
  & \phi = 0, ~ \forall \left(\mathbf{x},t\right) \in \Omega \times \left(0, +\infty\right), \\
  & \mathbf{n} \cdot \mathbf{A} = 0, ~ \forall \left(\mathbf{x},t\right) \in \partial \Omega \times \left(0, +\infty\right).
\end{split}\right.
\end{equation}
Under the Weyl gauge in Eq.~\ref{eq:WeylGauge}, the time-dependent Ginzburg-Landau equations can be reduced to
\begin{equation}\label{eq:GaugeTransformedTDGLEqs}
\left\{\begin{split}
\eta {\partial \psi \over \partial t} = & - \left( {i\over\kappa} \nabla + \mathbf{A} \right)^2 \psi + \left(1 - \left| \psi \right|^2\right) \psi, ~ \forall \left(\mathbf{x},t\right) \in \Omega \times \left(0, +\infty\right), \\
\sigma {\partial \mathbf{A} \over \partial t} = & - \nabla \times \left( \nabla \times \mathbf{A} - \mathbf{H} \right) + {1\over2i\kappa} \left( \psi^* \nabla \psi - \psi \nabla \psi^* \right) \\
& - \left| \psi \right|^2 \mathbf{A}, ~ \forall \left(\mathbf{x},t\right) \in \Omega \times \left(0, +\infty\right), \\
\end{split}\right.
\end{equation}
with the boundary conditions reduced to
\begin{equation}\label{eq:GaugeTransformedTDGLEqsBndConds}
\left\{\begin{split}
& \mathbf{n} \cdot \nabla \psi + \gamma \psi = 0, ~ \forall \left(\mathbf{x},t\right) \in \partial\Omega \times \left(0, +\infty\right), \\
& \mathbf{n} \cdot \mathbf{A} = 0, ~ \forall \left(\mathbf{x},t\right) \in \partial\Omega \times \left(0, +\infty\right), \\
& \mathbf{n} \times \left( \nabla \times \mathbf{A} - \mathbf{H} \right) = \mathbf{0}, ~ \forall \left(\mathbf{x},t\right) \in \partial\Omega \times \left(0, +\infty\right). \\
\end{split}\right.
\end{equation}
The equations in Eqs.~\ref{eq:GaugeTransformedTDGLEqs} and \ref{eq:GaugeTransformedTDGLEqsBndConds} are referred to as the gauged time-dependent Ginzburg-Landau equations. They describe the evolution of the pair $\left(\psi, \mathbf{A}\right)$ from the initial conditions expressed as
\begin{equation}
\left\{\begin{split}
  & \left[\psi\right]_{t=0} = \psi_0, ~ \forall \mathbf{x} \in \Omega, \\
  & \left[\mathbf{A}\right]_{t=0} = \mathbf{A}_0, ~ \forall \mathbf{x} \in \Omega,
\end{split}\right.
\end{equation}
where $\psi_0$ and $\mathbf{A}_0$ are the known distribution defined on $\Omega$ at $t=0$. Usually, $\psi_0$ is set as the distribution of the order parameter in the Meissner state with $\left| \psi_0 \right| = 1$, and $\mathbf{A}_0$ is set as $\mathbf{0}$.

The Weyl gauge in Eq.~\ref{eq:WeylGauge} is incompatible with the Coulomb gauge of $\nabla\cdot\mathbf{A}=0$ for $\forall \left(\mathbf{x},t\right) \in \Omega \times \left(0, +\infty\right)$. Therefore, the Coulomb gauge cannot be imposed on Eq.~\ref{eq:GaugeTransformedTDGLEqs}. If $\omega > 0$, every solution of the time-dependent Ginzburg-Landau equations is attracted to a set of stationary solutions, satisfying the Coulomb gauge with divergence-free vector potential \cite{Fleckinger1998,KaperTakac1997}.

\subsubsection{Split and gauged time-dependent Ginzburg-Landau equations}\label{sec:SplitTDGLEqus}

By splitting the order parameter into its real and imaginary parts, the gauged time-dependent Ginzburg-Landau equations can be expressed with real variables. Then, the adjoint analysis of the topology optimization model can be implemented on the spaces of real functions, and the adjoint sensitivity of optimization objective is real valued. This is consistent with the real function property of the design variable \cite{DengJCP2018}. Therefore, splitting the complex order parameter can ensure the self-consistency of the adjoint analysis which will be presented in Section~\ref{sec:TOOPAdjAnalysisLowTempTypeIISupercond}.

By setting the order parameter as $\psi = \psi_r + i \psi_i$, the gauged time-dependent Ginzburg-Landau equations can be transformed into
\begin{equation}\label{eq:TransformedGaugeTransformedTDGLEqs}
\left\{\begin{split}
& \left.\begin{split}
\eta {\partial \left(\psi_r + i \psi_i\right) \over \partial t} = & - \left( {i\over\kappa} \nabla + \mathbf{A} \right)^2 \left(\psi_r + i \psi_i\right) + I_d \left[ 1 - \left(\psi_r^2 + \psi_i^2\right) \right] \\
& \left(\psi_r + i \psi_i\right), ~ \forall \left(\mathbf{x},t\right) \in \Omega \times \left(0, +\infty\right), \\
\end{split}\right. \\
& \left.\begin{split}
\sigma {\partial \mathbf{A} \over \partial t} = & - \nabla \times \left[ w_p \left( \nabla \times \mathbf{A} - \mathbf{H} \right) \right] + {1\over2i\kappa} [ \left(\psi_r - i \psi_i\right) \nabla \left(\psi_r + i \psi_i\right) \\
& - \left(\psi_r + i \psi_i\right) \nabla \left(\psi_r - i \psi_i\right) ] - \left(\psi_r^2 + \psi_i^2\right) \mathbf{A}, \\
& ~ \forall \left(\mathbf{x},t\right) \in \Omega \times \left(0, +\infty\right), \\
\end{split}\right. \\
\end{split}\right.
\end{equation}
with the transformed boundary conditions expressed as
\begin{equation}\label{eq:TransformedGaugeTransformedTDGLEqsBndConds}
\left\{\begin{split}
& \mathbf{n} \cdot \nabla \left(\psi_r + i \psi_i\right) + \gamma \left(\psi_r + i \psi_i\right) = 0, ~ \forall \left(\mathbf{x},t\right) \in \partial\Omega \times \left(0, +\infty\right), \\
& \mathbf{n} \cdot \mathbf{A} = 0, ~ \forall \left(\mathbf{x},t\right) \in \partial\Omega \times \left(0, +\infty\right), \\
& \mathbf{n} \times \left( \nabla \times \mathbf{A} - \mathbf{H} \right) = \mathbf{0}, ~ \forall \left(\mathbf{x},t\right) \in \partial\Omega \times \left(0, +\infty\right), \\
\end{split}\right.
\end{equation} 
where $\psi_r$ and $\psi_i$ are the real and imaginary parts of the order parameter, respectively; $I_d$ is an indicator with the value of $1$ corresponding to the superconducting material, and it is added to implement the material interpolation of the Landau free energy in topology optimization; and $w_p$ is a weight parameter with the value of $1$ corresponding to the superconducting material, and it is added to implement the material interpolation of the magnetic energy and impose the third boundary condition in Eq.~\ref{eq:TransformedGaugeTransformedTDGLEqsBndConds} by using the penalization approach in topology optimization. 
By splitting Eqs.~\ref{eq:TransformedGaugeTransformedTDGLEqs} and \ref{eq:TransformedGaugeTransformedTDGLEqsBndConds} into real and imaginary parts, the split and gauged time-dependent Ginzburg-Landau equations can be derived as
\begin{equation}\label{eq:SplitGaugeTransformedTDGLEqs}
\left\{\begin{split}
& \left\{\begin{split}
\eta {\partial \psi_r \over \partial t} = & \left[ \kappa^{-1} \nabla \cdot \left( \kappa^{-1} \nabla \psi_r\right) + \kappa^{-1} \nabla \psi_i \cdot \mathbf{A} + \kappa^{-1} \left( \mathbf{A} \cdot \nabla \right) \psi_i - \mathbf{A}^2 \psi_r \right] \\
& + I_d \left[ 1 - \left(\psi_r^2 + \psi_i^2\right) \right] \psi_r, ~ \forall \left(\mathbf{x},t\right) \in \Omega \times \left(0, +\infty\right), \\
\eta {\partial \psi_i \over \partial t} = & \left[ \kappa^{-1} \nabla \cdot \left( \kappa^{-1} \nabla \psi_i \right) - \kappa^{-1} \nabla \psi_r \cdot \mathbf{A} - \kappa^{-1} \left( \mathbf{A} \cdot \nabla \right) \psi_r - \mathbf{A}^2 \psi_i \right] \\
& + I_d \left[ 1 - \left(\psi_r^2 + \psi_i^2\right) \right] \psi_i, ~ \forall \left(\mathbf{x},t\right) \in \Omega \times \left(0, +\infty\right), \\
\end{split}\right. \\
& \sigma {\partial \mathbf{A} \over \partial t} = - \nabla \times \left[ w_p \left( \nabla \times \mathbf{A} - \mathbf{H} \right) \right] + \kappa^{-1} \left( \psi_r \nabla \psi_i - \psi_i \nabla \psi_r \right) \\
& ~~~~~~~~~~ - \left(\psi_r^2 + \psi_i^2\right) \mathbf{A}, ~ \forall \left(\mathbf{x},t\right) \in \Omega \times \left(0, +\infty\right), \\
\end{split}\right.
\end{equation}
with the supercurrent expressed as
\begin{equation}
  \mathbf{j}_s = \kappa^{-1} \left( \psi_r \nabla \psi_i - \psi_i \nabla \psi_r \right) - \left(\psi_r^2 + \psi_i^2\right) \mathbf{A},
\end{equation}
the initial conditions expressed as
\begin{equation}
\left\{\begin{split}
  & \left[\psi_r\right]_{t=0} = \psi_{r0}, ~ \forall \mathbf{x} \in \Omega, \\
  & \left[\psi_i\right]_{t=0} = \psi_{i0}, ~ \forall \mathbf{x} \in \Omega, \\
  & \left[\mathbf{A}\right]_{t=0} = \mathbf{A}_0, ~ \forall \mathbf{x} \in \Omega,
\end{split}\right.
\end{equation}
and the split boundary conditions expressed as
\begin{equation}\label{eq:SplitGaugeTransformedTDGLEqsBndConds}
\left\{\begin{split}
& \mathbf{n} \cdot \nabla \psi_r + \gamma \psi_r = 0, ~ \forall \left(\mathbf{x},t\right) \in \partial\Omega \times \left(0, +\infty\right), \\
& \mathbf{n} \cdot \nabla \psi_i + \gamma \psi_i = 0, ~ \forall \left(\mathbf{x},t\right) \in \partial\Omega \times \left(0, +\infty\right), \\
& \mathbf{n} \cdot \mathbf{A} = 0, ~ \forall \left(\mathbf{x},t\right) \in \partial\Omega \times \left(0, +\infty\right), \\
& \mathbf{n} \times \left( \nabla \times \mathbf{A} - \mathbf{H} \right) = \mathbf{0}, ~ \forall \left(\mathbf{x},t\right) \in \partial\Omega \times \left(0, +\infty\right). \\
\end{split}\right.
\end{equation}

The cases of superconductor boundaries, composed of the interfaces between superconductor and dielectric/vacuum, are considered in this paper. Therefore, the surface tension $\gamma$ in Eqs.~\ref{eq:GaugeTransformedTDGLEqsBndConds}, \ref{eq:TransformedGaugeTransformedTDGLEqsBndConds} and \ref{eq:SplitGaugeTransformedTDGLEqsBndConds} is set as $0$ in the following sections.

\subsection{Material interpolations}\label{sec:MaterialInterpolation}

\begin{figure}[!htbp]
  \centering
  \subfigure[]
  {\includegraphics[width=0.9\textwidth]{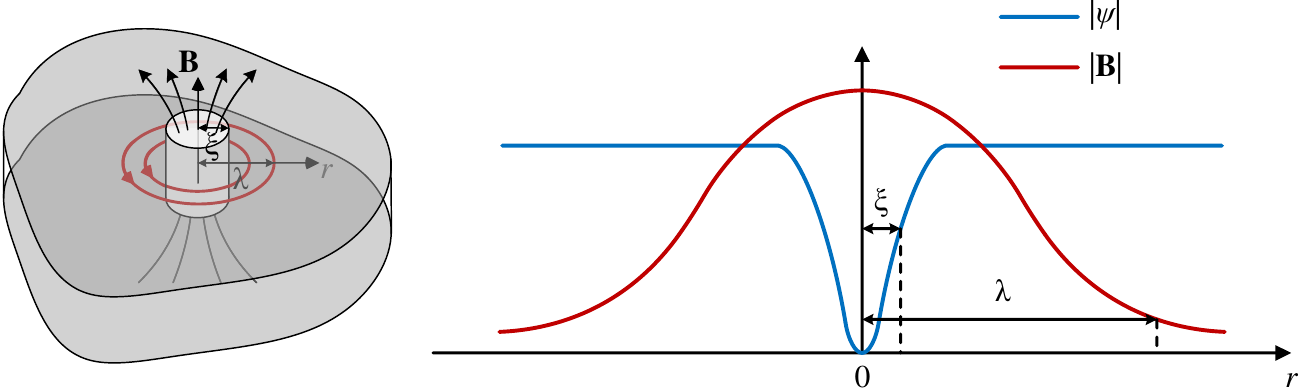}}
  \subfigure[]
  {\includegraphics[width=0.9\textwidth]{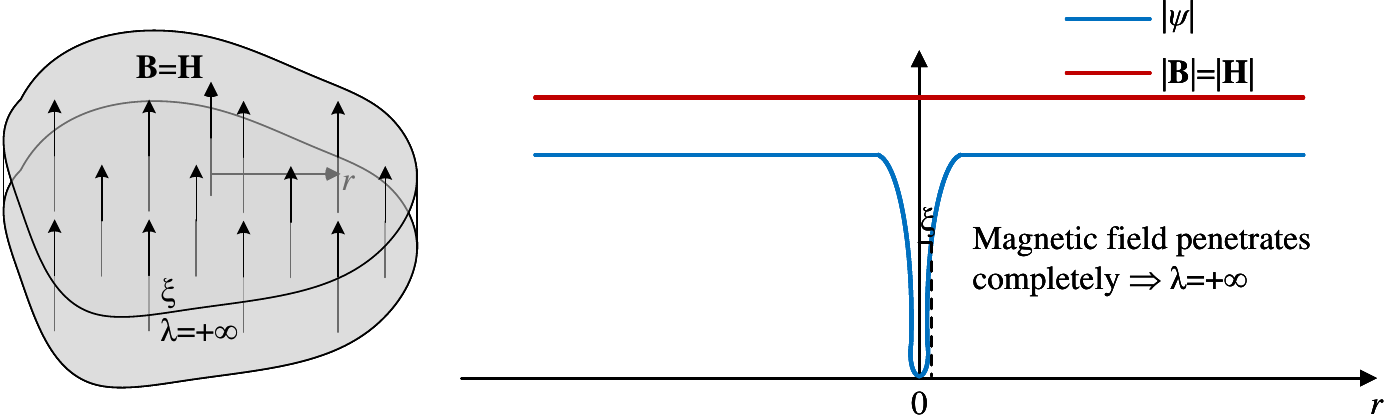}}
  \caption{Sketches for the Cooper-pair density and magnetic induction in the superconductor and dielectric/vacuum: (a) Cooper-pair density and magnetic induction in a flux line in the type-II superconductor; (b) Distribution of the Cooper-pair density and magnetic induction in the dielectric/vacuum. In the sketches, $r$ is the radial direction in the cross-section of the flux line; $\lambda$ and $\xi$ are the penetration depth and coherence length, respectively; and $\left|\psi\right|$ and $\left|\mathbf{B}\right|$ are the Cooper-pair density and magnetic-induction modulus, respectively.}\label{fig:StructureOfVortexLine}
\end{figure}

In topology optimization, the design domain is filled with both the superconducting material and dielectric/vacuum. A cross-section of a flux line in the superconductor is sketched in Fig.~\ref{fig:StructureOfVortexLine}a. 
Because the applied magnetic field can completely penetrate the dielectric/vacuum, the penetration depth is infinite in the dielectric/vacuum in Fig.~\ref{fig:StructureOfVortexLine}b. Therefore, the penetration depth satisfies
\begin{equation}
  \lambda = \left\{\begin{split}
  & \lambda_s, ~ \forall \mathbf{x} \in \Omega_s, \\
  & +\infty, ~ \forall \mathbf{x} \in \Omega\backslash\Omega_s,
  \end{split}\right.
\end{equation}
and the coherence length satisfies
\begin{equation}
  \xi = \left\{\begin{split}
  & \xi_s, ~ \forall \mathbf{x} \in \Omega_s, \\
  & \xi_d, ~ \forall \mathbf{x} \in \Omega\backslash\Omega_s,
  \end{split}\right.
\end{equation}
where $\Omega_s\subset\Omega$ is the subdomain occupied by the superconductor and $\Omega\setminus\Omega_s$ is the subdomain occupied by the dielectric/vacuum; $\xi_d$ is the coherence length in the dielectric/vacuum; $\lambda$ and $\xi$ are the penetration depth and coherence length defined in the design domain. Because the Ginzburg-Landau parameter is the ratio between the penetration depth and coherence length, it can be extended and defined in the dielectric/vacuum with the value of positive infinity. Therefore, the Ginzburg-Landau parameter defined as $\kappa = \lambda/\xi$ satisfies
\begin{equation}\label{eq:ParameterKappa}
  \kappa^{-1} = \left\{\begin{split}
  & \kappa_s^{-1}, ~ \forall \mathbf{x} \in \Omega_s, \\
  & 0, ~ \forall \mathbf{x} \in \Omega\backslash\Omega_s,
  \end{split}\right.
\end{equation}
where $\kappa_s$ equal to $\lambda_s/\xi_s$ is the Ginzburg-Landau parameter of the superconducting material.

Because the Landau free energy does not exist in the Gibbs free energy of the dielectric/vacuum, it can be extended and defined in the dielectric/vacuum with the value of zero. Therefore, the indicator $I_d$ in Eq.~\ref{eq:TransformedGaugeTransformedTDGLEqs} is interpolated to satisfy
\begin{equation}\label{eq:ParameterIdentification}
  I_d = \left\{\begin{split}
  & 1, ~ \forall \mathbf{x} \in \Omega_s, \\
  & 0, ~ \forall \mathbf{x} \in \Omega\backslash\Omega_s,
  \end{split}\right.
\end{equation}
with $I_d=1$ and $I_d=0$ representing the indicators of the subdomains occupied by the superconductor and dielectric/vacuum, respectively.

In order to impose the fourth boundary condition in Eq.~\ref{eq:SplitGaugeTransformedTDGLEqsBndConds} at the implicit material interface in topology optimization, the weight parameter in Eq.~\ref{eq:TransformedGaugeTransformedTDGLEqs} is extended and defined in the dielectric/vacuum domain with the value of positive infinity. This corresponds to the penalization of the magnetic energy included in the Gibbs free energy, where the applied magnetic field completely penetrates the dielectric/vacuum and it is cancelled by the magnetic induction of the vector potential in the superconductor. Therefore, the weight parameter is interpolated to satisfy
\begin{equation}\label{eq:ParameterWeight}
  w_p = \left\{\begin{split}
  & 1, ~ \forall \mathbf{x} \in \Omega_s, \\
  & +\infty, ~ \forall \mathbf{x} \in \Omega\backslash\Omega_s.
  \end{split}\right.
\end{equation}
In numerical computation, $w_p$ is approximately set as a finite and large value in the dielectric/vacuum domain, to simultaneously ensure the numerical stability and approximation accuracy.

In sum, material interpolations in topology optimization can be implemented on the Ginzburg-Landau parameter, indicator and weight parameter. They are implemented by using the $q$-parameter scheme expressed as \cite{PeterssonIJNMF2003}
\begin{equation}\label{eq:MaterialInterpolation}
\left.\begin{split}
& {\kappa^{-1}\left(\rho_p\right)} = \kappa_s^{-1} + {\left(\kappa_d^{-1} - \kappa_s^{-1} \right)} q {1-\rho_p \over q+\rho_p}, \\
& I_d \left(\rho_p\right) = I_{ds} + {\left( I_{dd} - I_{ds} \right)} q {1-\rho_p \over q + \rho_p}, \\
& w_p \left(\rho_p\right) = w_{ps} + {\left(w_{pd} - w_{ps}\right)} q {1-\rho_p \over q+\rho_p}, \\
\end{split}\right\} ~ q \in \left( 0, 1 \right],
\end{equation}
where $\kappa_d$ is the Ginzburg-Landau parameter of the dielectric/vacuum; $I_{ds}$ and $I_{dd}$ are the indicators of the superconducting and dielectric/vacuum domains, respectively;  $w_{ps}$ and $w_{pd}$ are the weight parameters of the superconducting and dielectric/vacuum domains, respectively; $q$ is the parameter used to tune the convexity of the material interpolation and it is valued in $\left( 0, 1 \right]$; and $\rho_p\in\left[0,1\right]$ is the material density with $1$ and $0$ representing the superconductor and dielectric/vacuum, respectively. According to Eqs.~\ref{eq:ParameterKappa}, \ref{eq:ParameterIdentification} and \ref{eq:ParameterWeight}, the fixed values of the related parameters in the material interpolations in Eq.~\ref{eq:MaterialInterpolation} are listed in Tab.~\ref{tab:ParametersForMaterialInterpolation}. The value of $\kappa_s^{-1}$ is determined based on the property of the chosen superconducting material, and that of the convexity parameter $q$ is determined based on numerical experiments.

\begin{table}[!htbp]
\centering
\begin{tabular}{ccccc}
  $\kappa_d^{-1}$ & $I_{ds}$ & $I_{dd}$ & $w_{ps}$ & $w_{pd}$ \\
  \midrule
  $0 $            & $1$    & $0$    & $1$      & $10^4$
\end{tabular}
\caption{Parameters and their values in the material interpolations in Eq.~\ref{eq:MaterialInterpolation}, where $\kappa_d^{-1}=0$ is equivalent to $\kappa_d=+\infty$.}\label{tab:ParametersForMaterialInterpolation}
\end{table}

\begin{figure}[!htbp]
  \centering
  \subfigure[]
  {\includegraphics[height=0.35\textwidth]{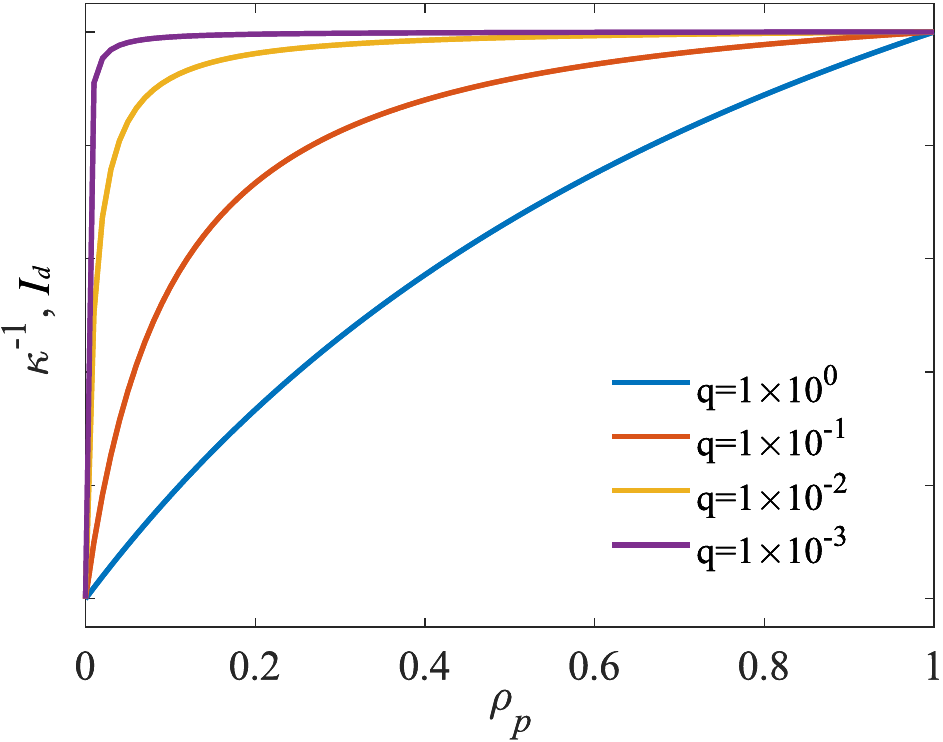}}\hspace{2em}
  \subfigure[]
  {\includegraphics[height=0.352\textwidth]{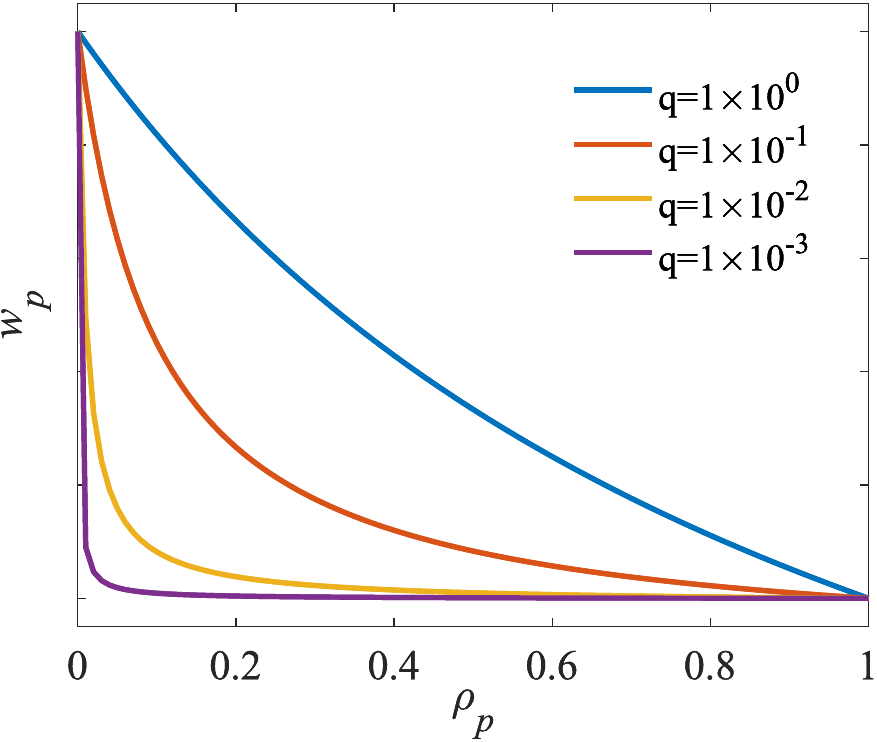}}
  \caption{Graphs of the material interpolations in Eq.~\ref{eq:MaterialInterpolation} for different values of $q$ used to tune the convexity: (a) Graphs for $\kappa^{-1}$ and $I_d$; (b) Graphs for $w_p$.}\label{fig:MaterialInterpolationCurves}
\end{figure}

It is key to make the material interpolation of the Ginzburg-Landau parameter and indicator satisfy that the values of the material density $1$ and $0$ represent the superconductor and dielectric/vacuum, respectively. This can make the corresponding material interpolations be convex as shown in Fig.~\ref{fig:MaterialInterpolationCurves}a and decrease the sensitivity of the solution of the split and gauged time-dependent Ginzburg-Landau equations in Eq.~\ref{eq:SplitGaugeTransformedTDGLEqs} to the variation of the material density in the iterative solution procedure of the topology optimization model, and further ensure the robustness of the iterative solution algorithm that will be introduced in Section~\ref{sec:NumericalImplementationLowTempTypeIISupercond}. To ensure the consistency, the same scheme is adopted for the material interpolation of the weight parameter as shown by the curves in Fig.~\ref{fig:MaterialInterpolationCurves}b, where the curves are concave because of $w_{pd} \gg w_{ps}$.

\subsection{Material density} \label{sec:MaterialDensityTDGLEqs}

The material density in the material interpolations introduced in Section~\ref{sec:MaterialInterpolation} can be derived by regularizing a design variable defined on the design domain of topology optimization. The regularization procedure is implemented by sequentially imposing a PDE filter, piecewise homogenization and threshold projection on the design variable, where the piecewise homogenization is implemented to avoid the gradient of the filtered design variable and ensure the robustness of the iterative algorithm used to solve the topology optimization model.

The PDE filter is used to remove the unreasonable tiny islands in the obtained structure and ensure the feature size for manufacturability. It is implemented by solving a PDE in the form of the Helmholtz equation:
\begin{equation}\label{eq:PDEFilter}
  \left\{\begin{split}
  & - \nabla \cdot \left( r_\rho^2 \nabla \rho_f \right) + \rho_f = \rho, ~\forall \mathbf{x} \in \Omega, \\
  & \mathbf{n} \cdot \nabla \rho_f = 0, ~\forall \mathbf{x} \in \partial \Omega, \\
\end{split}
\right.
\end{equation}
where $\rho$ is the design variable; $\rho_f$ is the filtered design variable; and $r_\rho$ is the filter radius set based on the requirement of the feature size. 

The piecewise homogenization is implemented on the filtered design variable to make the material density be a piecewise constant function \cite{DengProcA2016}. By dividing the design domain into non-overlapping pieces satisfying
\begin{equation}\label{eq:PieceDivision}
\left\{\begin{split}
  & \bigcup_{n=1}^N \Omega_n = \Omega, \\
  & \Omega_l \bigcap \Omega_m = \emptyset, ~ l \neq m, ~ \forall l \in \left\{1,2,\cdots N\right\}, ~ \forall m \in \left\{1,2,\cdots N\right\},
\end{split}\right.
\end{equation}
the piecewise homogenization of the filtered design variable can be implemented as
\begin{equation}\label{eq:PiecewiseHomogenization}
  \rho_e = {1\over\left|\Omega_n\right|} \int_{\Omega_n} \rho_f \,\mathrm{d}\Omega, ~ \forall\mathbf{x}\in\Omega_n, ~ \forall n \in \left\{1, 2, \cdots N\right\},
\end{equation}
where $\rho_e$ is the piecewisely homogenized design variable; $N$ is the number of the non-overlapping pieces; $\Omega_n$, $\Omega_l$ and $\Omega_m$ are the $n$-th, $l$-th and $m$-th pieces of $\Omega$, respectively; and $\left|\Omega_n\right| = \int_{\Omega_n} 1 \,\mathrm{d}\Omega$ is the volume of $\Omega_n$.

The threshold projection is further implemented on the piecewisely homogenized design variable to remove gray pieces and obtain clear structural boundaries. It is implemented as \cite{GuestIJNME2004} 
\begin{equation}\label{eq:ThresholdProjection}
  \rho_p = { \tanh\left(\beta_p \xi_p\right) + \tanh\left(\beta_p \left(\rho_e-\xi_p\right)\right) \over \tanh\left(\beta_p \xi_p\right) + \tanh\left(\beta_p \left(1-\xi_p\right)\right)}, ~ \xi_p \in \left(0,1\right), ~ \beta_p \in \left[1,+\infty \right),
\end{equation}
where $\rho_p$ is the projected design variable and it is the material density; and $\xi_p$ and $\beta_p$ are the projection parameters. 

Because the filtered design variable is piecewisely homogenized, $\rho_p$ is piecewise constant. The terms $\psi_i \mathbf{A} \cdot \nabla \left( \kappa^{-1} \tilde{\psi}_r \right)$ and $- \psi_r \mathbf{A} \cdot \nabla \left( \kappa^{-1} \tilde{\psi}_i \right)$ of Eq.~\ref{equ:VariationalFormsSplitTDGLOrderParameter} in the appendix can be transformed into $\kappa^{-1} \psi_i \mathbf{A} \cdot \nabla \tilde{\psi}_r$ and $- \kappa^{-1} \psi_r \mathbf{A} \cdot \nabla \tilde{\psi}_i$, where $\kappa$ is interpolated as that in Eq.~\ref{eq:MaterialInterpolation}. Therefore, the piecewise homogenization can avoid the gradient of the filtered design variable, and this is one key point to ensure the robustness of the iterative algorithm used to solve the topology optimization model.

\subsection{Objective and constraint} \label{sec:ObjectiveConstraintTDGLEqs}

In a type-II superconductor under an applied magnetic field, the mixed state of the superconductivity can be lost, when supercurrent density exceeds a critical value. Beyond this point, the Lorentz force overcomes the pinning force. By optimizing geometry of the superconductor to minimize supercurrents, this breakdown can be avoided.
Therefore, the goal of topology optimization is to find the structural geometry that can enhance the robustness of the mixed state. 
This can be achieved by using geometric features in the optimized topology to pin and confine flux lines and minimize the Lorentz force density.

The supercurrents flowing in the type-II superconductor generate the magnetic fluxes trying to cancel and expel the applied magnetic field. This creates a Lorentz force density acting on the flux lines. The Lorentz force density is expressed as
\begin{equation}\label{eq:LorentzForceDensity}
  \mathbf{f}_L = \mathbf{j}_s \times \mathbf{H},
\end{equation}
where $\mathbf{f}_L$ is the Lorentz force density. Even the magnetic field is below the critical value for the second-order phase transition at the upper critical magnetic field, flux lines can be unpinned and move due to the Lorentz force exceeding the pinning force. Further, non-zero resistance can be caused with resulting in the loss the superconductivity. From Eq.~\ref{eq:LorentzForceDensity}, it can be concluded that minimizing the Lorentz force density is equivalent to minimizing the supercurrent density. Therefore, by setting a finite and large enough value for the terminal time in Eq.~\ref{eq:SplitGaugeTransformedTDGLEqs} to ensure the full evolution of the order parameter and vector potential, the optimization objective is set to minimize the well-posed least-square form of the supercurrent density in the design domain:
\begin{equation}\label{eq:ObjFunction}
\begin{split}
J & = \int_{T_m}^{T_t} \left\| \left[ \mathbf{j}_s \right]_{\forall\mathbf{x}\in\Omega_s} \right\|_2^2 \,\mathrm{d}t \\
& = \int_{T_m}^{T_t} \int_\Omega \left|\left[ \mathbf{j}_s \right]_{\forall\mathbf{x}\in\Omega}\right|^2 \,\mathrm{d}\Omega \mathrm{d}t \\
& = \int_0^{T_t} \left[ \int_\Omega \left|\left[ \mathbf{j}_s \right]_{\forall\mathbf{x}\in\Omega}\right|^2 \,\mathrm{d}\Omega H \left( t-T_m \right) \right] \mathrm{d}t \\
& = \int_0^{T_t} \int_\Omega \left|\left[ \mathbf{j}_s \right]_{\forall\mathbf{x}\in\Omega}\right|^2 H \left( t-T_m \right) \,\mathrm{d}\Omega \mathrm{d}t,
\end{split}
\end{equation}
with
\begin{equation}
\begin{split}
\left[\mathbf{j}_s \right]_{\forall\mathbf{x}\in\Omega} = \kappa^{-1}\left(\rho_p\right) \left( \psi_r \nabla \psi_i - \psi_i \nabla \psi_r \right) - I_d\left(\rho_p\right) \left(\psi_r^2 + \psi_i^2\right) \mathbf{A},
\end{split}
\end{equation}
where $J$ is the optimization objective; ${T_t}$ is the terminal time; $T_m$ is a time point within $\left(0,T_t\right)$ and it is chosen as the time point with the order parameter reaching the steady state; $H \left( t-T_m \right)$ is the Heaviside function of time and it is expressed as
\begin{equation}\label{equ:HeavisideFunctionTime}
  H \left( t-T_m \right) = 
  \left\{\begin{split}
  & 1, ~ t \in \left(T_m, +\infty\right), \\
  & 0, ~ t \in \left(-\infty, T_m\right].
  \end{split}\right.
\end{equation}
According to equivalence of norms,
\begin{equation}\label{eq:EquivalentNorms}
\left\{\begin{split}
  & \exists ~ C_1 \in \mathbb{R}~ \mathrm{and} ~ C_2 \in \mathbb{R}, ~ \mathrm{with} ~ C_1 > 0, ~ C_2 > 0 ~ \mathrm{and} ~ C_1 \leq C_2, \\
  & \mathrm{such~that} ~ C_1 \left\|\left[ \mathbf{j}_s \right]_{\forall\mathbf{x}\in\Omega}\right\|_2 \leq \left\|\left[ \mathbf{j}_s \right]_{\forall\mathbf{x}\in\Omega}\right\|_\infty \leq C_2 \left\|\left[ \mathbf{j}_s \right]_{\forall\mathbf{x}\in\Omega}\right\|_2
\end{split}\right.
\end{equation}
is satisfied with
\begin{equation}\label{eq:2AndInftyNorms}
\left\{\begin{split}
  & \left\|\left[ \mathbf{j}_s \right]_{\forall\mathbf{x}\in\Omega}\right\|_2 = \left(\int_\Omega \left|\left[ \mathbf{j}_s \right]_{\forall\mathbf{x}\in\Omega}\right|^2 \,\mathrm{d}\Omega \right)^{1\over2}, \\
  & \left\|\left[ \mathbf{j}_s \right]_{\forall\mathbf{x}\in\Omega}\right\|_\infty = \sup_{\forall \mathbf{x} \in \Omega} \left|\left[ \mathbf{j}_s \right]_{\forall\mathbf{x}\in\Omega}\right|,
\end{split}\right.
\end{equation} 
where $C_1$ and $C_2$ are real-valued constants; $\mathbb{R}$ is the real number field; $\left\|\cdot\right\|_2$ and $\left\|\cdot\right\|_\infty$ are the $2$-norm and $\infty$-norm of a vector variable, respectively. Therefore, minimizing the objective function in Eq.~\ref{eq:ObjFunction} is equivalent to minimizing the maximal value of the supercurrent density in a type-II superconductor, and this can avoid the excess of the critical current density.

The topology optimization problem is regularized by using the volume fraction constraint of the superconducting material. In the volume fraction constraint, the volume fraction of the superconducting material is constrained as a specified value:
\begin{equation}\label{eq:VolumeFractionConstr}
\left| v - v_0 \right| \leq 1\times10^{-3},
\end{equation}
with
\begin{equation}\label{eq:VolumeFraction}
  v = {1\over\left|\Omega\right|} \int_\Omega \rho_p \,\mathrm{d}\Omega,
\end{equation}
where $v_0\in\left(0,1\right)$ is the specified volume fraction of the superconducting material; and $\left|\Omega\right|=\int_\Omega 1 \,\mathrm{d}\Omega$ is the volume of the design domain. The volume fraction constraint is essential, because it can avoid the unreasonable zero minima caused by the zero volume of superconducting material in the design domain.

\subsection{Topology optimization model}\label{sec:TOOPModelLowTempTypeIISupercond}

Based on the above introduction of the split and gauged time-dependent Ginzburg-Landau equations, material interpolations, and optimization objective and constraint, the topology optimization model for low-temperature type-II superconductors can be constructed as
\begin{equation}\label{eq:TOOPModelLowTempTypeIISuperconductor}
\left\{\begin{split}
& \text{Find}~\rho\left(\mathbf{x}\right)\in\left[0,1\right] ~ \text{with} ~ \forall\mathbf{x}\in\Omega \\
& \text{to minimize} ~ {J \over J_0} ~ \text{for} ~ J = \int_0^{T_t} \int_\Omega \left|\left[ \mathbf{j}_s \right]_{\forall\mathbf{x}\in\Omega}\right|^2 H \left( t-T_m \right) \,\mathrm{d}\Omega \mathrm{d}t \\
& \text{with} \left\{\begin{split}
& \left[\mathbf{j}_s \right]_{\forall\mathbf{x}\in\Omega} = \kappa^{-1}\left(\rho_p\right) \left( \psi_r \nabla \psi_i - \psi_i \nabla \psi_r \right) - I_d \left(\rho_p\right) \left(\psi_r^2 + \psi_i^2\right) \mathbf{A} \\
& H \left( t-T_m \right) = 
  \left\{\begin{split}
  & 1, ~ t \in \left(T_m, +\infty\right) \\
  & 0, ~ t \in \left(-\infty, T_m\right]
  \end{split}\right. \\
\end{split}\right.\\
& \text{constrained by} \\
& \left\{\begin{split}
& \left\{\begin{split}
& \left\{\begin{split}
\eta {\partial \psi_r \over \partial t} = & {\kappa^{-1}\left(\rho_p\right)} \nabla \cdot \left( {\kappa^{-1}\left(\rho_p\right)} \nabla \psi_r\right) + {\kappa^{-1}\left(\rho_p\right)} \nabla \psi_i \cdot \mathbf{A} \\
& + {\kappa^{-1}\left(\rho_p\right)} \left( \mathbf{A} \cdot \nabla \right) \psi_i - \mathbf{A}^2 \psi_r + I_d \left(\rho_p\right) \left[ 1 - \left(\psi_r^2 + \psi_i^2\right) \right] \psi_r, \\
& \forall \left(\mathbf{x},t\right) \in \Omega \times \left(0, T_t\right) \\
\eta {\partial \psi_i \over \partial t} = & {\kappa^{-1}\left(\rho_p\right)} \nabla \cdot \left( {\kappa^{-1}\left(\rho_p\right)} \nabla \psi_i \right) - {\kappa^{-1}\left(\rho_p\right)} \nabla \psi_r \cdot \mathbf{A} \\
& - {\kappa^{-1}\left(\rho_p\right)} \left( \mathbf{A} \cdot \nabla \right) \psi_r - \mathbf{A}^2 \psi_i + I_d \left(\rho_p\right) \left[ 1 - \left(\psi_r^2 + \psi_i^2\right) \right] \psi_i, \\
& \forall \left(\mathbf{x},t\right) \in \Omega \times \left(0, T_t\right) \\
\end{split}\right.\\
& \sigma {\partial \mathbf{A} \over \partial t} = - \nabla \times \left[ w_p \left(\rho_p\right) \left( \nabla \times \mathbf{A} - \mathbf{H} \right) \right] + {\kappa^{-1}\left(\rho_p\right)} \left( \psi_r \nabla \psi_i - \psi_i \nabla \psi_r \right) \\
& ~~~~~~~~~~ - \left(\psi_r^2 + \psi_i^2\right) \mathbf{A}, ~ \forall \left(\mathbf{x},t\right) \in \Omega \times \left(0, T_t\right) \\
\end{split}\right. \\
& \left\{\begin{split}
  & \left[\psi_r\right]_{t=0} = \psi_{r0}, ~ \forall \mathbf{x} \in \Omega \\
  & \left[\psi_i\right]_{t=0} = \psi_{i0}, ~ \forall \mathbf{x} \in \Omega \\
  & \left[\mathbf{A}\right]_{t=0} = \mathbf{A}_0, ~ \forall \mathbf{x} \in \Omega
\end{split}\right. \\
& \left\{\begin{split}
& \mathbf{n} \cdot \nabla \psi_r = 0, ~ \forall \left(\mathbf{x},t\right) \in \partial\Omega \times \left(0, T_t\right) \\
& \mathbf{n} \cdot \nabla \psi_i = 0, ~ \forall \left(\mathbf{x},t\right) \in \partial\Omega \times \left(0, T_t\right) \\
& \mathbf{n} \cdot \mathbf{A} = 0, ~ \forall \left(\mathbf{x},t\right) \in \partial\Omega \times \left(0, T_t\right) \\
& \mathbf{n} \times \left( \nabla \times \mathbf{A} - \mathbf{H} \right) = \mathbf{0}, ~ \forall \left(\mathbf{x},t\right) \in \partial\Omega \times \left(0, T_t\right) \\
\end{split}\right.
\end{split}\right.\\
& \left\{\begin{split}
& {\kappa^{-1}\left(\rho_p\right)} = \kappa_s^{-1} + {\left(\kappa_d^{-1} - \kappa_s^{-1} \right)} q {1-\rho_p \over q+\rho_p} \\
& I_d \left(\rho_p\right) = I_{ds} + \left( I_{dd} - I_{ds} \right) q {1-\rho_p \over q + \rho_p} \\
& w_p \left(\rho_p\right) = w_{ps} +\left(w_{pd} - w_{ps}\right) q {1-\rho_p \over q+\rho_p} \\ 
\end{split}\right. \\
& \left\{\begin{split}
& - \nabla \cdot \left( r_\rho^2 \nabla \rho_f \right) + \rho_f = \rho, ~\forall \mathbf{x} \in \Omega \\
& \mathbf{n} \cdot \nabla \rho_f = 0, ~\forall \mathbf{x} \in \partial \Omega \\
\end{split}
\right. \\
& \left\{\begin{split}
& \rho_e = {1\over\left|\Omega_n\right|} \int_{\Omega_n} \rho_f \,\mathrm{d}\Omega, ~ \forall\mathbf{x}\in\Omega_n, ~ \forall n \in \left\{1, 2, \cdots N\right\} \\
& \left\{\begin{split}
  & \bigcup_{n=1}^N \Omega_n = \Omega \\
  & \Omega_l \bigcap \Omega_m = \emptyset, ~ l \neq m, ~ \forall l \in \left\{1,2,\cdots N\right\}, ~ \forall m \in \left\{1,2,\cdots N\right\}
\end{split}\right.
\end{split}\right. \\
& \rho_p = { \tanh\left(\beta_p \xi_p\right) + \tanh\left(\beta_p \left(\rho_e-\xi_p\right)\right) \over \tanh\left(\beta_p \xi_p\right) + \tanh\left(\beta_p \left(1-\xi_p\right)\right)}, ~ \xi_p \in \left(0,1\right), ~ \beta_p \in \left(0,1 \right] \\
& \left| v - v_0 \right| \leq 1\times10^{-3} ~ \text{with} ~ v = {1\over\left|\Omega\right|} \int_\Omega \rho_p \,\mathrm{d}\Omega
\end{split}\right.
\end{equation}
where $J_0$ is the objective value corresponding to the initial distribution of the design variable in the iterative procedure employed to solve the topology optimization model; and the optimization objective is normalized by $J_0$.

\subsection{Adjoint analysis}\label{sec:TOOPAdjAnalysisLowTempTypeIISupercond}

The topology optimization model in Eq.~\ref{eq:TOOPModelLowTempTypeIISuperconductor} can be solved by using a gradient-based iterative procedure, where the adjoint sensitivities are used to determine the relevant gradient information. The continuous adjoint analysis is implemented for the design objective and volume fraction constraint to derive the adjoint sensitivities \cite{DengJCP2011}. The self-consistency of the adjoint analysis is ensured by splitting the complex order parameter in Section~\ref{sec:SplitTDGLEqus} and implement it on the functional spaces of real functions \cite{DengJCP2018}. Details for the adjoint analysis have been provided in Section~\ref{sec:DetailsAdjointAnalysis} of the appendix.

Based on the continuous adjoint analysis method, the adjoint sensitivity of the optimization objective in Eq.~\ref{eq:ObjFunction} can be derived as
\begin{equation}\label{eq:AdjSensDesignObj}
\begin{split}
\delta J = - \int_\Omega \rho_{fa} \delta \rho \,\mathrm{d}\Omega, ~ \forall \delta \rho \in \mathcal{L}^2\left(\Omega\right),
\end{split}
\end{equation} 
where $\delta$ is the operator for the first-order variational of a functional; $\rho_{fa}$ is the adjoint variable of the filtered design variable and it can be derived by solving the following adjoint equations in variational formulations; and $\mathcal{L}^2 \left(\Omega\right)$ represents the second-order Lebesgue space defined on $\Omega$. The variational formulations for the adjoint equations of the split and gauged time-dependent Ginzburg-Landau equations are
\begin{equation}\label{equ:VariationalFormsAdjEquForOrderParameter} 
\left\{\begin{split}
& \left\{\begin{split}
& \text{Find} ~ \mathbf{A}_a \in \left(\mathcal{H}\left(\left(0,T_t\right);\mathcal{H}\left(\Omega\right)\right)\right)^3 ~ \text{with} ~ \left[\mathbf{A}_a\right]_{t=T_t} = \mathbf{0}~ \mathrm{at} ~ \forall\mathbf{x}\in\Omega \\
& \text{for} ~ \forall \tilde{\mathbf{A}}_a \in \left(\mathcal{H}\left(\left(0,T_t\right);\mathcal{H}\left(\Omega\right)\right)\right)^3, ~ \text{such~that} \\
& \int_0^{T_t} \int_\Omega \left[ {\partial \left|\left[ \mathbf{j}_s \right]_{\forall\mathbf{x}\in\Omega}\right|^2 \over \partial \mathbf{A}} \cdot \tilde{\mathbf{A}}_a + {\partial \left|\left[ \mathbf{j}_s \right]_{\forall\mathbf{x}\in\Omega}\right|^2 \over \partial \nabla \times \mathbf{A}} \cdot \left( \nabla \times \tilde{\mathbf{A}}_a \right) \right] H \left(t-T_m\right) \\
& - \sigma {\partial \mathbf{A}_a \over \partial t} \cdot \tilde{\mathbf{A}}_a + w_p \left( \nabla \times \tilde{\mathbf{A}}_a \right) \cdot \left( \nabla \times \mathbf{A}_a \right) + \left(\psi_r^2 + \psi_i^2\right) \tilde{\mathbf{A}}_a \\
& \cdot \mathbf{A}_a + \tilde{\mathbf{A}}_a \cdot \left[ \psi_i \nabla \left( \kappa^{-1} \psi_{ra} \right) - \psi_r \nabla \left( \kappa^{-1} \psi_{ia} \right) \right] \\
& - \kappa^{-1} \left[ \left( \tilde{\mathbf{A}}_a \cdot \nabla \right) \psi_i \psi_{ra} - \left( \tilde{\mathbf{A}}_a \cdot \nabla \right) \psi_r \psi_{ia} \right] \\
& + 2 \mathbf{A} \cdot \tilde{\mathbf{A}}_a \left( \psi_r \psi_{ra} + \psi_i \psi_{ia} \right) \, \mathrm{d}\Omega\mathrm{d}t = 0,
\end{split}\right. \\
& \left\{\begin{split}
& \left\{\begin{split}
& \text{Find} ~ \psi_{ra} \in \mathcal{H}\left(\left(0,T_t\right);\mathcal{H}\left(\Omega\right)\right) ~ \text{with} ~ \left[\psi_{ra}\right]_{t=T_t} = 0 ~ \mathrm{at} ~ \forall \mathbf{x} \in \Omega \\
& \text{for} ~ \forall \tilde{\psi}_{ra} \in \mathcal{H}\left(\left(0,T_t\right);\mathcal{H}\left(\Omega\right)\right), ~ \text{such that} \\ 
& \int_0^{T_t} \int_\Omega \left( {\partial \left|\left[ \mathbf{j}_s \right]_{\forall\mathbf{x}\in\Omega}\right|^2 \over \partial \psi_r} \tilde{\psi}_{ra} + {\partial \left|\left[ \mathbf{j}_s \right]_{\forall\mathbf{x}\in\Omega}\right|^2 \over \partial \nabla \psi_r} \cdot \nabla \tilde{\psi}_{ra} \right) H \left(t-T_m\right) \\
& - \eta {\partial \psi_{ra} \over \partial t} \tilde{\psi}_{ra} + \kappa^{-2} \nabla \tilde{\psi}_{ra} \cdot \nabla \psi_{ra} + \kappa^{-1} \left( \mathbf{A} \cdot \nabla \right) \tilde{\psi}_{ra} \psi_{ia} + \kappa^{-1} \psi_i \nabla \tilde{\psi}_{ra} \\
& \cdot \mathbf{A}_a + \Big[ \mathbf{A}^2 \psi_{ra} - I_d \psi_{ra} + 2 I_d \psi_r \left( \psi_r \psi_{ra} + \psi_i \psi_{ia} \right) + I_d \left(\psi_r^2 + \psi_i^2\right) \psi_{ra} \\
& - \mathbf{A} \cdot \nabla \left( \kappa^{-1} \psi_{ia} \right) - \kappa^{-1} \nabla \psi_i \cdot \mathbf{A}_a + 2 \psi_r \mathbf{A} \cdot \mathbf{A}_a \Big] \tilde{\psi}_{ra} \,\mathrm{d}\Omega\mathrm{d}t = 0,
\end{split}\right. \\
& \left\{\begin{split}
& \text{Find} ~ \psi_{ia} \in \mathcal{H}\left(\left(0,T_t\right);\mathcal{H}\left(\Omega\right)\right) ~ \text{with} ~ \left[\psi_{ia}\right]_{t=T_t} = 0 ~ \mathrm{at} ~ \forall \mathbf{x} \in \Omega \\
& \text{for} ~ \forall \tilde{\psi}_{ia} \in \mathcal{H}\left(\left(0,T_t\right);\mathcal{H}\left(\Omega\right)\right), ~ \text{such that} \\
& \int_0^{T_t} \int_\Omega \left( {\partial \left|\left[ \mathbf{j}_s \right]_{\forall\mathbf{x}\in\Omega}\right|^2 \over \partial \psi_i} \tilde{\psi}_{ia} + {\partial \left|\left[ \mathbf{j}_s \right]_{\forall\mathbf{x}\in\Omega}\right|^2 \over \partial \nabla \psi_i} \cdot \nabla \tilde{\psi}_{ia} \right) H \left(t-T_m\right) \\
& - \eta {\partial \psi_{ia} \over \partial t} \tilde{\psi}_{ia} + \kappa^{-2} \nabla \tilde{\psi}_{ia} \cdot \nabla \psi_{ia} - \kappa^{-1} \left( \mathbf{A} \cdot \nabla \right) \tilde{\psi}_{ia} \psi_{ra} - \kappa^{-1} \psi_r \nabla \tilde{\psi}_{ia} \\
& \cdot \mathbf{A}_a + \Big[ \mathbf{A}^2 \psi_{ia} - I_d \psi_{ia} + 2 I_d \psi_i \left( \psi_r \psi_{ra} + \psi_i \psi_{ia} \right) + I_d \left(\psi_r^2 + \psi_i^2\right) \psi_{ia} \\
& + \mathbf{A} \cdot \nabla \left( \kappa^{-1} \psi_{ra} \right) + \kappa^{-1} \nabla \psi_r \cdot \mathbf{A}_a + 2 \psi_i \mathbf{A} \cdot \mathbf{A}_a \Big] \tilde{\psi}_{ia} \,\mathrm{d}\Omega\mathrm{d}t = 0;
\end{split}\right.
\end{split}\right.
\end{split}\right.
\end{equation}
and the variational formulation for the adjoint equation of the PDE filter is
\begin{equation}\label{equ:VariationalFormsAdjEquForPDEFilter} 
\left\{\begin{split}
& \text{Find} ~ \rho_{fa} \in \mathcal{H}\left(\Omega\right) ~ \text{for} ~ \forall \tilde{\rho}_{fa} \in \mathcal{H}\left(\Omega\right), ~ \text{such~that} \\
& \int_0^{T_t} \Bigg\{ \sum_{n=1}^N \int_{\Omega_n} {\partial \rho_p \over \partial \rho_e} {\partial \rho_e \over \partial \rho_f} \Bigg[ {\partial \left|\left[ \mathbf{j}_s \right]_{\forall\mathbf{x}\in\Omega}\right|^2 \over \partial \rho_p} H \left(t-T_m\right) \tilde{\rho}_{fa} + {\partial\kappa^{-1} \over \partial \rho_p} \\
& \big\{ 2 \kappa^{-1} \left( \nabla \psi_r \cdot \nabla \psi_{ra} + \nabla \psi_i \cdot \nabla \psi_{ia} \right) + \mathbf{A} \cdot \left( \psi_i \nabla \psi_{ra} - \psi_r \nabla \psi_{ia} \right) \\
& - \left[ \left( \mathbf{A} \cdot \nabla \right) \psi_i \psi_{ra} - \left( \mathbf{A} \cdot \nabla \right) \psi_r \psi_{ia} \right] - \left( \psi_r \nabla \psi_i - \psi_i \nabla \psi_r \right) \cdot \mathbf{A}_a \big\} \tilde{\rho}_{fa} \\
& - {\partial I_d \over \partial \rho_p} \left[ 1 - \left(\psi_r^2 + \psi_i^2\right) \right] \left( \psi_r \psi_{ra} + \psi_i \psi_{ia} \right) \tilde{\rho}_{fa} + {\partial w_p \over \partial \rho_p} \left( \nabla \times \mathbf{A} - \mathbf{H} \right) \\
& \cdot \left( \nabla \times \mathbf{A}_a \right) \tilde{\rho}_{fa} \Bigg] \,\mathrm{d}\Omega \Bigg\} \mathrm{d}t + \int_\Omega r_\rho^2 \nabla \tilde{\rho}_{fa} \cdot \nabla \rho_{fa} + \rho_{fa} \tilde{\rho}_{fa} \,\mathrm{d}\Omega = 0,
\end{split}\right.
\end{equation}
where $\psi_{ra}$, $\psi_{ia}$, $\mathbf{A}_a$ and $\rho_{fa}$ are the adjoint variables of $\psi_r$, $\psi_i$, $\mathbf{A}$ and $\rho_f$, respectively; $\tilde{\psi}_{ra}$, $\tilde{\psi}_{ia}$, $\tilde{\mathbf{A}}_a$ and $\tilde{\rho}_{fa}$ are the test functions of $\psi_{ra}$, $\psi_{ia}$, $\mathbf{A}_a$ and $\rho_{fa}$, respectively; $\mathcal{H} \left(\Omega\right)$ represents the first-order Sobolev space defined on $\Omega$; $\mathcal{H}\left(\left(0,T_t\right);\mathcal{H}\left(\Omega\right)\right)$ represents the Hilbert space defined as
\begin{equation}
\begin{split}
  \mathcal{H}\left(\left(0,T_t\right);\mathcal{H}\left(\Omega\right)\right) = \bigg\{ u\left(t,\mathbf{x}\right) : &  \left[u\left(t,\mathbf{x}\right)\right]_{\mathbf{x}=\mathbf{x}_0} \in \mathcal{H}\left(\left(0,T_t\right)\right), \\
  & \left[u \left(t,\mathbf{x}\right)\right]_{t=t_0} \in \mathcal{H}\left(\Omega\right), \\
  & {\partial u \over \partial t} \in \mathcal{L}^2\left(\left(0,T_t\right); \mathcal{H}\left(\Omega\right)\right); \\
  & \forall \left( \mathbf{x}_0, t_0\right) \in \Omega \times \left(0,T_t\right) \bigg\},
\end{split}
\end{equation}
with $\mathcal{L}^2\left(\left(0,T_t\right); \mathcal{H}\left(\Omega\right)\right)$ defined as
\begin{equation}
\begin{split}
  \mathcal{L}^2\left(\left(0,T_t\right); \mathcal{H}\left(\Omega\right)\right) = \big\{ u\left(t,\mathbf{x}\right) : & \left[u\left(t,\mathbf{x}\right)\right]_{\mathbf{x}=\mathbf{x}_0} \in \mathcal{L}^2\left(\left(0,T_t\right)\right), \\
  & \left[u \left(t,\mathbf{x}\right)\right]_{t=t_0} \in \mathcal{H}\left(\Omega\right); \\
  & \forall \left( \mathbf{x}_0, t_0\right) \in \Omega \times \left(0,T_t\right) \big\},
\end{split}
\end{equation}
and $\mathcal{H} \left(\left(0,T_t\right)\right)$ and $\mathcal{L}^2 \left(\left(0,T_t\right)\right)$ representing the first-order Sobolev space and second-order Lebesgue space defined on $\left(0,T_t\right)$, respectively.

The adjoint sensitivity of the volume fraction in Eq.~\ref{eq:VolumeFraction} is
\begin{equation}\label{eq:AdjSensVolumeFraction}
\begin{split}
\delta v = - \int_\Omega \rho_{fa} \delta \rho \,\mathrm{d}\Omega, ~ \forall \delta \rho \in \mathcal{L}^2\left(\Omega\right),
\end{split}
\end{equation}
where the adjoint variable $\rho_{fa}$ can be derived by solving the variational formulation for the adjoint equation of the PDE filter as
\begin{equation}\label{eq:VariationalFormAdjEquVolumeFraction}
\left\{\begin{split}
& \text{Find} ~ \rho_{fa} \in \mathcal{H}\left(\Omega\right) ~ \text{for} ~ \forall \tilde{\rho}_{fa} \in \mathcal{H}\left(\Omega\right), ~ \text{such~that} \\
& \sum_{n=1}^N \int_{\Omega_n} {\partial \rho_p \over \partial \rho_e} {\partial \rho_e \over \partial \rho_f} \tilde{\rho}_{fa} \,\mathrm{d}\Omega + \int_\Omega r_\rho^2 \nabla \tilde{\rho}_{fa} \cdot \nabla \rho_{fa} + \rho_{fa} \tilde{\rho}_{fa} \,\mathrm{d}\Omega = 0.
\end{split}\right.
\end{equation}

After the derivation of the adjoint sensitivities in Eqs. \ref{eq:AdjSensDesignObj} and \ref{eq:AdjSensVolumeFraction}, the design variable can be evolved iteratively to solve the topology optimization model for low-temperature type-II superconductors.

\subsection{Numerical implementation}\label{sec:NumericalImplementationLowTempTypeIISupercond}

\begin{figure}[!htbp]
  \centering
  \subfigure[]
  {\includegraphics[height=0.2\textwidth]{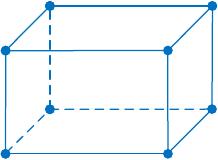}}\hspace{4em}
  \subfigure[]
  {\includegraphics[height=0.2\textwidth]{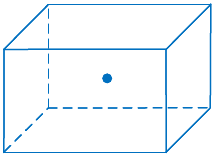}}
  \caption{Sketches for the first-order nodal elements and zeroth-order discontinuous element in the hexahedron shapes: (a) First-order nodal element; (b) Zeroth-order discontinuous element.}\label{fig:NodalEdgeElementSketches}
\end{figure}

The topology optimization model in Eq.~\ref{eq:TOOPModelLowTempTypeIISuperconductor} is solved by using an iterative procedure described as the algorithm in Tab.~\ref{tab:IterativeProcedureSurfaceFlow}, where a loop is included for the iterative solution. The algorithm is implemented in the Matlab environment combined with the finite element software COMSOL Multiphysics. The nodal element based finite element method is utilized to solve the variational formulations of the relevant PDEs and adjoint equations for the order parameter, vector potential and PDE filter in Eqs. \ref{equ:VariationalFormsAdjEquForOrderParameter}, \ref{equ:VariationalFormsAdjEquForPDEFilter}, \ref{eq:VariationalFormAdjEquVolumeFraction} and \ref{equ:VariationalFormsSplitTDGLOrderParameter} \cite{ReddyMcGraw2006}. 
The first-order nodal elements in the hexahedron shape sketched in Fig.~\ref{fig:NodalEdgeElementSketches}a are used to discretize the order parameter, vector potential, filtered design variable and their adjoint variables. Especially, the piecewise homogenization of the filtered design variable is implemented by using the zeroth-order discontinuous elements sketched in Fig.~\ref{fig:NodalEdgeElementSketches}b, where the non-overlapping pieces are the finite elements, i.e., Eq.~\ref{eq:PieceDivision} is implemented on every element of the meshes used to discretize the design domain. The 5-step backward differentiation formula (BDF) is utilized to discretize the time derivatives in the variational formulations \cite{AscherPetzoldSIAM1998}. The split and gauged time-dependent Ginzburg-Landau equations are initial-value problems, for which the time discretization is implemented from $0$ to $T_t$. Meanwhile, the adjoint equations of the split and gauged time-dependent Ginzburg-Landau equations are terminal-value problem, for which the time discretization is implemented in the reverse direction from $T_t$ to $0$.

In the iterative procedure, the radius of the PDE filter is set as triple of the element size; the projection parameter $\beta_p$ with the initial value of $1$ is doubled after every $30$ iterations; the loop is stopped when the maximal iteration number is reached, or if the averaged variation of the normalized optimization objective in continuous 5 iterations and the residual of the volume fraction constraint are simultaneously less than the specified tolerance $10^{-3}$ chosen to be much less than $1$. The design variable is updated by using the method of moving asymptotes \cite{SvanbergIntJNumerMethodsEng1987}.

In the numerical implementation, the smoothed Heaviside function is used to approximate the Heaviside function in Eq.~\ref{equ:HeavisideFunctionTime} by using the third-order polynomial of $t-T_m$:
\begin{equation}\label{eq:ApproximatedHeavisideFunction}
  H_s\left( t-T_m, h_t\right) = \left\{\begin{split}
  & 1, ~ t \in \left[T_m + h_t, T_t \right), \\
  & {1\over2} + {3 \over 4} \left({t-T_m \over h_t}\right) - {1 \over 4}\left({t-T_m \over h_t}\right)^3, ~ t \in \left(T_m - h_t, T_m + h_t \right), \\
  & 0, ~ t \in \left(0, T_m - h_t \right],
  \end{split}\right.
\end{equation}
where $H_s\left( t-T_m, h_t\right)$ is the smoothed Heaviside function; $h_t$ is the size of the support set and it is set as $\Delta t/4$ with $\Delta t = {T_t / N_t}$ denoting the specified time step in the discretization of the time derivatives and $N_t$ denoting a positive integer chosen to be large enough to ensure the numerical accuracy. 

In the calculation of the objective value, the time-dependent equation of an auxiliary variable as 
\begin{equation}\label{eq:AuxiliaryEqForJ}
  {\partial u \over \partial t} = \left|\left[ \mathbf{j}_s \right]_{\forall\mathbf{x}\in\Omega}\right|^2 H_s\left( t-T_m, h_t\right), ~ \forall \left(\mathbf{x},t\right) \in \Omega \times \left(0, T_t\right)
\end{equation}
is introduced and its variational formulation
\begin{equation}
\left\{\begin{split}
  & \text{Find} ~ u \in \mathcal{H} \left(\left(0,T_t\right); \mathcal{L}^2\left(\Omega\right)\right) ~\text{for}~ \forall \tilde{u} \in \mathcal{L}^2 \left(\left(0,T_t\right); \mathcal{L}^2\left(\Omega\right)\right), \\
  & \text{such~that} \int_\Omega \left[ {\partial u \over \partial t} - \left|\left[ \mathbf{j}_s \right]_{\forall\mathbf{x}\in\Omega}\right|^2 H_s\left( t-T_m, h_t\right) \right] \tilde{u} \,\mathrm{d}\Omega = 0
\end{split}\right. 
\end{equation}
is solved to implement the time integration in Eq.~\ref{eq:ObjFunction} and derive the time integration as $\left[u\right]_{t=T}$, where $u$ is the auxiliary variable; $\mathcal{H} \left(\left(0,T_t\right); \mathcal{L}^2\left(\Omega\right)\right)$ is the functional space defined as
\begin{equation}
\begin{split}
  \mathcal{H} \left(\left(0,T_t\right); \mathcal{L}^2\left(\Omega\right)\right) = \bigg\{ u\left(t,\mathbf{x}\right) : &  \left[u\left(t,\mathbf{x}\right)\right]_{\mathbf{x}=\mathbf{x}_0} \in \mathcal{H}\left(\left(0,T_t\right)\right), \\
  & \left[u \left(t,\mathbf{x}\right)\right]_{t=t_0} \in \mathcal{L}^2\left(\Omega\right); \\
  & \forall \left( \mathbf{x}_0, t_0\right) \in \Omega \times \left(0,T_t\right) \bigg\}
\end{split}
\end{equation}
and $\mathcal{L}^2 \left(\left(0,T_t\right); \mathcal{L}^2\left(\Omega\right)\right)$ is the functional space defined as
\begin{equation}
\begin{split}
  \mathcal{L}^2 \left(\left(0,T_t\right); \mathcal{L}^2\left(\Omega\right)\right) = \bigg\{ u\left(t,\mathbf{x}\right) : &  \left[u\left(t,\mathbf{x}\right)\right]_{\mathbf{x}=\mathbf{x}_0} \in \mathcal{L}^2\left(\left(0,T_t\right)\right), \\
  & \left[u \left(t,\mathbf{x}\right)\right]_{t=t_0} \in \mathcal{L}^2\left(\Omega\right); \\
  & \forall \left( \mathbf{x}_0, t_0\right) \in \Omega \times \left(0,T_t\right) \bigg\}.
\end{split}
\end{equation}
Then, the objective value can be derived by further implementing the spatial integration of $\left[u\right]_{t=T_t}$ on $\Omega$:
\begin{equation}\label{eq:AuxiliaryIntForJ}
  J = \int_\Omega \left[u\right]_{t=T_t} \,\mathrm{d}\Omega.
\end{equation}

In the solution of the variational formulation for the adjoint equation in Eq.~\ref{equ:VariationalFormsAdjEquForPDEFilter}, the time-dependent equation of the auxiliary variable as
\begin{equation}\label{equ:TimeIntegrationTerm}
\begin{split}
{\partial u \over \partial t} = & {\partial \rho_p \over \partial \rho_e} {\partial \rho_e \over \partial \rho_f} \Bigg\{ {\partial \left|\left[ \mathbf{j}_s \right]_{\forall\mathbf{x}\in\Omega}\right|^2 \over \partial \rho_p} H_s\left( t-T_m, h_t \right) \tilde{\rho}_{fa} \\
& + {\partial\kappa^{-1} \over \partial \rho_p} \big\{ 2 \kappa^{-1} \left( \nabla \psi_r \cdot \nabla \psi_{ra} + \nabla \psi_i \cdot \nabla \psi_{ia} \right) + \mathbf{A} \\
& \cdot \left( \psi_i \nabla \psi_{ra} - \psi_r \nabla \psi_{ia} \right) - \left[ \left( \mathbf{A} \cdot \nabla \right) \psi_i \psi_{ra} - \left( \mathbf{A} \cdot \nabla \right) \psi_r \psi_{ia} \right] \\
& - \left( \psi_r \nabla \psi_i - \psi_i \nabla \psi_r \right) \cdot \mathbf{A}_a \big\} \tilde{\rho}_{fa} - {\partial I_d \over \partial \rho_p} \left[ 1 - \left(\psi_r^2 + \psi_i^2\right) \right] \\
& \left( \psi_r \psi_{ra} + \psi_i \psi_{ia} \right) \tilde{\rho}_{fa} + {\partial w_p \over \partial \rho_p} \left( \nabla \times \mathbf{A} - \mathbf{H} \right) \cdot \left( \nabla \times \mathbf{A}_a \right) \tilde{\rho}_{fa} \Bigg\}, \\
& \forall \left(\mathbf{x},t\right) \in \Omega \times \left(0, T_t\right)
\end{split}
\end{equation}
is introduced, and its variational formulation 
\begin{equation}
\left\{\begin{split}
& \text{Find} ~ u \in \mathcal{H} \left(\left(0,T_t\right); \mathcal{L}^2\left(\Omega\right)\right) ~\text{for}~ \forall\tilde{u} \in \mathcal{L}^2 \left(\left(0,T_t\right); \mathcal{L}^2\left(\Omega\right)\right), ~ \text{such~that}  \\
& \int_\Omega {\partial u \over \partial t} \tilde{u} \,\mathrm{d}\Omega - \sum_{n=1}^N \int_{\Omega_n} {\partial \rho_p \over \partial \rho_e} {\partial \rho_e \over \partial \rho_f} \Bigg\{ {\partial \left|\left[ \mathbf{j}_s \right]_{\forall\mathbf{x}\in\Omega}\right|^2 \over \partial \rho_p} H_s\left( t-T_m, h_t \right) \tilde{\rho}_{fa} \\
& + {\partial\kappa^{-1} \over \partial \rho_p} \big\{ 2 \kappa^{-1} \left( \nabla \psi_r \cdot \nabla \psi_{ra} + \nabla \psi_i \cdot \nabla \psi_{ia} \right) + \mathbf{A} \cdot \left( \psi_i \nabla \psi_{ra} - \psi_r \nabla \psi_{ia} \right) \\
& - \left[ \left( \mathbf{A} \cdot \nabla \right) \psi_i \psi_{ra} - \left( \mathbf{A} \cdot \nabla \right) \psi_r \psi_{ia} \right] - \left( \psi_r \nabla \psi_i - \psi_i \nabla \psi_r \right) \cdot \mathbf{A}_a \big\} \tilde{\rho}_{fa} \\
& - {\partial I_d \over \partial \rho_p} \left[ 1 - \left(\psi_r^2 + \psi_i^2\right) \right] \left( \psi_r \psi_{ra} + \psi_i \psi_{ia} \right) \tilde{\rho}_{fa} + {\partial w_p \over \partial \rho_p} \left( \nabla \times \mathbf{A} - \mathbf{H} \right) \\
& \cdot \left( \nabla \times \mathbf{A}_a \right) \tilde{\rho}_{fa} \Bigg\} \tilde{u} \,\mathrm{d}\Omega = 0
\end{split}\right. 
\end{equation}
is solved to derive the time integration as $\left[u\right]_{t=T_t}$. Then, the adjoint variable $\rho_{fa}$ in Eq.~\ref{equ:VariationalFormsAdjEquForPDEFilter} can be derived by numerically solving
\begin{equation}
\left\{\begin{split}
& \text{Find} ~ \rho_{fa} \in \mathcal{H}\left(\Omega\right) ~ \text{for} ~ \forall \tilde{\rho}_{fa} \in \mathcal{H}\left(\Omega\right), ~ \text{such~that} \\
& \int_\Omega \left[u\right]_{t=T_t} \tilde{\rho}_{fa} + r_\rho^2 \nabla \tilde{\rho}_{fa} \cdot \nabla \rho_{fa} + \rho_{fa} \tilde{\rho}_{fa} \,\mathrm{d}\Omega = 0.
\end{split}\right.
\end{equation}

\begin{table}[!htbp]
\centering
\begin{tabular}{l}
  \hline
  \textbf{Algorithm}: iterative solution of Eq.~\ref{eq:TOOPModelLowTempTypeIISuperconductor} \\
  \hline
  Set $\psi_{r0}$, $\psi_{i0}$, $\mathbf{A}_0$, $T_t$, $T_m$ and $q$;\\
  Set $\left\{
  \begin{array}{l}
    \rho \leftarrow v_0 \\
    n_{\max} \leftarrow 315 \\
    n_i \leftarrow 1 \\
    \xi_p \leftarrow 0.5 \\
    \beta_p \leftarrow 1
  \end{array}
  \right.$; \\
  \textbf{loop} \\
          \hspace{1em} Solve Eq.~\ref{eq:PDEFilter} to derive $\rho_f$ by filtering $\rho$; \\
          \hspace{1em} Piecewisely homogenize $\rho_f$ to derive $\rho_e$ based on Eq.~\ref{eq:PiecewiseHomogenization}; \\
          \hspace{1em} Project $\rho_e$ to derive $\rho_p$ based on Eq.~\ref{eq:ThresholdProjection}; \\
          \hspace{1em} Solve $\psi_r$, $\psi_i$ and $\mathbf{A}$ from Eq. \ref{equ:VariationalFormsSplitTDGLOrderParameter}; \\
          \hspace{1em} Compute $J$ in Eq.~\ref{eq:ObjFunction} by solving Eqs. \ref{eq:AuxiliaryIntForJ} and \ref{eq:AuxiliaryEqForJ}; \\
          \hspace{1em} Solve $\mathbf{A}_a$, $\psi_{ra}$ and $\psi_{ia}$ from Eq. \ref{equ:VariationalFormsAdjEquForOrderParameter}; \\
          \hspace{1em} Solve $\rho_{fa}$ from Eq.~\ref{equ:VariationalFormsAdjEquForPDEFilter} with the auxiliary solved from Eq.~\ref{equ:TimeIntegrationTerm}; \\
          \hspace{1em} Compute $\delta J $ from Eq.~\ref{eq:AdjSensDesignObj}; \\
          \hspace{1em} Solve $\rho_{fa}$ from Eq.~\ref{eq:VariationalFormAdjEquVolumeFraction};\\
          \hspace{1em} Compute $\delta v$ from Eq.~\ref{eq:AdjSensVolumeFraction}; \\
          \hspace{1em} Update $\rho$ based on $\delta J$ and $\delta v$; \\
          \hspace{1em} \textbf{if} $n_i==1$ \\
          \hspace{2em} $J_0 \leftarrow J$; \\
          \hspace{1em} \textbf{end} \textbf{if} \\
          \hspace{1em} \textbf{if} $\mod\left(n_i,30\right)==0$ \\
          \hspace{2em} $\beta_p \leftarrow 2\beta_p$; \\
          \hspace{1em} \textbf{end} \textbf{if} \\
          \hspace{1em} \textbf{if} $ \left( n_i==n_{\max} \right) $ or $\left\{
          \begin{array}{l}
            \beta_p == 2^{10} \\
            {1\over5}\sum_{m=0}^4 \left| J_{n_i} - J_{n_i-m} \right|\Big/J_0 \leq 10^{-3} \\
            \left|v-v_0\right| \leq 10^{-3}
          \end{array}
          \right.$ \\
          \hspace{2em} break; \\
          \hspace{1em} \textbf{end} \textbf{if} \\
          \hspace{1em} $n_i \leftarrow n_i+1$ \\
  \textbf{end} \textbf{loop} \\
  \hline
\end{tabular}
\caption{Pseudocode used to solve the topology optimization model in Eq.~\ref{eq:TOOPModelLowTempTypeIISuperconductor}. In the iterative solution loop, $n_i$ is the loop-index, $n_{\max}$ is the maximal value of $n_i$, $J_{n_i}$ is the value of $J$ in the $n_i$-th iteration, and $\mod$ is the operator used to take the remainder. In this paper, the terminal value $2^{10}$ of the projection parameter $\beta_p$ is used to make the material interface to be clear enough.}\label{tab:IterativeProcedureSurfaceFlow}
\end{table}

\section{Results and discussion} \label{sec:ResultsDiscussionTOOPSuperCond}

\begin{figure}[!htbp]
  \centering
  \includegraphics[width=0.7\textwidth]{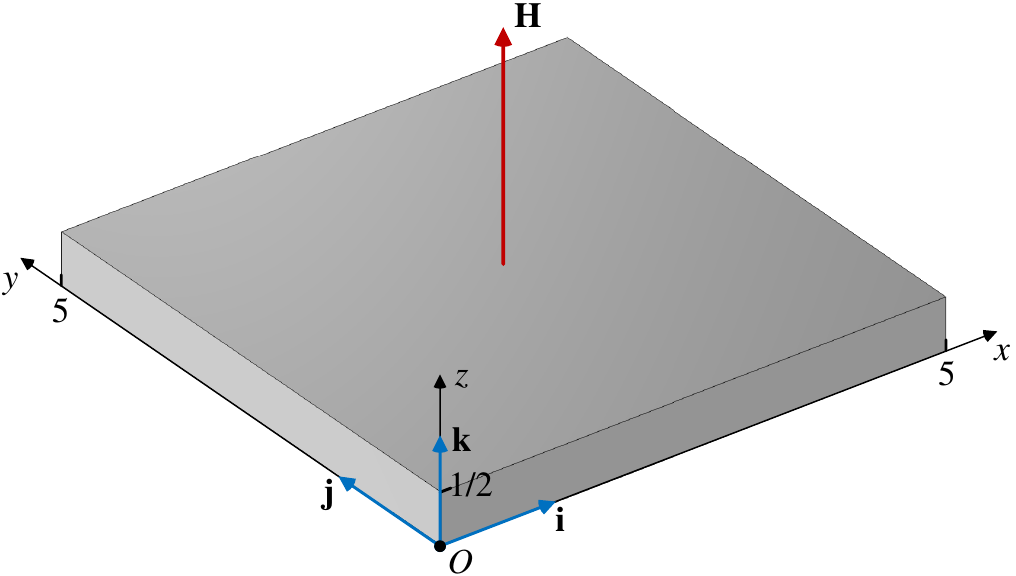}
  \caption{Sketch for the design domain. In the sketch, the sizes of the design domain are set as $5$, $5$ and $1/2$ in the unit of the penetration depth for the width, length and thickness, respectively; and $\mathbf{i}$, $\mathbf{j}$ and $\mathbf{k}$ represent the unitary orientational vectors of the three coordinate axes.}\label{fig:SketchDesignDom}
\end{figure}

In this section, numerical results are presented to demonstrate topology optimization of low-temperature type-II superconductors. The design domain is set as sketched in Fig.~\ref{fig:SketchDesignDom}, where the size unit is the penetration depth and the Cartesian system is defined as $O$-$xyz$ with $\mathbf{i}$, $\mathbf{j}$ and $\mathbf{k}$ representing the unitary orientational vectors of the three coordinate axes, respectively. The structured meshes with cubic elements are used to discretize the design domain. The optimization parameters, and initial values of the order parameter and vector potential are listed in Tab.~\ref{tab:NumericalOptimizationParameters}. In Tab.~\ref{tab:NumericalOptimizationParameters}, the $q$ parameter in the material interpolations is chosen as $1\times10^{-2}$ based on numerical experiments; $T_m$ is chosen as $45$ which is large enough to reach the steady state of the order parameter; and the initial value of the order parameter is set corresponding to the Meissner state with $\left|\psi_0\right|=1$. In order to ensure monotonic convergence of the objective values in an iterative procedure, symmetry constraints are imposed to enforce the $D_{4h}$ symmetries of the material density in the design domain.

\begin{table}[!htbp]
\centering
\begin{tabular}{ccccccccccccc}
  $\kappa_s$ & $q$ & $r_\rho$ & $h_E$ & $\eta$ & $\sigma$ & $\gamma$ & $T_m$  & $T_t$ & $h_t$  & $\psi_{r0}$  & $\psi_{i0}$ & $\mathbf{A}_0$ \\
 \midrule
  $ 4 $      & $1\times10^{-2}$    & $3/10$   & $1/10$   & $1$    & $1$    & $0$      & $45$ & $50$  & $1/40$ & $1$ & $0$      & $\mathbf{0}$
\end{tabular}
\caption{Optimization parameters and initial values of the order parameter and vector potential used in the numerical solution of the topology optimization model, where $h_E$ is the size of the finite elements used to discretize the design domain. In reality, a dirty or technical Vanadium has a Ginzburg-Landau parameter around $4$, with a penetration depth and coherence length around $80$ nm and $20$ nm, respectively.}\label{tab:NumericalOptimizationParameters}
\end{table}

\begin{figure}[!htbp]
  \centering
  \subfigure[$v_0=0.1$]
  {\includegraphics[width=0.18\textwidth]{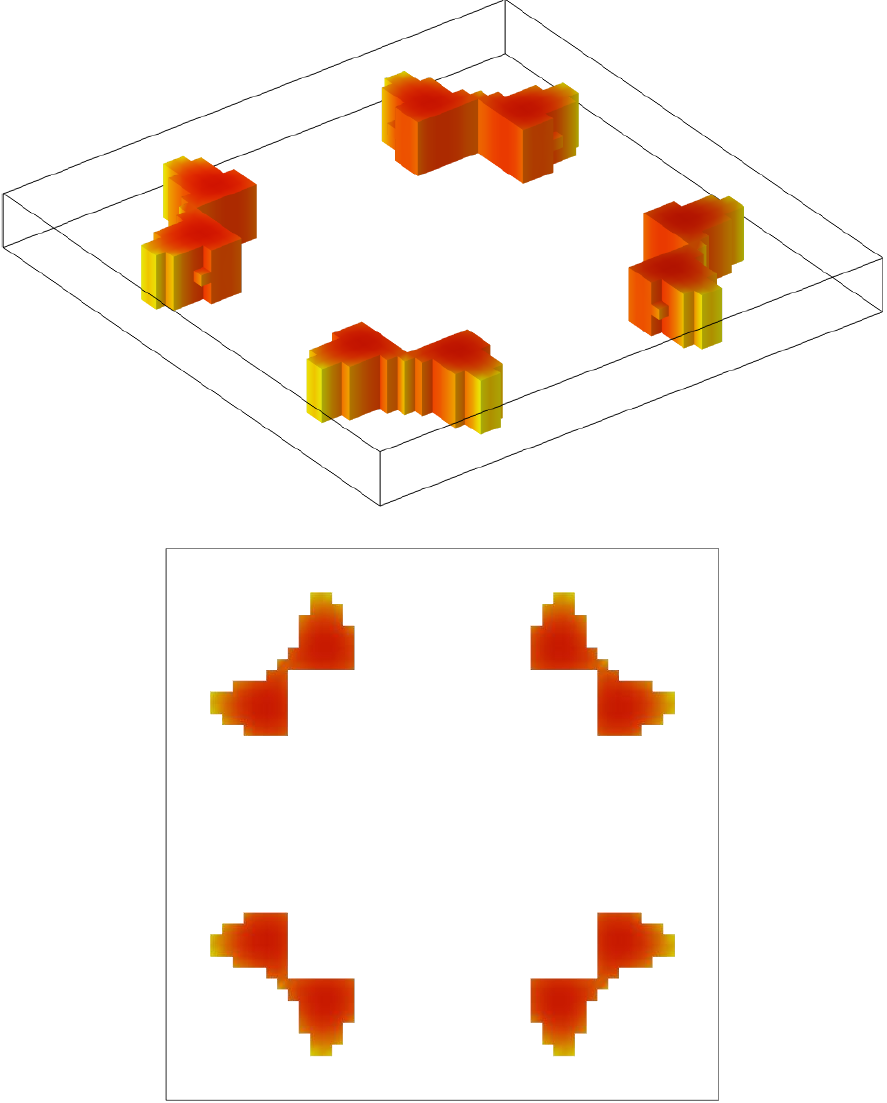}}
  \subfigure[$v_0=0.3$]
  {\includegraphics[width=0.18\textwidth]{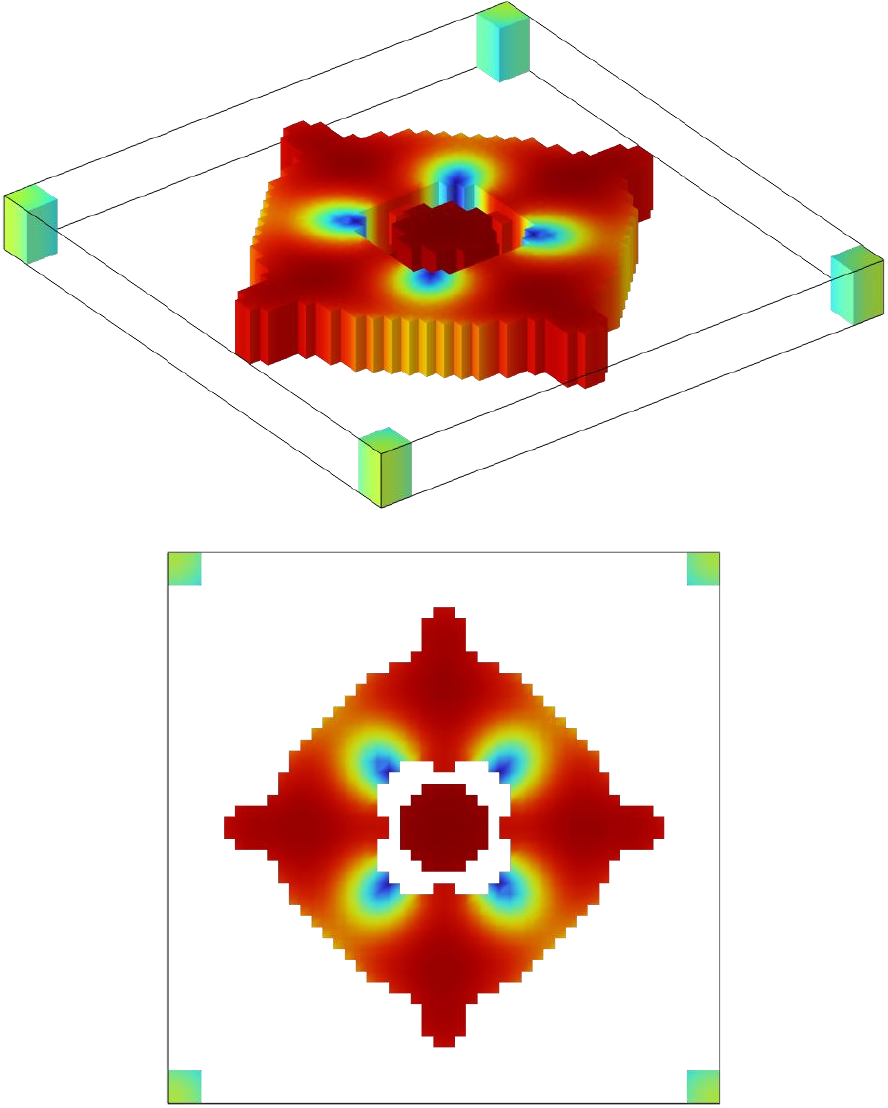}}
  \subfigure[$v_0=0.5$]
  {\includegraphics[width=0.18\textwidth]{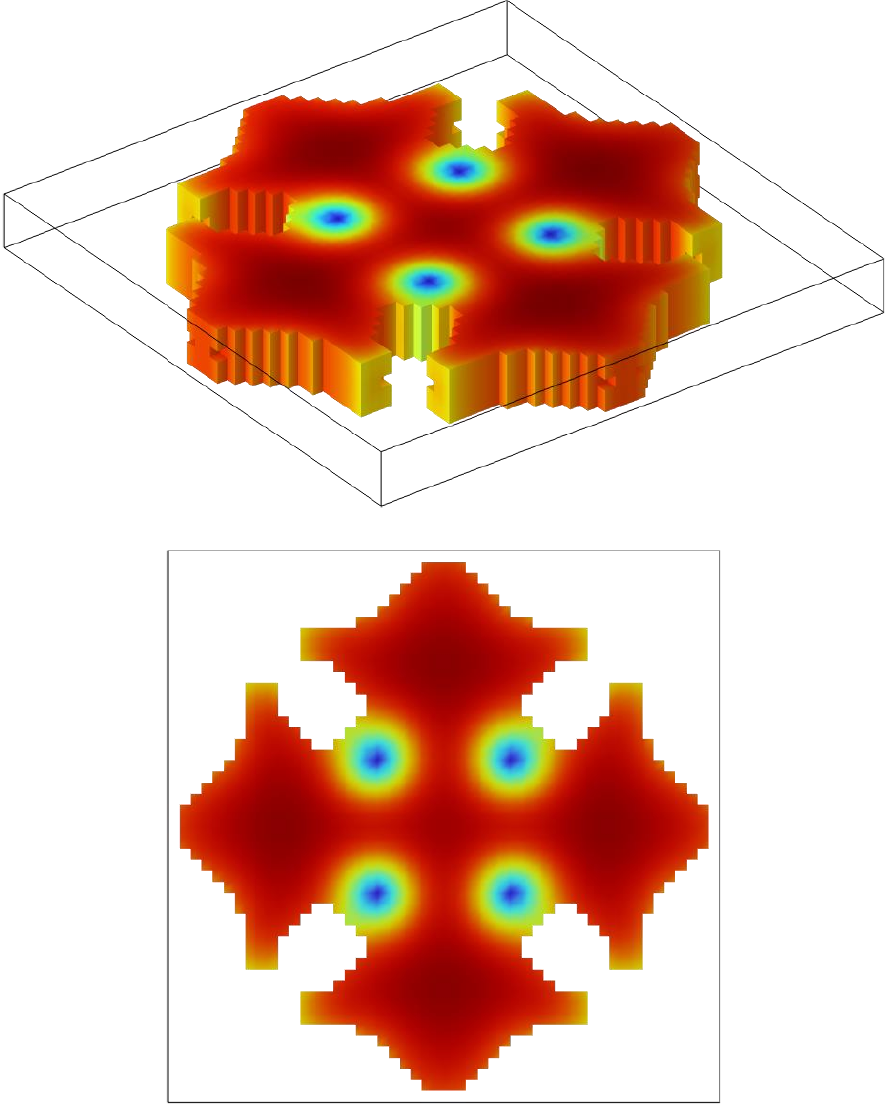}}
  \subfigure[$v_0=0.7$]
  {\includegraphics[width=0.18\textwidth]{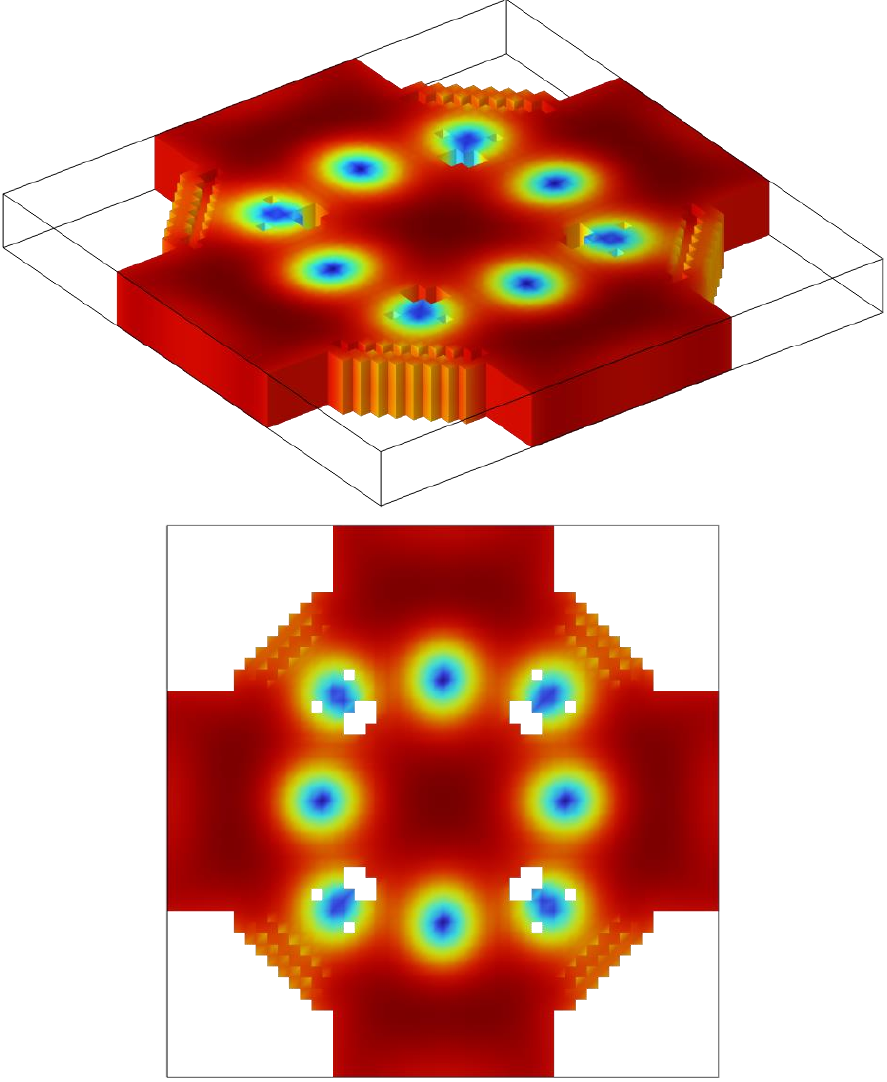}}
  \subfigure[$v_0=0.9$]
  {\includegraphics[width=0.18\textwidth]{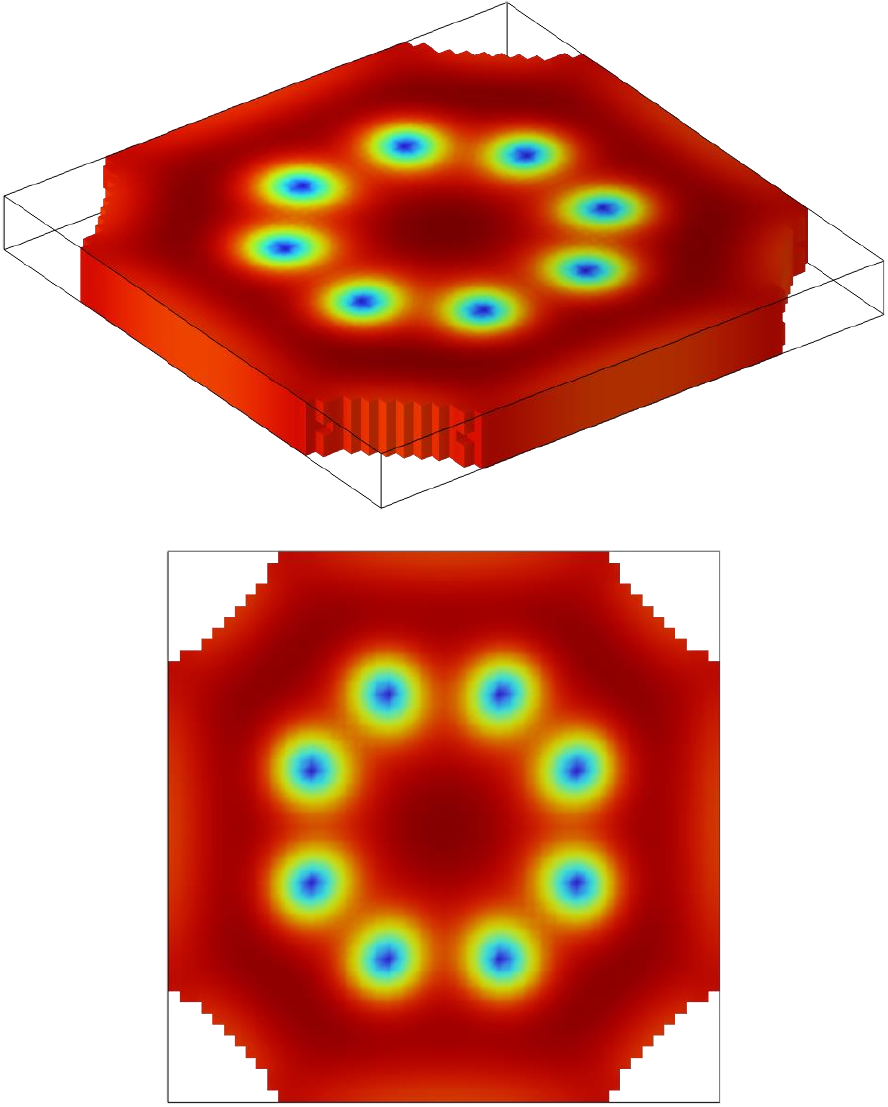}}
  \includegraphics[height=0.22\textwidth]{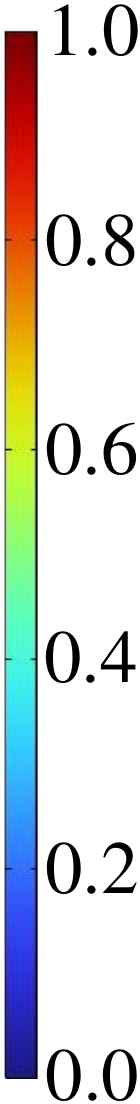} 
  \caption{Stereo and top views of the optimized topologies together with the modular distribution of the order parameter at the terminal time for different volume fractions in topology optimization of low-temperature type-II superconductors.}\label{fig:LHTS_vf_H3=12}
\end{figure}

\begin{figure}[!htbp]
  \centering
  \includegraphics[width=0.85\textwidth]{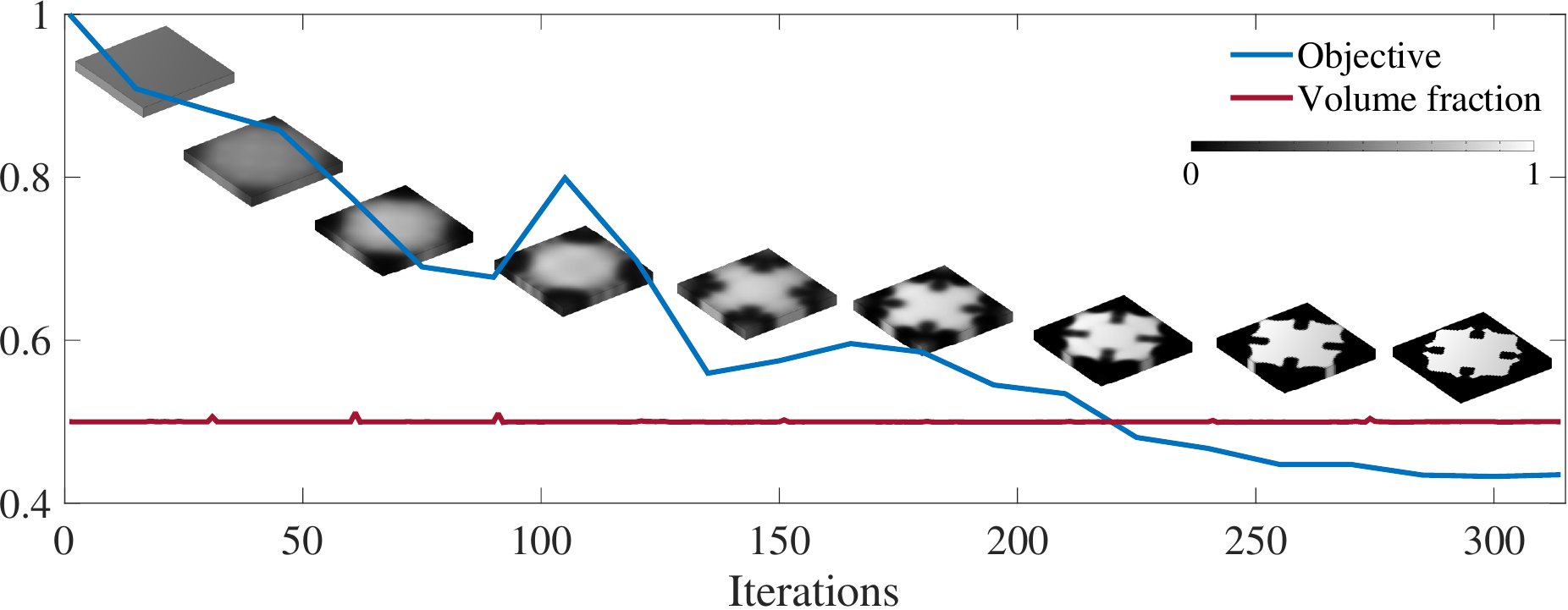}
  \caption{Convergence histories of the normalized objective values and volume fractions for the optimized topologies of low-temperature type-II superconductors in Fig.~\ref{fig:LHTS_vf_H3=12}c, where snapshots for the iterative evolutions of the material density are included.}\label{fig:LHTSConvergentHistories}
\end{figure}

\begin{figure}[!htbp]
  \centering
  \includegraphics[width=0.25\textwidth]{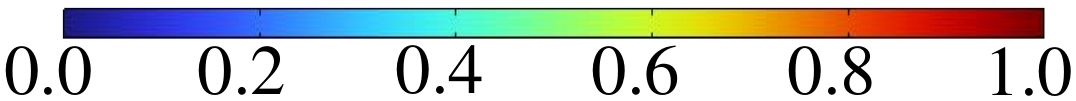} \vspace{0.3em} \\
  \subfigure[$v_0=0.1$]
  {\includegraphics[width=1\textwidth]{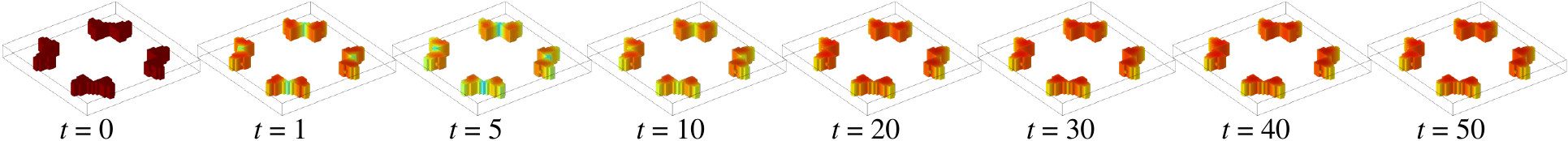}}
  \subfigure[$v_0=0.3$]
  {\includegraphics[width=1\textwidth]{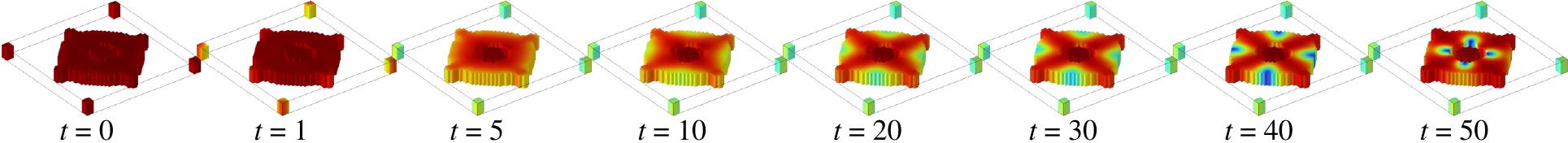}}
  \subfigure[$v_0=0.5$]
  {\includegraphics[width=1\textwidth]{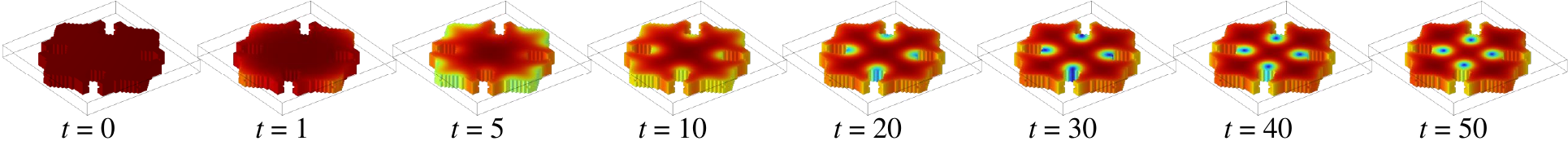}}
  \subfigure[$v_0=0.7$]
  {\includegraphics[width=1\textwidth]{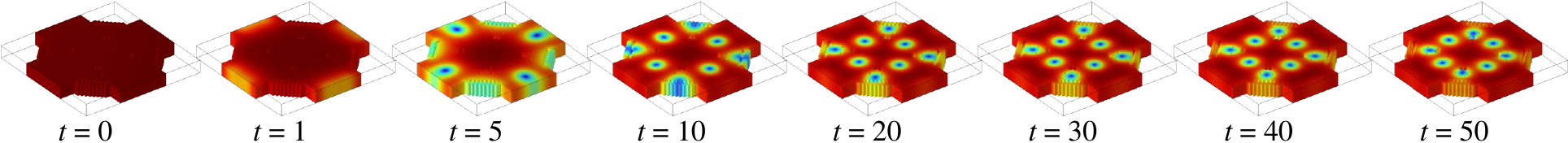}}
  \subfigure[$v_0=0.9$]
  {\includegraphics[width=1\textwidth]{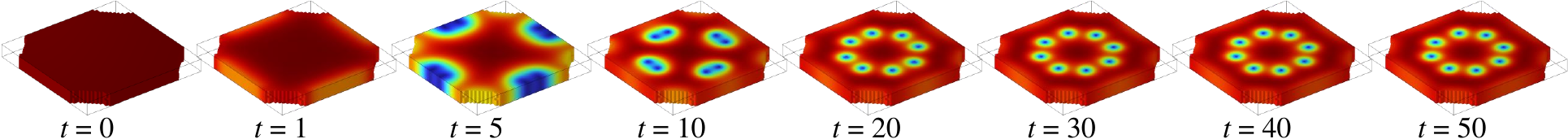}}
  \subfigure[Unoptimized]
  {\includegraphics[width=1\textwidth]{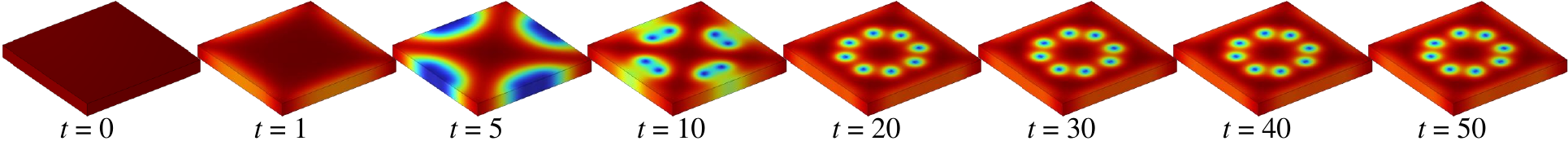}}
  \caption{Snapshots for the temporal evolution of the modular distribution of the order parameter at the terminal time in the optimized topologies in Fig.~\ref{fig:LHTS_vf_H3=12}(a-e) for low-temperature type-II superconductors with the volume fractions of $0.1$, $0.3$, $0.5$, $0.7$ and $0.9$, respectively, where snapshots for the temporal evolution of the order parameter in the unoptimized low-temperature type-II superconductor are included.}\label{fig:TimeEvolution_LTS_OrderParameter}
\end{figure}

Based on the above setting of the design domain and optimization parameters, the optimized topologies together with the modular distribution of the order parameter for different volume fractions are derived as shown in Fig.~\ref{fig:LHTS_vf_H3=12} for low-temperature type-II superconductors, where the applied magnetic field is set as $\mathbf{H} = 1.2\mathbf{k}$. 
The convergence histories of the objective values and volume fractions are provided in Fig.~\ref{fig:LHTSConvergentHistories} for the optimized topologies in Fig.~\ref{fig:LHTS_vf_H3=12}c, where snapshots for the iterative evolution of the material density are included. From monotonicity of the objective values and satisfication of the volume fraction constraints, it can be concluded that the iterative algorithm used to solve the topology optimization model is robust. Snapshots for the temporal evolution of the order-parameter modulus in the optimized topologies with different volume fractions are provided in Fig.~\ref{fig:TimeEvolution_LTS_OrderParameter}. In the temporal-evolution snapshots, the optimized topologies can effectively delay the entry of flux lines; the optimized topologies with smaller volume fraction can achieve more delay; and the entry of flux lines can even be completely prevented by the optimized topologies, when the volume fraction is set to be small enough.

\begin{figure}[!htbp]
  \centering
  \subfigure[$v_0=0.1$]
  {\includegraphics[width=0.3\textwidth]{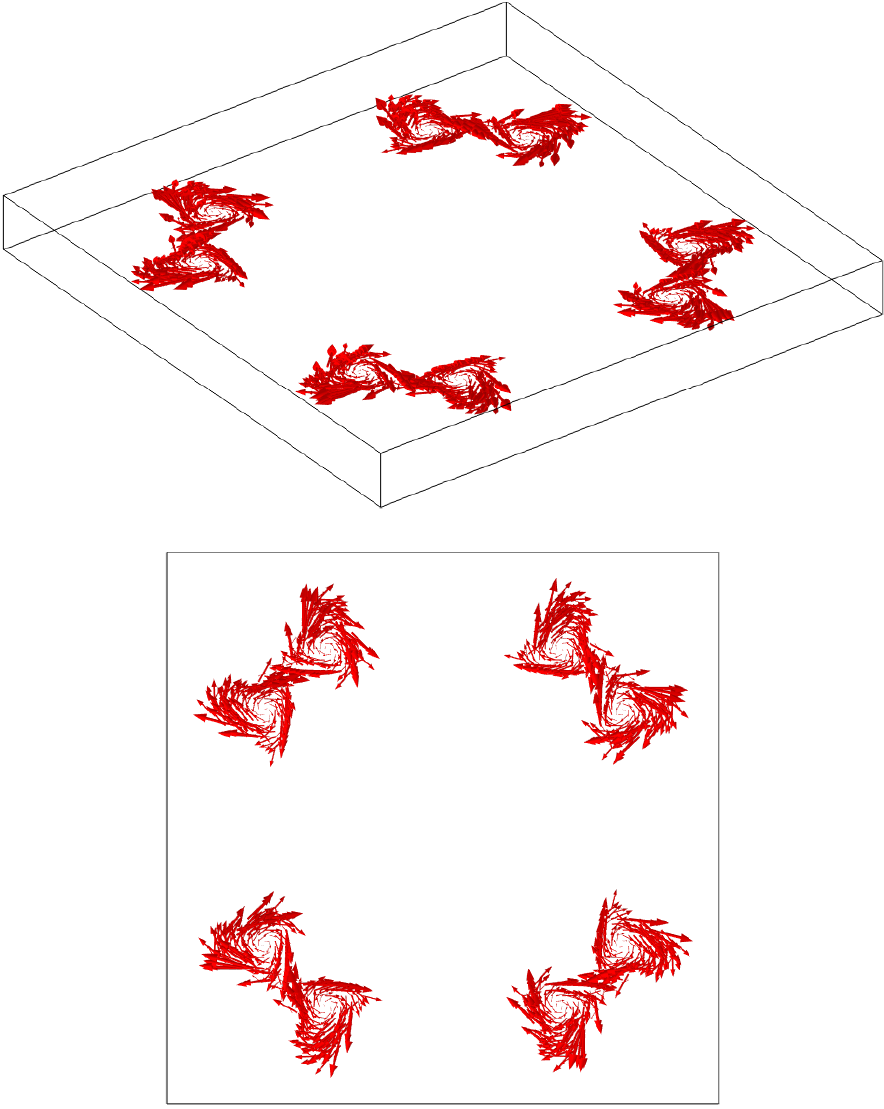}} \hspace{1em}
  \subfigure[$v_0=0.3$]
  {\includegraphics[width=0.3\textwidth]{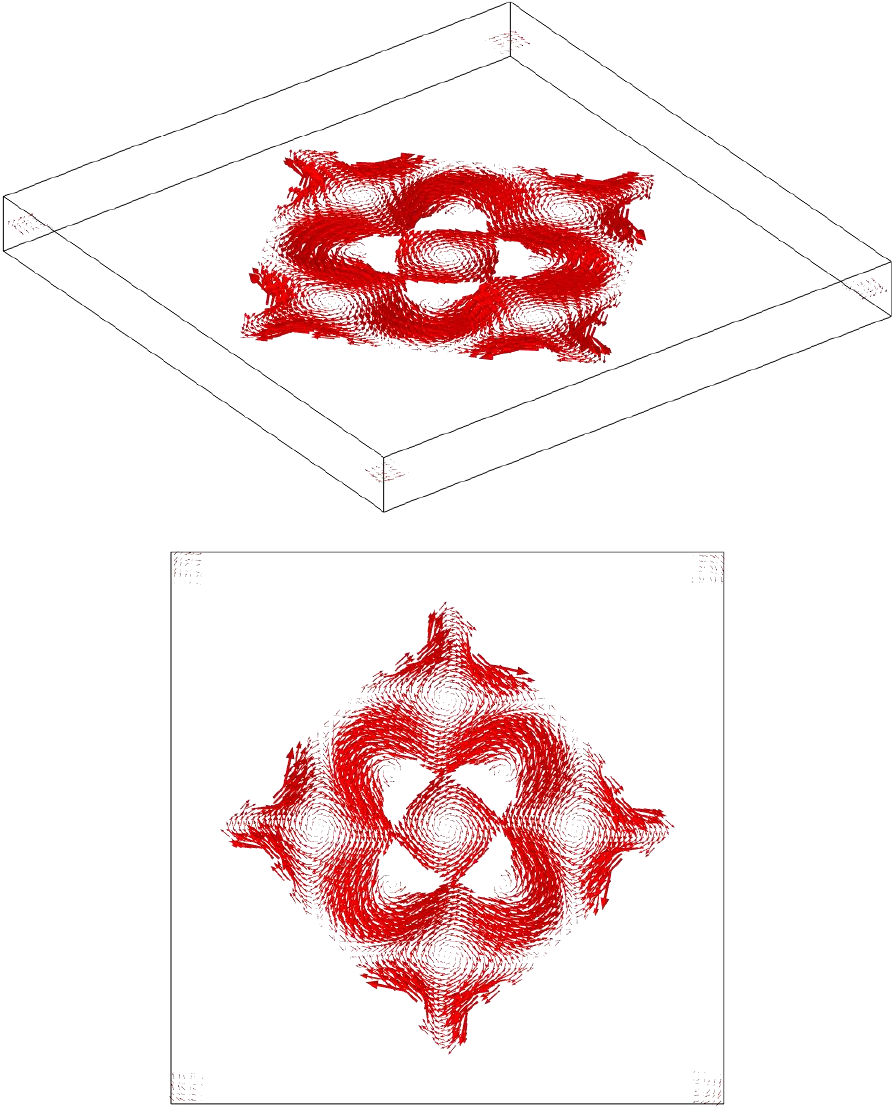}} \hspace{1em}
  \subfigure[$v_0=0.5$]
  {\includegraphics[width=0.3\textwidth]{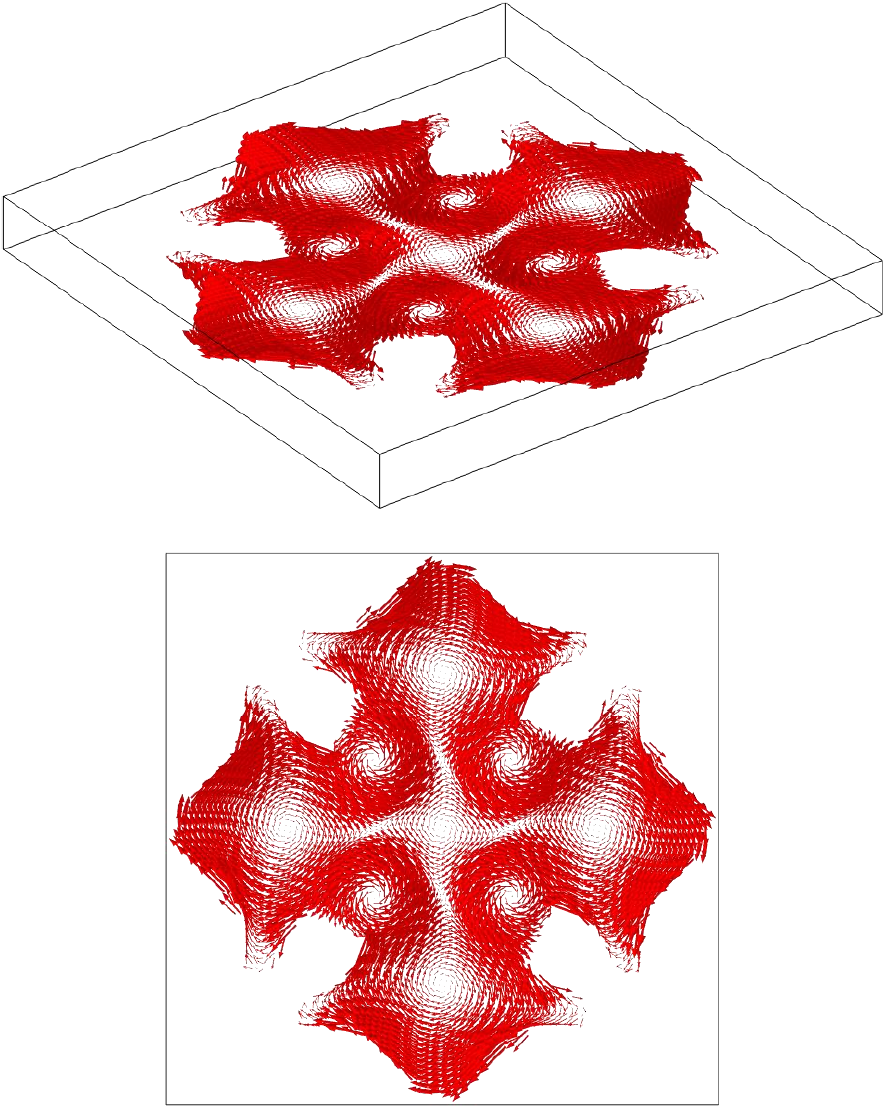}} \hspace{1em}
  \subfigure[$v_0=0.7$]
  {\includegraphics[width=0.3\textwidth]{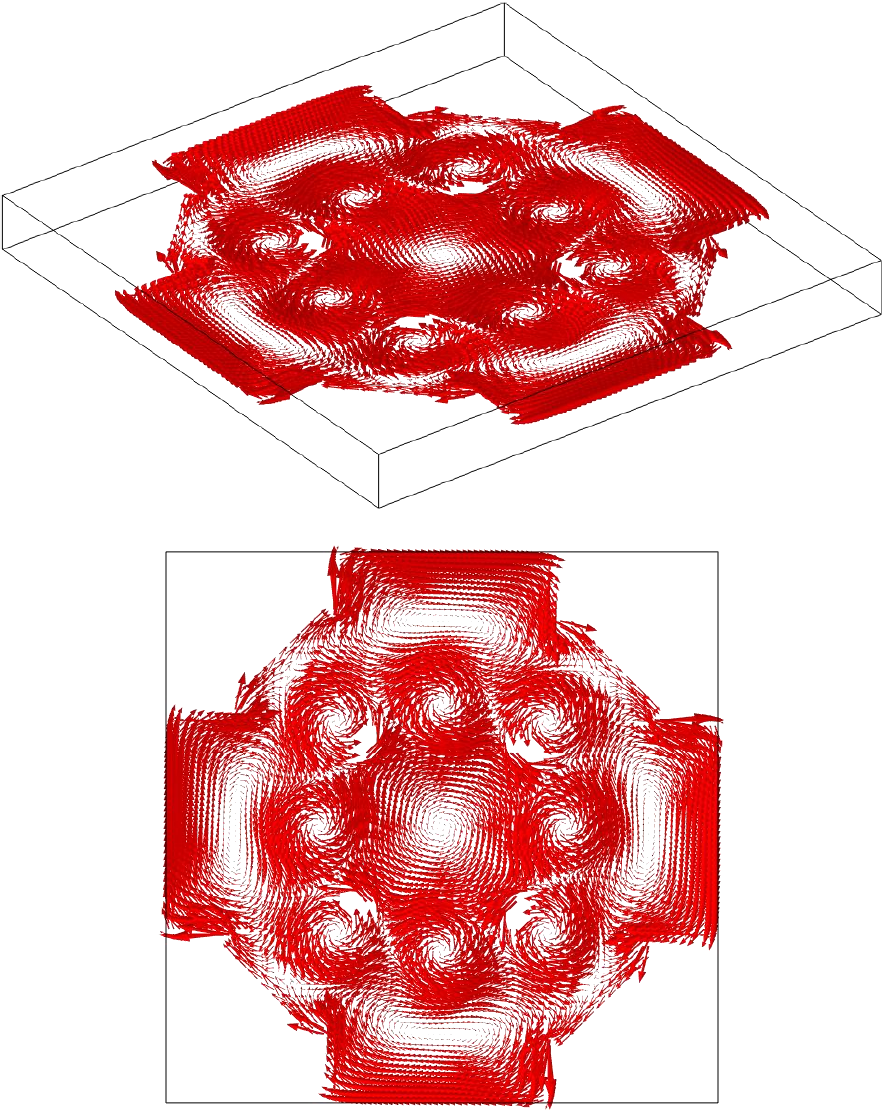}} \hspace{1em}
  \subfigure[$v_0=0.9$]
  {\includegraphics[width=0.3\textwidth]{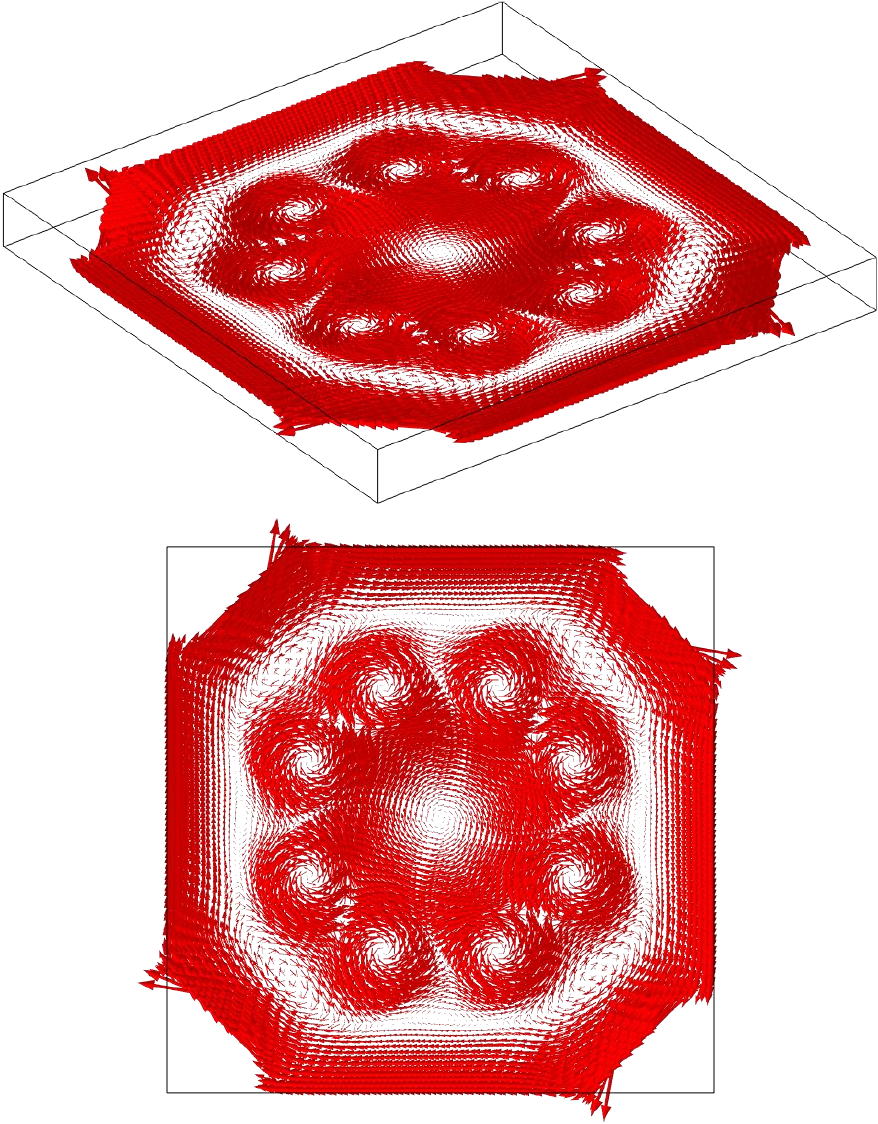}}
  \caption{Stereo and top views of the supercurrent distribution at the terminal time in the optimized topologies of low-temperature type-II superconductors with the volume fractions of $0.1$, $0.3$, $0.5$, $0.7$ and $0.9$, respectively.}\label{fig:LTS_vf_H3=12_Supercurrent}
\end{figure}

When the volume fraction is $0.1$, the feature size of the structures in the optimized topologies is close or smaller than the penetration depth, and the flux lines cannot be nucleated as shown in Figs. \ref{fig:LHTS_vf_H3=12}a and f, where the temporal evolution of the order parameters is provided in Fig.~\ref{fig:TimeEvolution_LTS_OrderParameter}a; and flux lines do not appear as shown by the supercurrent distribution in Fig.~\ref{fig:LTS_vf_H3=12_Supercurrent}a. Intermediate states between the Meissner states and normal states exist in the optimized topologies; and Meissner currents, also named as shielding or screening currents, exist at the interfaces between the superconductor and dielectric/vacuum. According to the Lenz's law, the Meissner currents induce the magnetic fields to cancel parts of the applied magnetic fields and the intermediate states are thereby achieved, where the induced magnetic fields can be derived from the vector potential by using Eq.~\ref{eq:MagneticInductionOfVectorPotential}.

\begin{figure}[!htbp]
  \centering
  \subfigure[]
  {\includegraphics[height=0.25\textwidth]{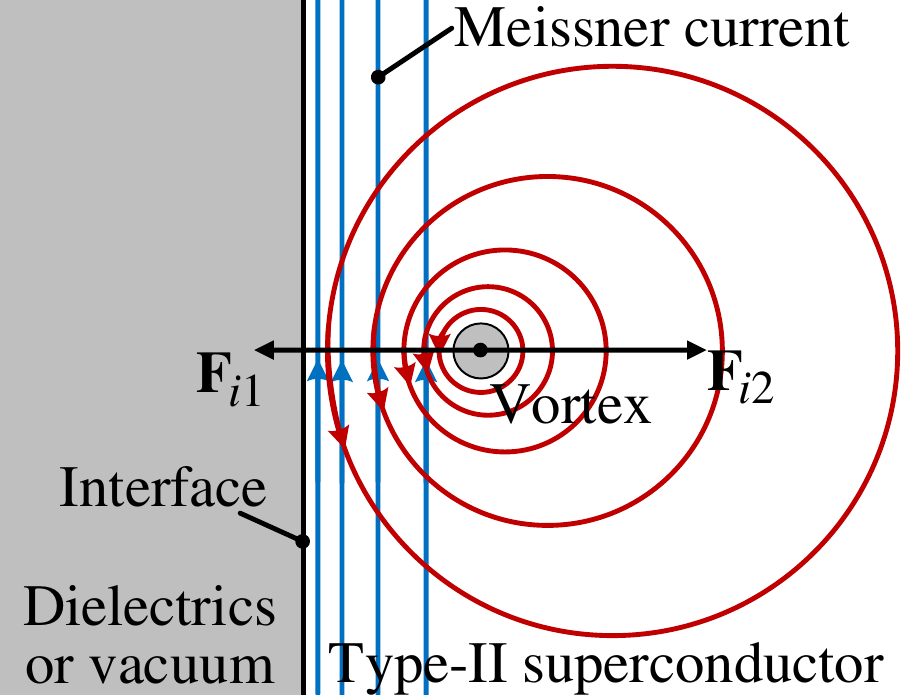}} \hspace{2.5em}
  \subfigure[]
  {\includegraphics[height=0.23\textwidth]{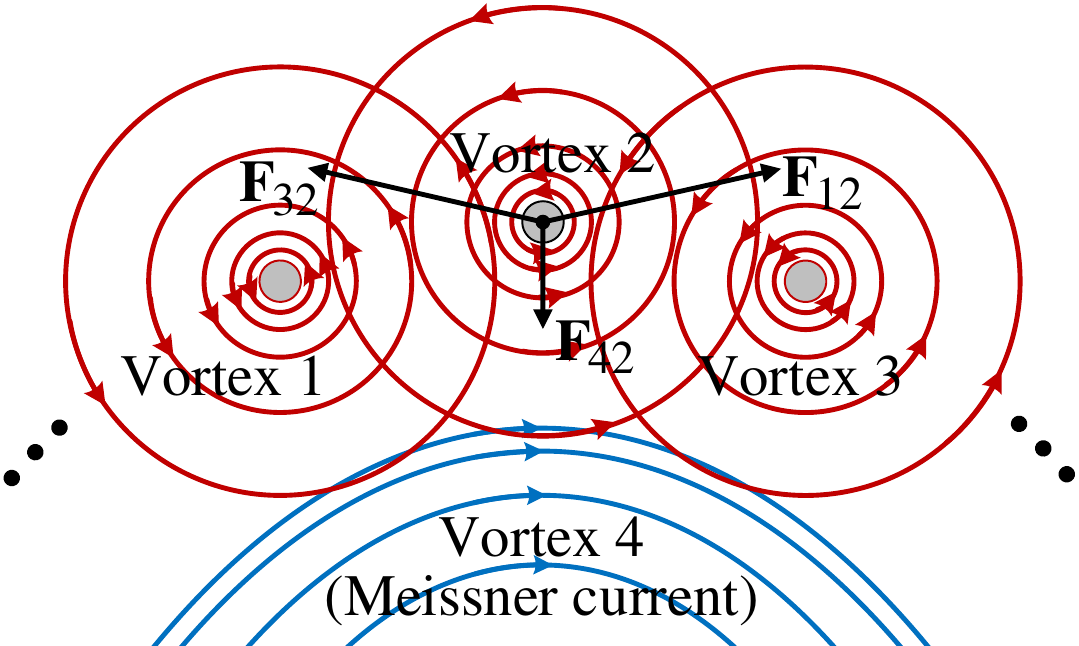}}
  \caption{Sketches for the forces imposed on flux lines: (a) Pushing and pulling forces imposed on a flux line by the interface between the low-temperature type-II superconductor and dielectric/vacuum, where $\mathbf{F}_{i1}$ is the pushing force and $\mathbf{F}_{i2}$ is the pulling force; (b) Repulsive forces imposed on a flux line by another two neighboring flux lines with the same orientation, and the attractive force imposed on the flux line by the Meissner-current vortex with the opposite orientation, where $\mathbf{F}_{12}$ is the repulsive force imposed on Vortex 2 by Vortex 1, $\mathbf{F}_{32}$ is the repulsive force imposed on Vortex 2 by Vortex 3, $\mathbf{F}_{42}$ is the attractive force imposed on Vortex 2 by Vortex 4, Vortices 1, 2 and 3 are the flux lines and they have the same orientation, and Vortex 4 is the Meissner-current vortex and it has the orientation opposite to that of the supercurrent vortices. In the sketches, the red and blue colored curves/lines with arrows represent the flux lines and Meissner currents, respectively.}\label{fig:SketchRepelPushForce}
\end{figure}

When the volume fraction is increased to $0.3$ and $0.5$, flux lines are nucleated as shown in Figs.~\ref{fig:LHTS_vf_H3=12}b$\&$c, where the temporal evolution of the order parameters is provided in Figs.~\ref{fig:TimeEvolution_LTS_OrderParameter}b$\&$c; and supercurrents distribute as shown in Figs.~\ref{fig:LTS_vf_H3=12_Supercurrent}b$\&$c. 
The distances among the flux lines in Figs.~\ref{fig:LHTS_vf_H3=12}b$\&$c are close to the penetration depth. Therefore, there is ignorable interaction between two neighboring flux lines. The interaction between a flux line and the material interface is also the dominant factor for the arrangements of the flux lines in the optimized topologies.
In the results with the volume fraction of $0.3$, the suppercurrents cannot pass through the normal cores of the flux lines, and hence they cannot flow along the parts of material interfaces with normal cores. Instead, they bypass the normal cores as shown in Fig.~\ref{fig:LTS_vf_H3=12_Supercurrent}b. Therefore, the normal cores of the flux lines are pinned at the material interfaces as shown in Fig.~\ref{fig:LHTS_vf_H3=12}b. As shown in Fig.~\ref{fig:TimeEvolution_LTS_OrderParameter}b, the flux lines enter the superconductors from the structural edges, which are one of the typical geometrical features be capable of introducing flux lines into the type-II superconductors. 
In the results with the volume fraction of $0.5$, the normal cores of the flux lines localize close to the material interfaces. As shown in Fig.~\ref{fig:TimeEvolution_LTS_OrderParameter}c, the flux lines enter from the concave corners or structural edges in the optimized topologies. Cooper pairs are bosonic pairs of electrons, condensing into a charged superfluid state that underlies superconductivity \cite{GinzburgLandau1950,BardeenCooperSchrieffer1957}. The supercurrent is compressed and high superfluid velocity exists between the material interfaces and normal cores of the flux lines. This corresponds to low Bernoulli pressure between the material interfaces and normal cores of the flux lines. Pushing forces are thereby imposed on the flux lines. Along with the material interfaces, there are Meissner currents and their direction is opposite to the supercurrents in the vortices. Therefore, the Meissner currents can result in the pulling forces on the flux lines. When the pushing forces are cancelled out by the pulling forces as sketched in Fig.~\ref{fig:SketchRepelPushForce}a, the flux lines arrive at the balance positions as shown in Fig.~\ref{fig:LTS_vf_H3=12_Supercurrent}c. The pulling and pushing forces induce the Bean-Livingston barriers \cite{BeanLivingstonPRL1964}. Therefore, the vortices are pinned and confined in the optimized configurations under the common interaction of the vortex-vortex repulsions and Bean-Livingston barriers.

When the volume fraction is further increased to $0.7$ and $0.9$, flux lines are nucleated as shown in Figs.~\ref{fig:LHTS_vf_H3=12}d$\&$e, where the temporal evolution of the order parameters is provided in Figs.~\ref{fig:TimeEvolution_LTS_OrderParameter}d$\&$e; and supercurrents distribute as shown in Figs.~\ref{fig:LTS_vf_H3=12_Supercurrent}d$\&$e. As shown in Figs.~\ref{fig:TimeEvolution_LTS_OrderParameter}d$\&$e, the flux lines enter the superconductors from the concave corners and structural edges in the optimized topologies with the volume fraction of $0.7$ and $0.9$, respectively. 
The flux lines in Figs.~\ref{fig:LHTS_vf_H3=12}d$\&$e are confined around the central vortices. The central vortices have the opposite orientations with the supercurrent vortices, because they are formed by the Meissner currents. Such arrangements of the flux lines in the optimized topologies are achieved based on the interaction among the flux lines and central vortices. Because the supercurrents have opposite direction in the region between two neighboring flux lines, the Bernoulli pressure of the superfluid is high in this region. A repulsive force is thereby imposed on the flux line by its neighboring flux line. Meanwhile, the central vortex of the Meissner currents and the circulating supercurrents of a flux line have the opposite orientation, and the supercurrents in the region between these two vortices have the same direction. Therefore, the Bernoulli pressure in that region is low; and the central vortex of the Meissner currents imposes attractive forces on the flux lines around it. As sketched in Fig.~\ref{fig:SketchRepelPushForce}b, a flux line bears the repulsive forces from its two neighboring supercurrent vortices, and it also bears the attractive force from the central vortex at the center of the superconductor. Based on the balance of the repulsive and attractive forces, the flux lines are confined around the central vortex.

\begin{figure}[!htbp]
  \centering
  \subfigure[$\mathbf{H}=0.4\mathbf{k}$]
  {\includegraphics[width=0.18\textwidth]{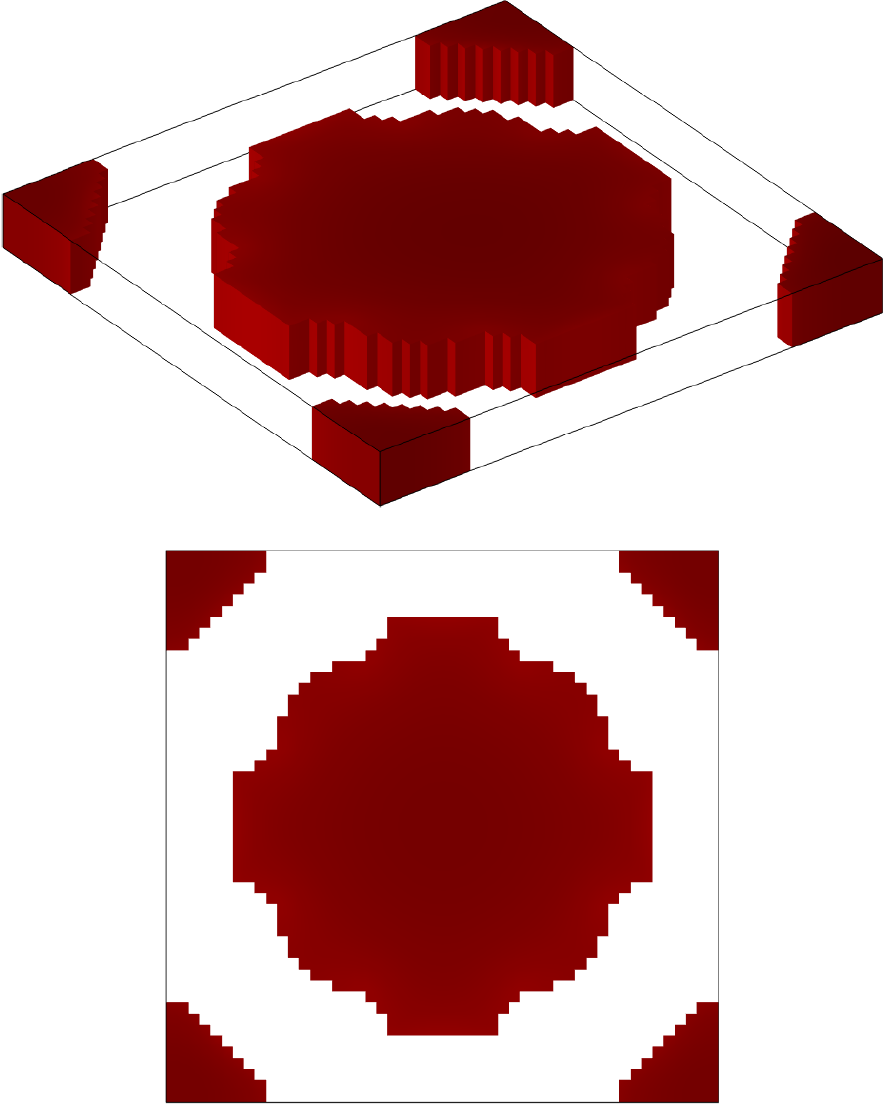}}
  \subfigure[$\mathbf{H}=0.8\mathbf{k}$]
  {\includegraphics[width=0.18\textwidth]{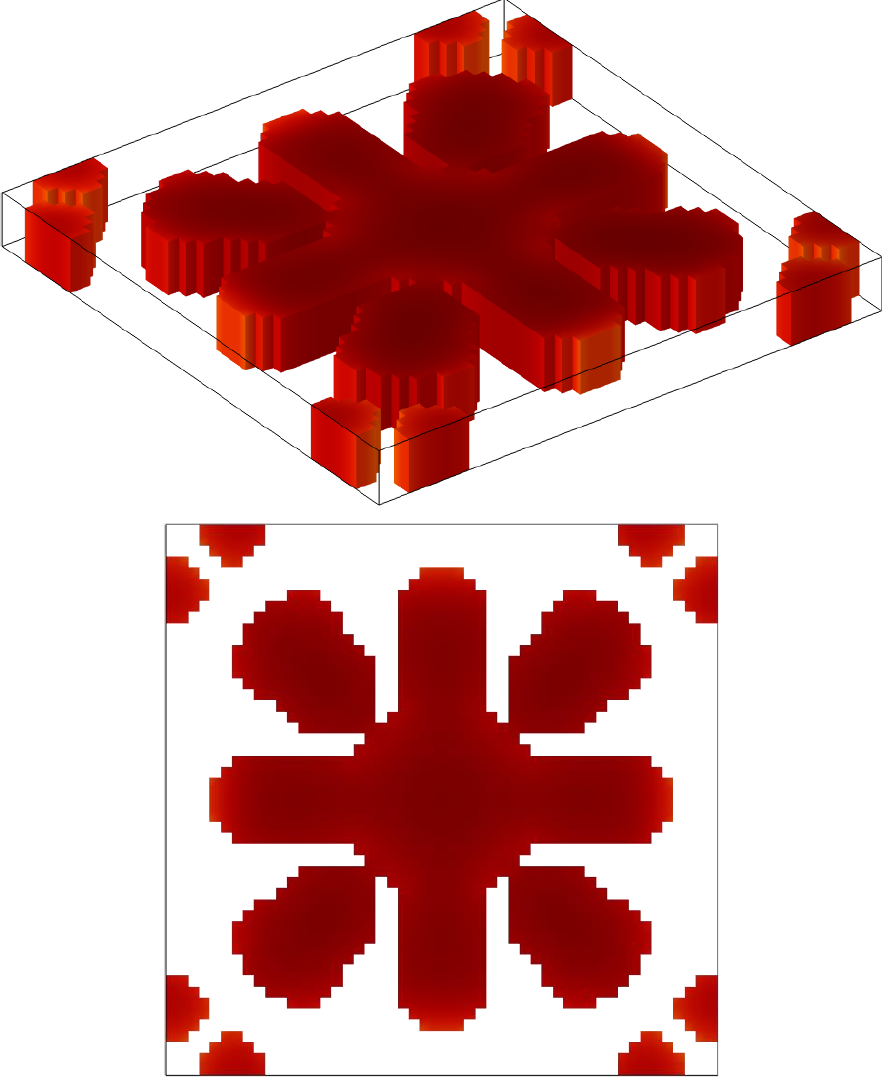}}
  \subfigure[$\mathbf{H}=1.2\mathbf{k}$]
  {\includegraphics[width=0.18\textwidth]{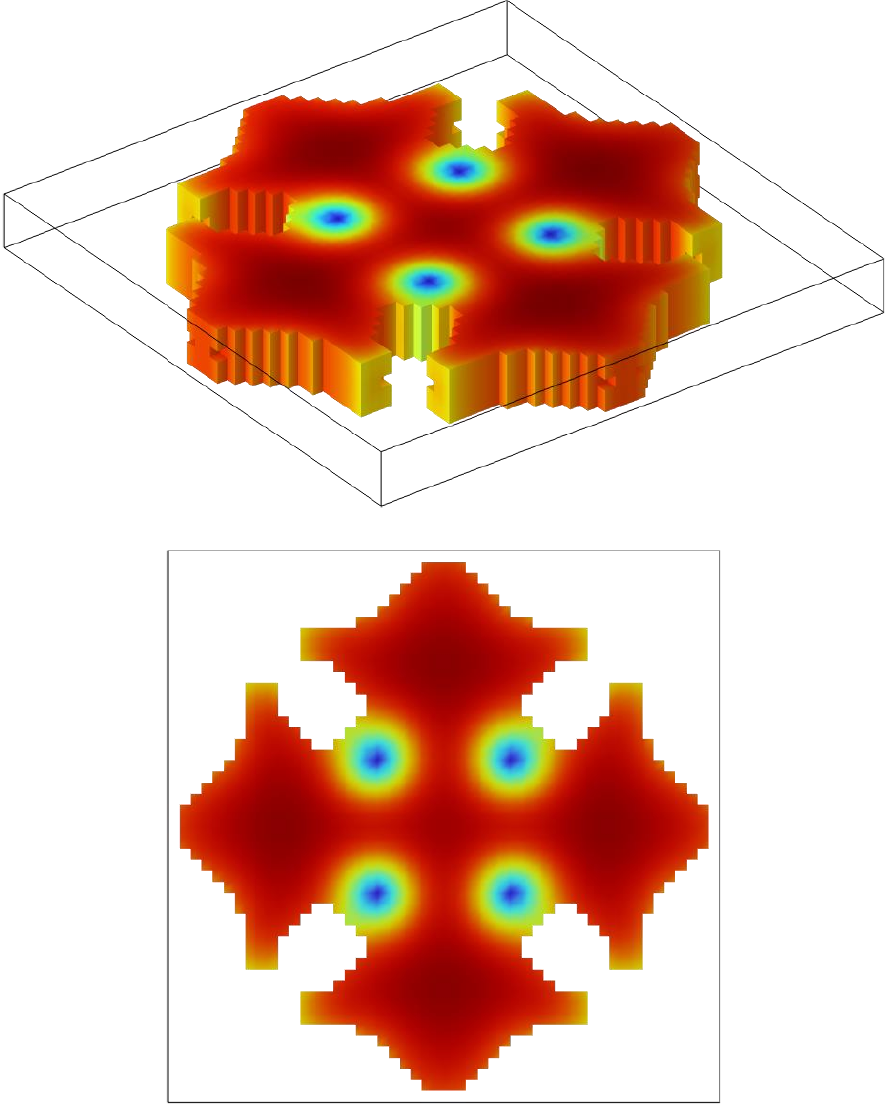}}
  \subfigure[$\mathbf{H}=1.6\mathbf{k}$]
  {\includegraphics[width=0.18\textwidth]{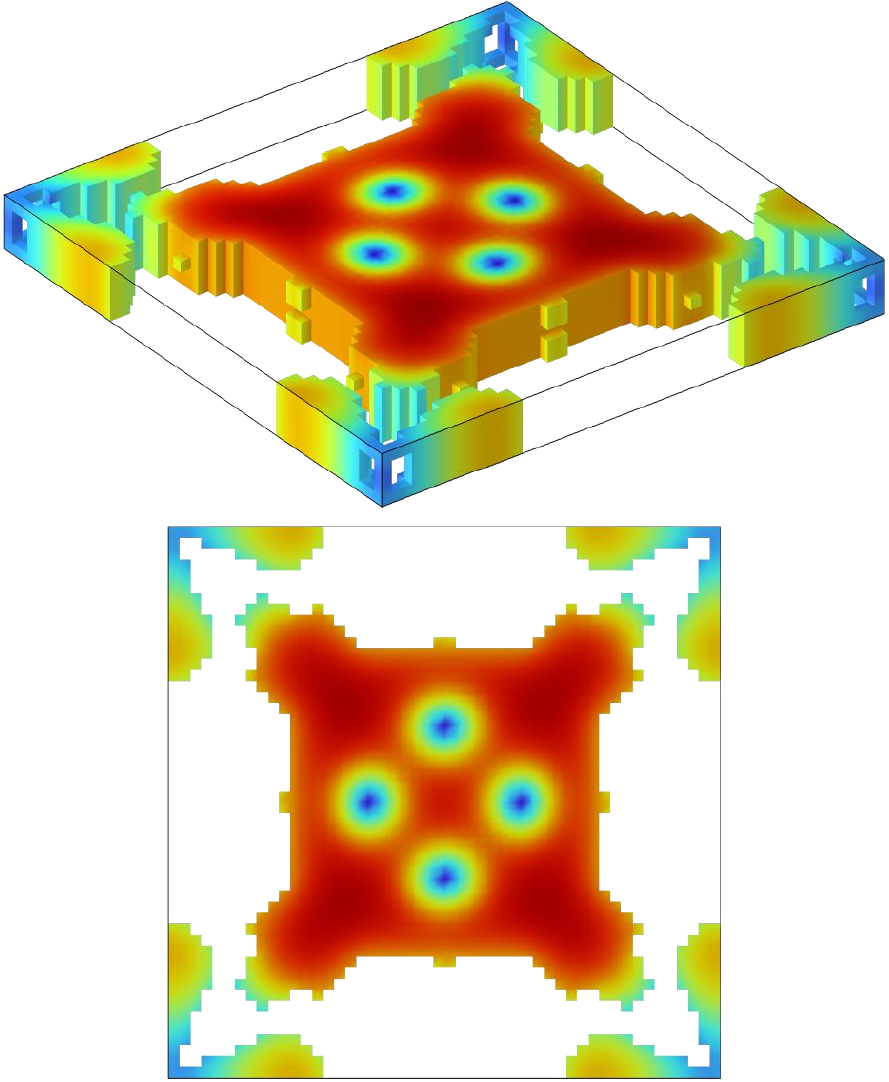}}
  \subfigure[$\mathbf{H}=2.0\mathbf{k}$]
  {\includegraphics[width=0.18\textwidth]{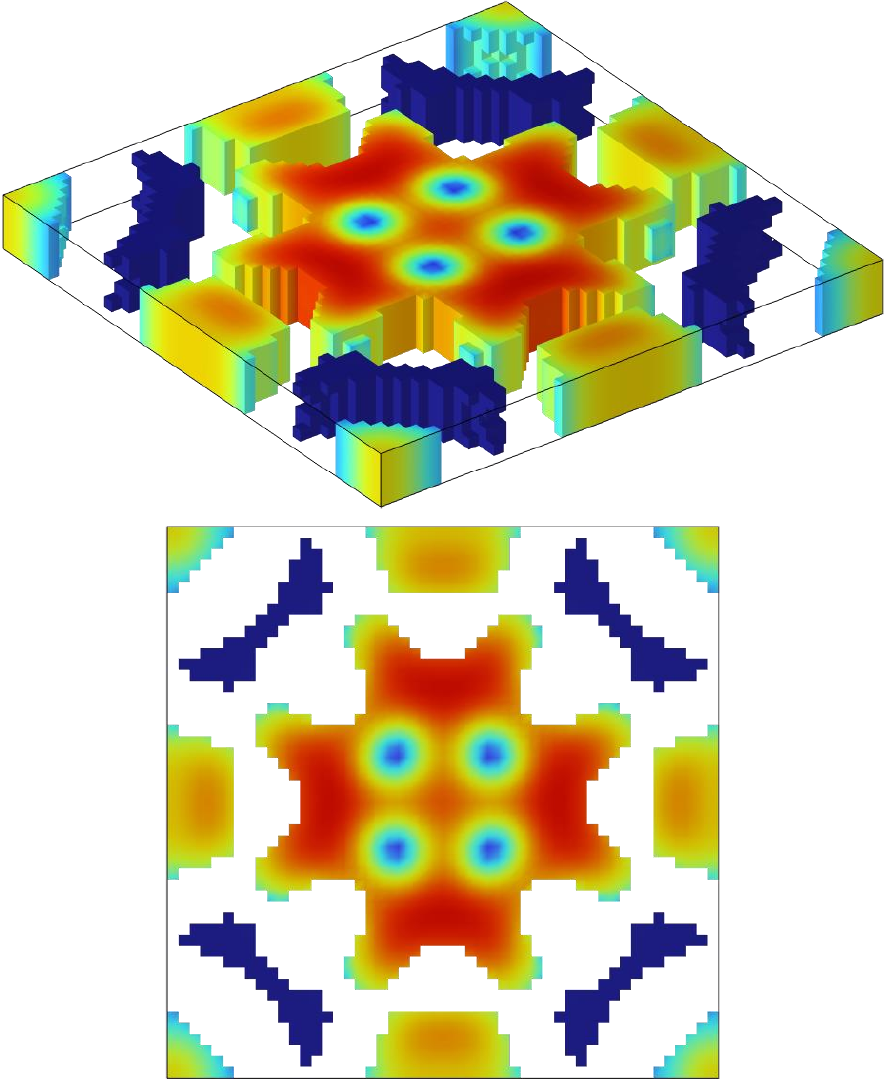}}
  \includegraphics[height=0.22\textwidth]{Figures/HTS_VF_Leg.pdf} 
  \caption{Stereo and top views of the optimized topologies for low-temperature type-II superconductors together with the modular distribution of the order parameter at the terminal time for different applied magnetic fields.}\label{fig:LHTS_H3_vf=05}
\end{figure}

Under different applied magnetic fields, the optimized topologies together with the modular distribution of the order parameter are obtained as shown in Fig.~\ref{fig:LHTS_H3_vf=05} for the low-temperature type-II superconductors, where the volume fraction is set as $0.5$. When the applied magnetic fields are $0.4\mathbf{k}$ and $0.8\mathbf{k}$, the superconductors are optimized into the shapes that can try to prevent the penetration of applied magnetic fields. Therefore, superheated Meissner states can be achieved in the optimized topologies in Figs.~\ref{fig:LHTS_H3_vf=05}a$\&$b, where the applied magnetic fields penetrates the surface layers of the superconductors and set up the Meissner currents. According to Lenz's law, the Meissner currents generate the magnetic fields in the orientation opposite to that of the applied magnetic fields, to minimize the magnetic energy in the interior of the superconductors. When the applied magnetic field is increased to $1.2\mathbf{k}$, flux lines enter the optimized topologies of the superconductors as shown in Fig.~\ref{fig:LHTS_H3_vf=05}c, and the applied magnetic fields penetrate the superconductors from the normal cores of the flux lines. Along with further increasing the applied magnetic field to $1.6\mathbf{k}$, regions with the intermediate state presents along with the flux lines in the optimized low-temperature type-II superconductor as shown in Fig.~\ref{fig:LHTS_H3_vf=05}d. When the applied magnetic field reaches $2.0\mathbf{k}$, the regions in the normal state, together with the regions in the mixed states with flux lines, appear in the optimized topologies as shown in Fig.~\ref{fig:LHTS_H3_vf=05}e. The regions in the normal state are separated and disconnected with that in the mixed states. They can be penetrated completely by the applied magnetic field. It can be regarded that the second-order phase transition at the upper critical magnetic field has started to occur under the applied magnetic field of $2.0\mathbf{k}$. In the optimized topologies under the applied magnetic fields of $1.2\mathbf{k}$, $1.6\mathbf{k}$ and $2.0\mathbf{k}$, the supercurrent density is minimized to decrease the density of the flux lines, which are confined around the center of the low-temperature type-II superconductors. 

\begin{figure}[!htbp]
  \centering
  \subfigure[$\kappa_s=3.0$]
  {\includegraphics[width=0.18\textwidth]{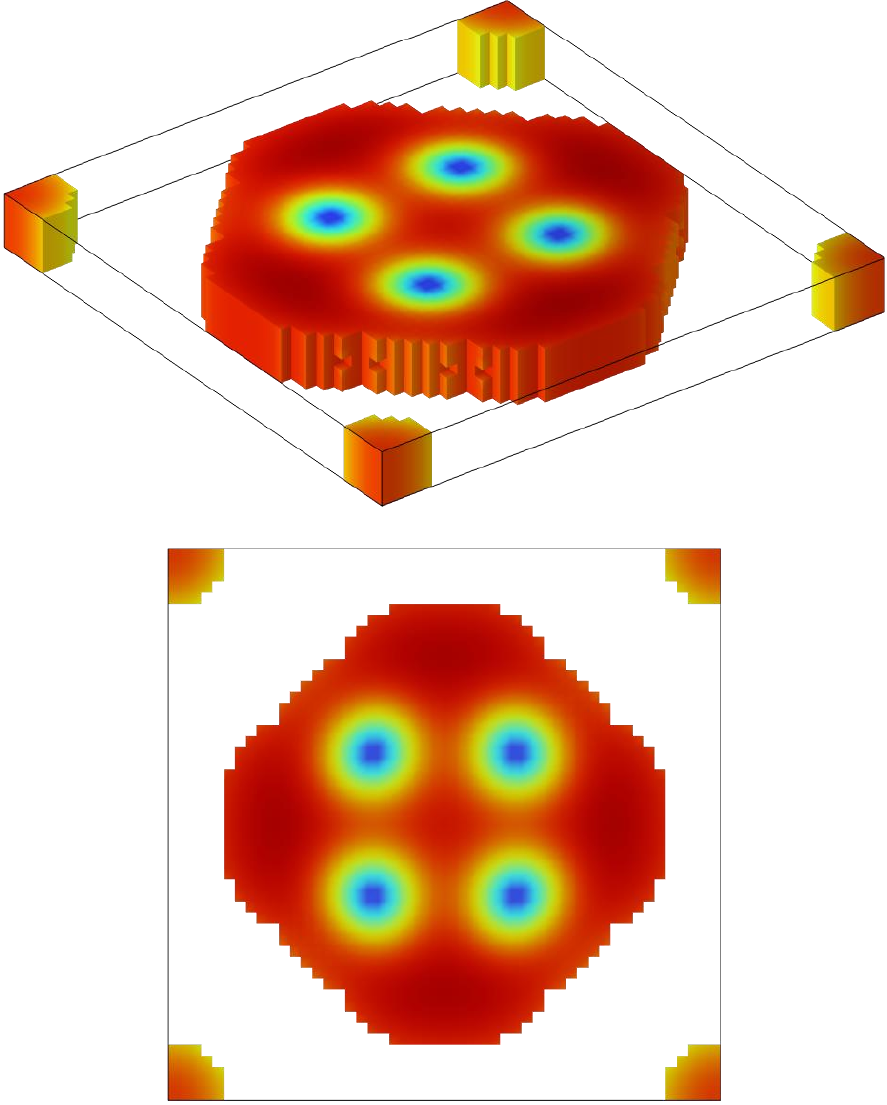}}
  \subfigure[$\kappa_s=3.5$]
  {\includegraphics[width=0.18\textwidth]{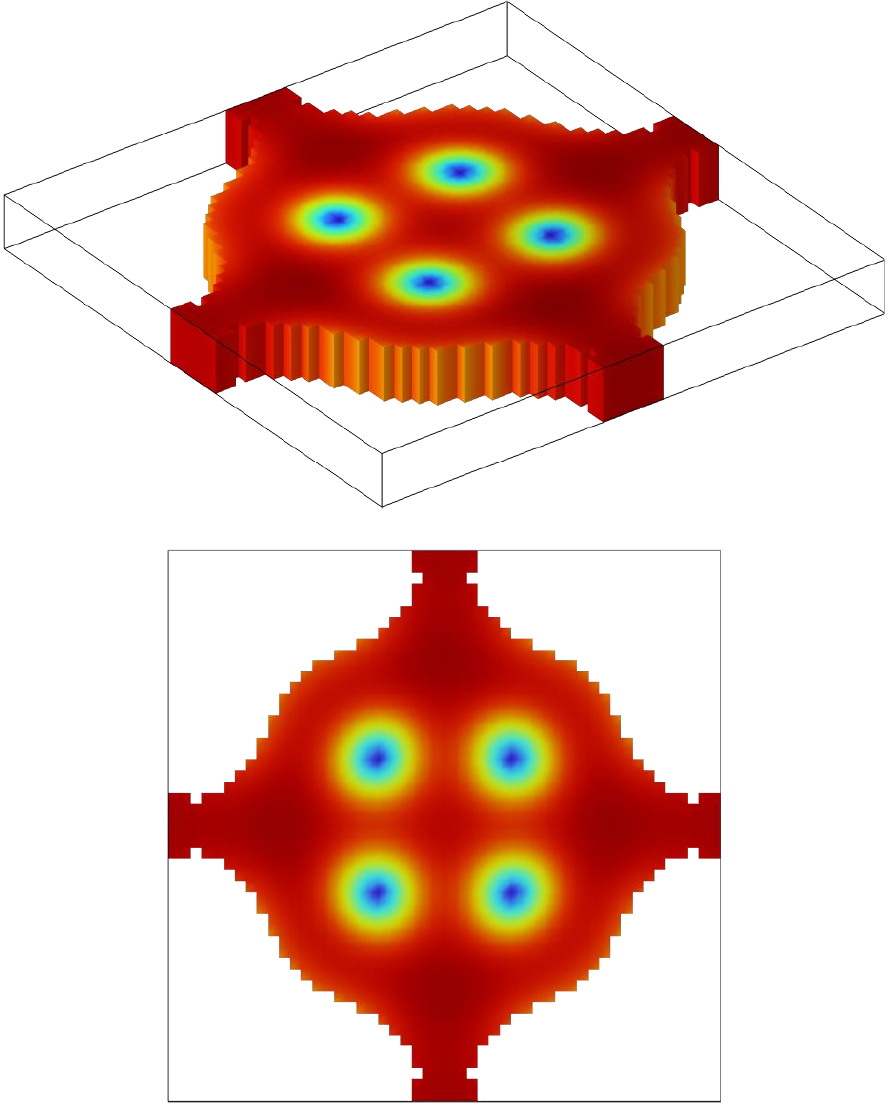}}
  \subfigure[$\kappa_s=4.0$]
  {\includegraphics[width=0.18\textwidth]{Figures/LTS_H3=12_vf=05.pdf}}
  \subfigure[$\kappa_s=4.5$]
  {\includegraphics[width=0.18\textwidth]{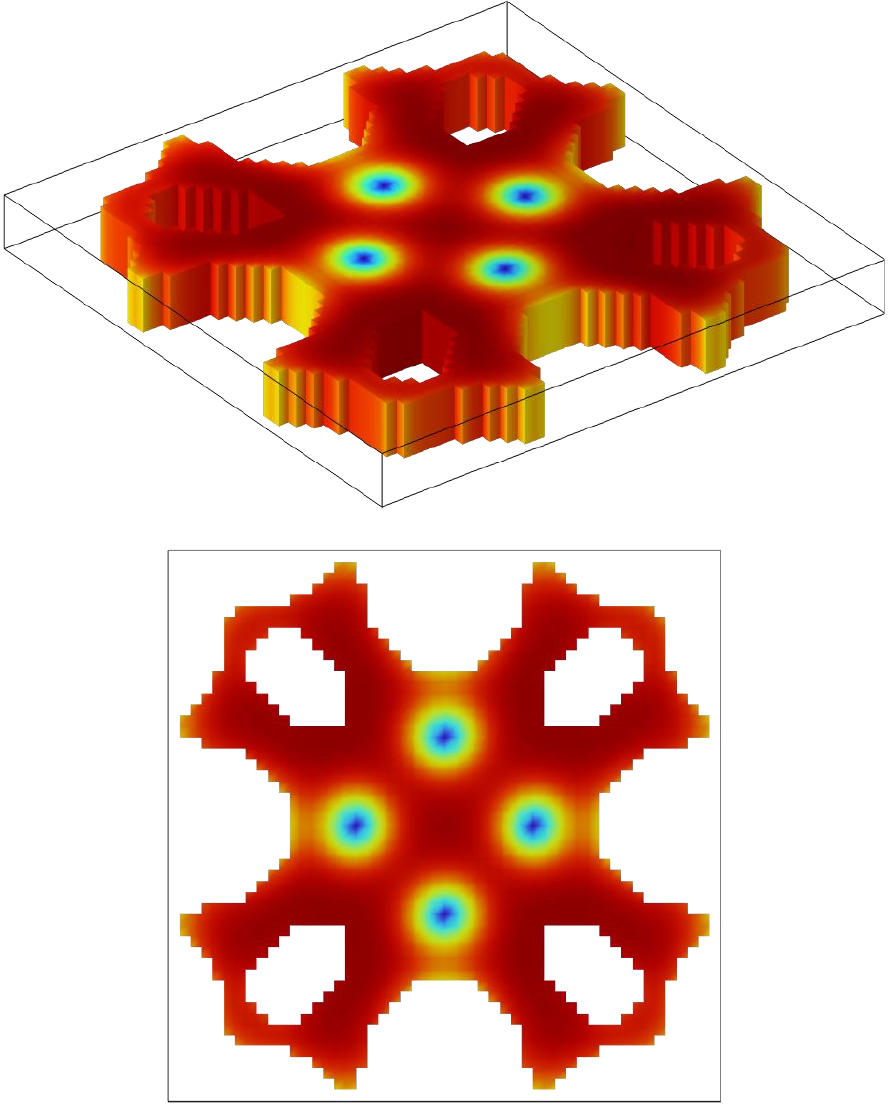}}
  \subfigure[$\kappa_s=5.0$]
  {\includegraphics[width=0.18\textwidth]{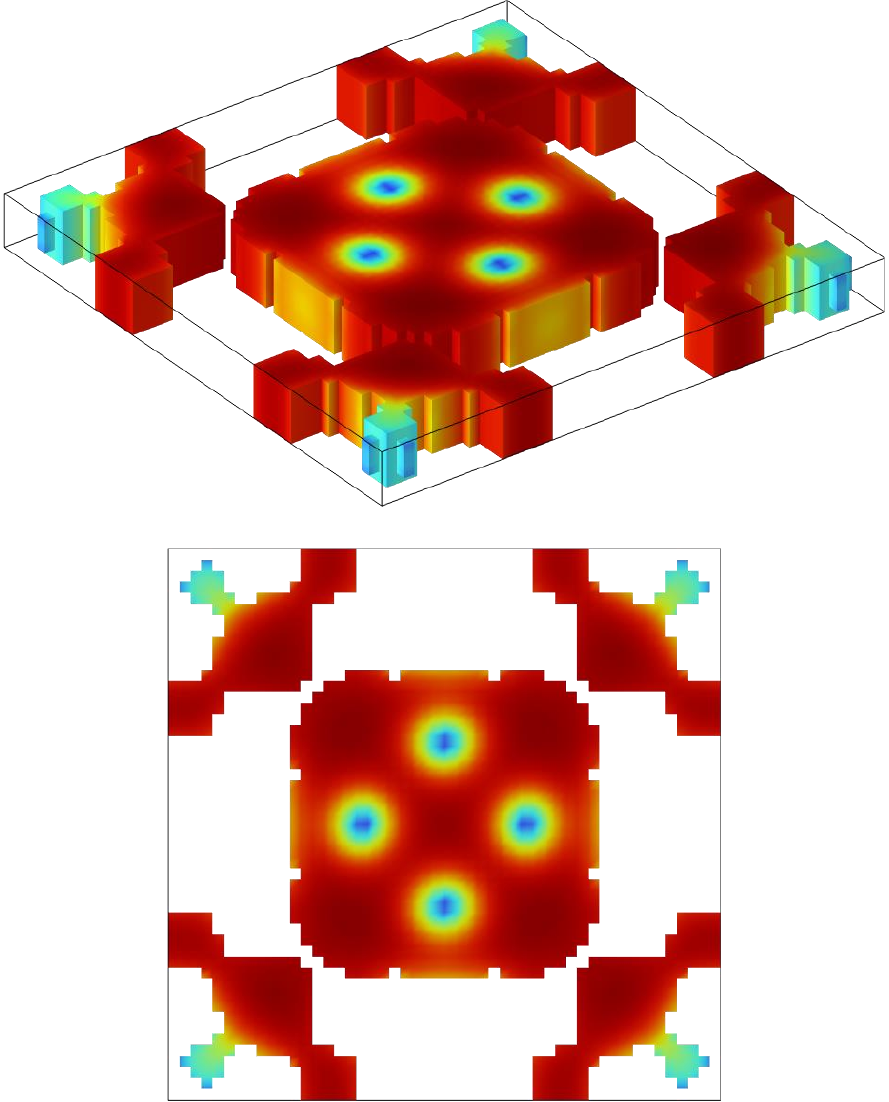}}
  \includegraphics[height=0.22\textwidth]{Figures/HTS_VF_Leg.pdf} 
  \caption{Stereo and top views of the optimized topologies for low-temperature type-II superconductors together with the modular distribution of the order parameter at the terminal time for different Ginzburg-Landau parameters.}\label{fig:LHTSGLparameter}
\end{figure}

For different Ginzburg-Landau parameters, the optimized topologies together with the modular distribution of the order parameter are obtained as shown in Fig.~\ref{fig:LHTSGLparameter} for the low-temperature type-II superconductors, where the volume fraction of the superconducting material is $0.5$ and the applied magnetic field is $1.2\mathbf{k}$. Because the size of the design domain is nondimensionalized by the penetration depth of the superconducting material, the coherence length decreases along with the increase of the Ginzburg-Landau parameter. Therefore, the diameters of the flux lines in Fig.~\ref{fig:LHTSGLparameter} decrease along with the increase of the Ginzburg-Landau parameter. In the obtained results, four flux lines exist in every optimized topologies; the distance between two neighboring flux lines is around the penetration depth and the interaction between them is ignorable. The flux lines are confined in the optimized topologies based on the common interaction of cortex-vortex repulsions and Bean-Livingston barriers. 

To check the optimized performance, the obtained topologies in Fig.~\ref{fig:LHTS_H3_vf=05} are cross compared by computing the values of the optimization objective in Eq.~\ref{eq:ObjFunction} for different applied magnetic fields. The computed values of the optimization objective have been listed in Tab.~\ref{tab:CrossComparisonLHTS_H3_vf=05}. From the optimized entries noted in bold in every row of Tab.~\ref{tab:CrossComparisonLHTS_H3_vf=05}, the optimized performance of the obtained topologies in Fig.~\ref{fig:LHTS_H3_vf=05} can be confirmed. 

\begin{table}[!htbp]
\centering
\begin{tabular}{l|ccccc}
  \toprule
        & Fig.~\ref{fig:LHTS_H3_vf=05}a
        & Fig.~\ref{fig:LHTS_H3_vf=05}b
        & Fig.~\ref{fig:LHTS_H3_vf=05}c
        & Fig.~\ref{fig:LHTS_H3_vf=05}d
        & Fig.~\ref{fig:LHTS_H3_vf=05}e \\
  \midrule
  $\mathbf{H}=0.4\mathbf{k}$ & $\mathbf{0.4378}$ & $0.5319$ & $0.5607$ & $0.5014$ & $0.4548$ \\
  \midrule
  $\mathbf{H}=0.8\mathbf{k}$ & $2.0623$ & $\mathbf{0.8026}$ & $1.7793$ & $1.5588$ & $0.8834$ \\
  \midrule
  $\mathbf{H}=1.2\mathbf{k}$ & $2.3089$ & $1.5656$ & $\mathbf{1.5066}$ & $1.8013$ & $1.5361$ \\
  \midrule
  $\mathbf{H}=1.6\mathbf{k}$ & $1.9658$ & $1.8186$ & $1.4299$ & $\mathbf{1.1795}$ & $1.2452$ \\
  \midrule
  $\mathbf{H}=2.0\mathbf{k}$ & $1.7426$ & $1.5941$ & $1.1777$ & $1.1321$ & $\mathbf{0.7026}$ \\
  \bottomrule
\end{tabular}
\caption{Cross comparison of the objective values corresponding to different applied magnetic fields for the optimized topologies in Fig.~\ref{fig:LHTS_H3_vf=05}, where the optimized entries in every row of the tables are noted in bold.}\label{tab:CrossComparisonLHTS_H3_vf=05}
\end{table}

\begin{figure}[!htbp]
  \centering
  \subfigure[$C_{4h}$]
  {\includegraphics[width=0.18\textwidth]{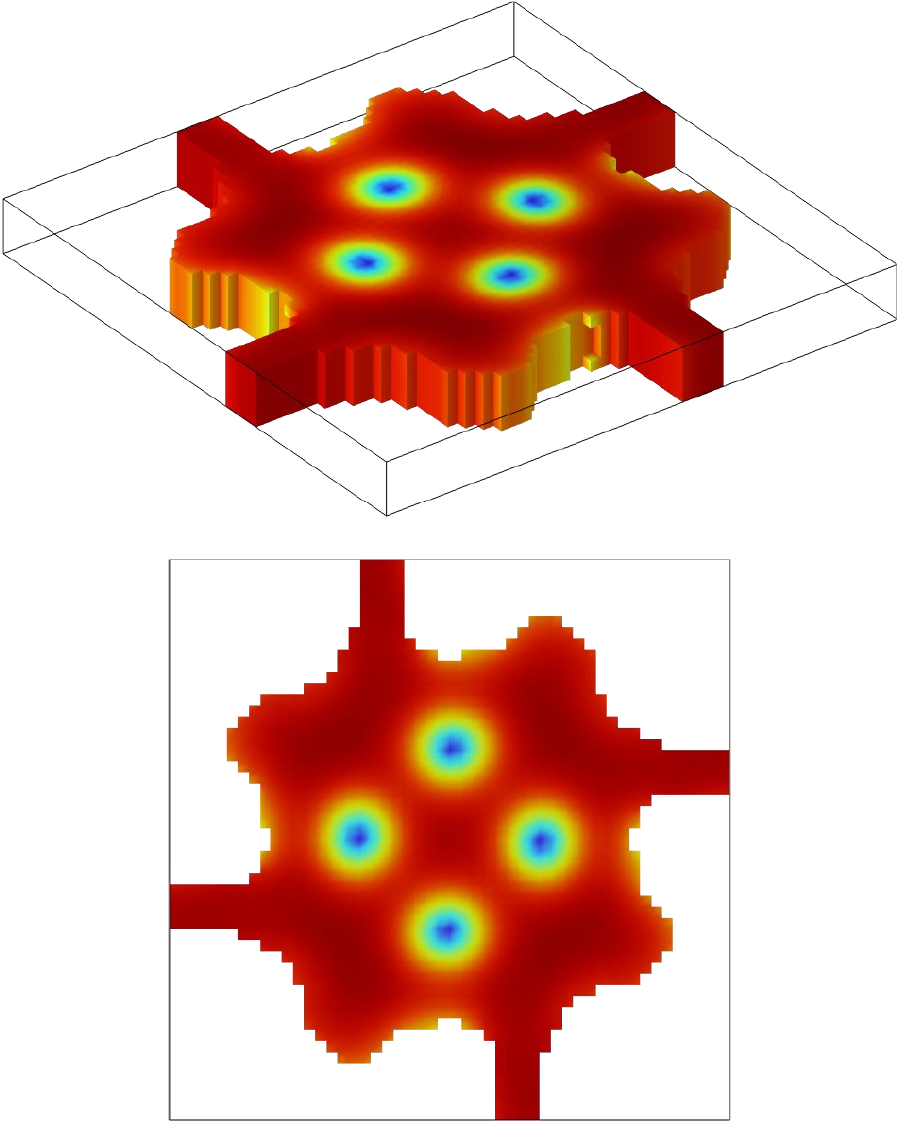}}
  \subfigure[$C_{2h}$]
  {\includegraphics[width=0.18\textwidth]{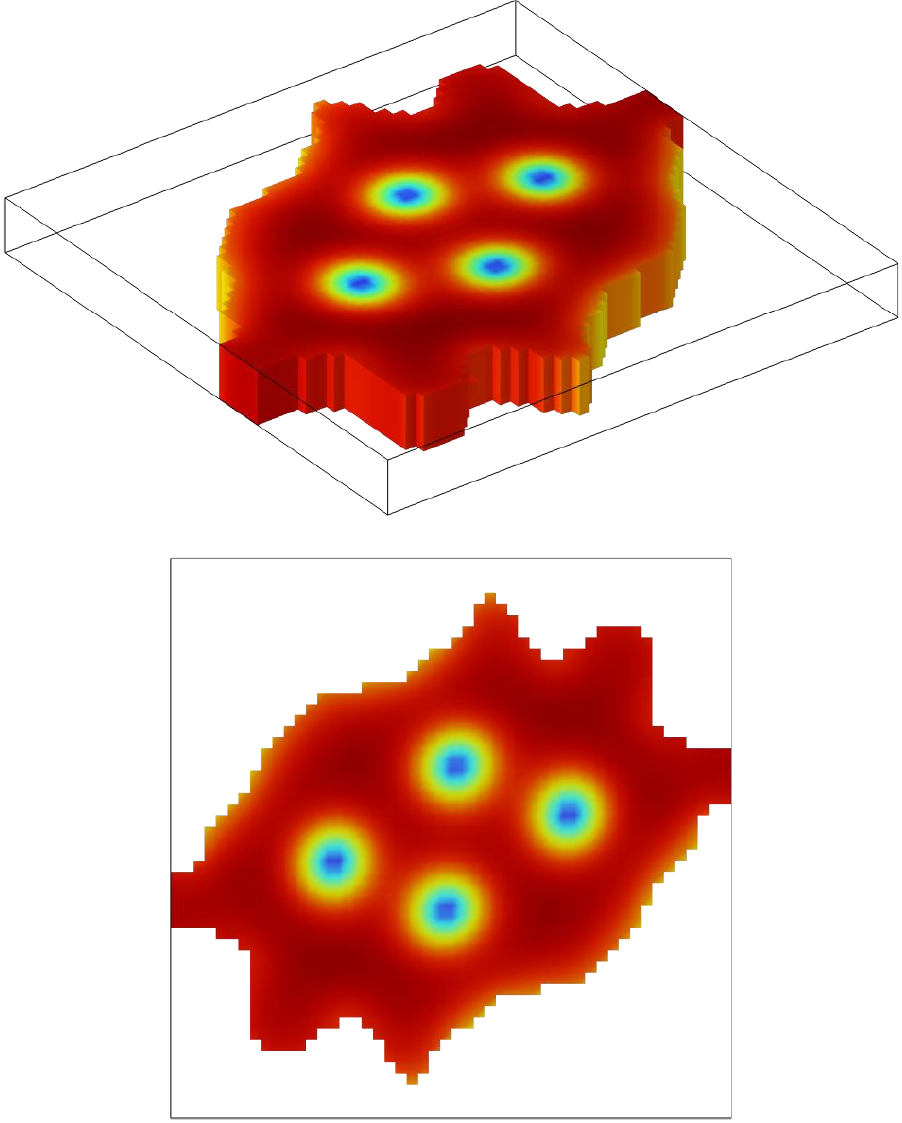}}
  \subfigure[$C_{2v}$]
  {\includegraphics[width=0.18\textwidth]{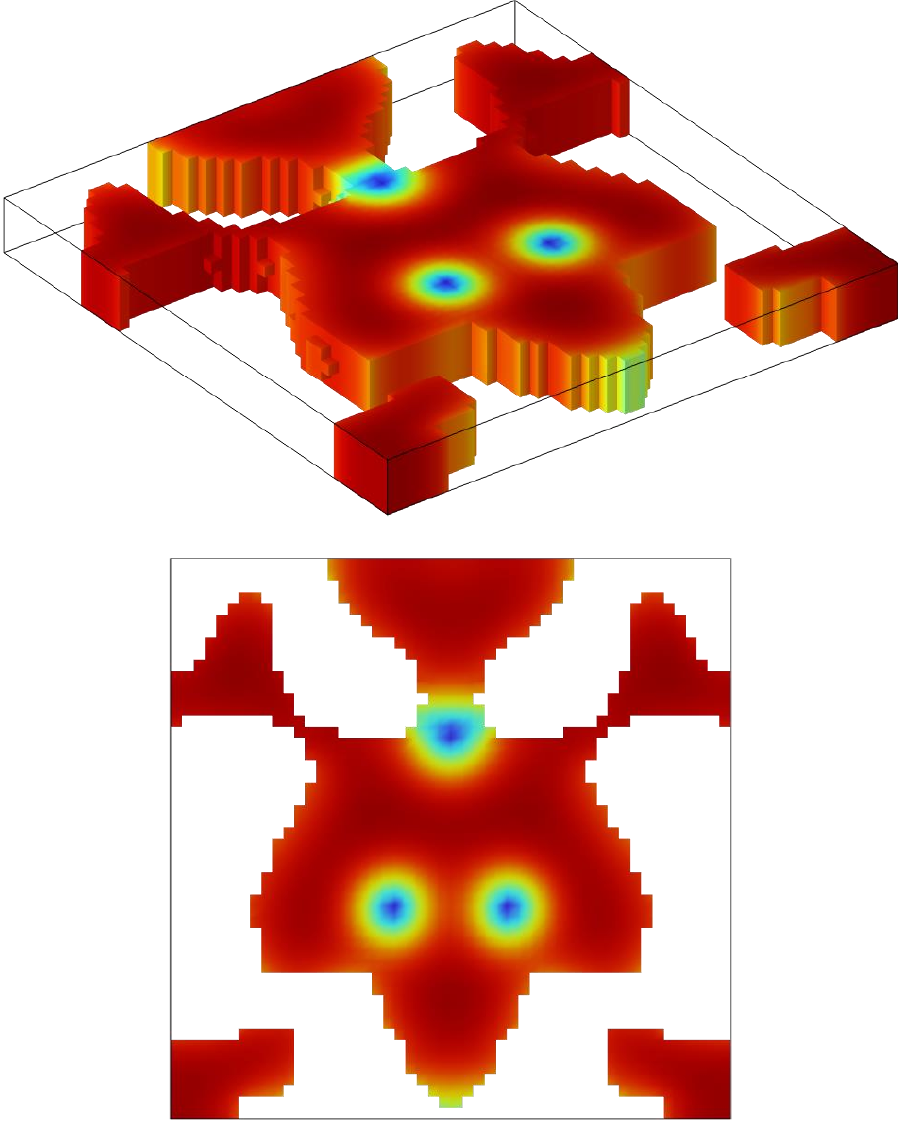}}
  \subfigure[$C_1$]
  {\includegraphics[width=0.18\textwidth]{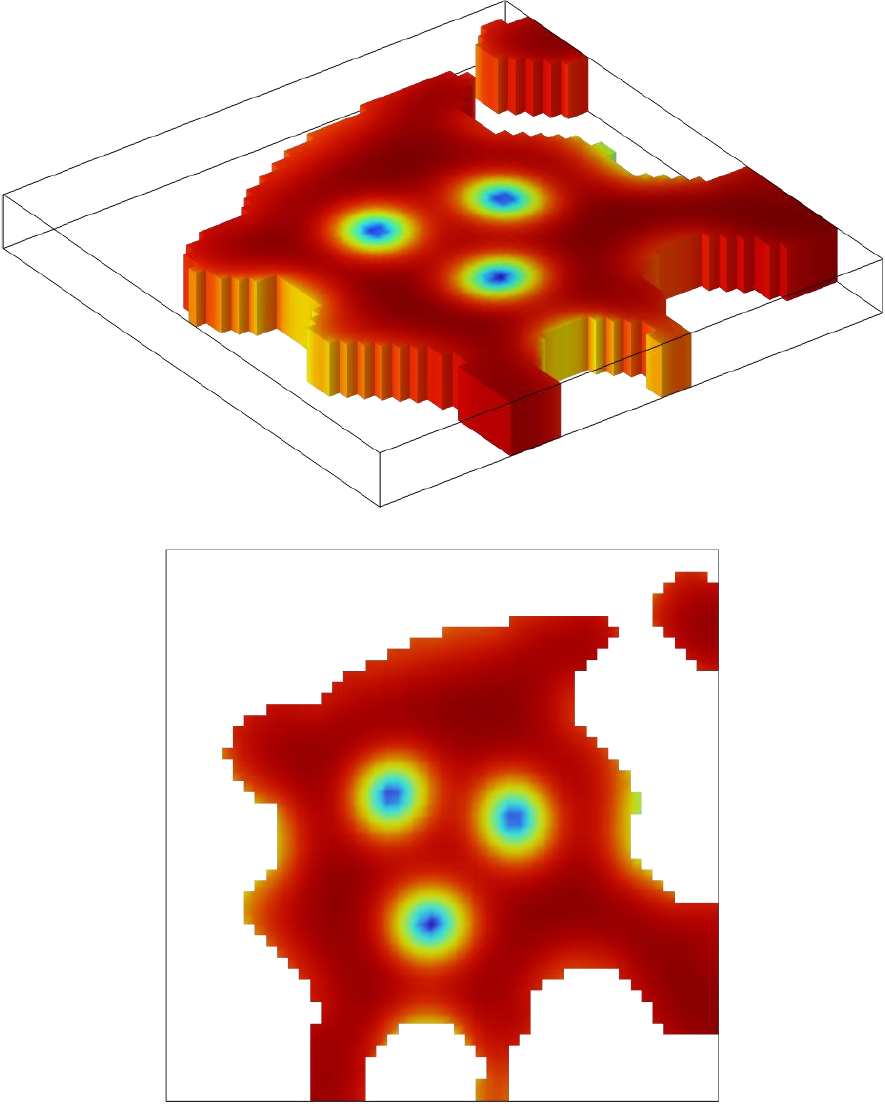}}
  {\includegraphics[height=0.22\textwidth]{Figures/HTS_VF_Leg.pdf}}
  \caption{Stereo and top views of the optimized topologies together with the modular distribution of the order parameter at the terminal time for the low-temperature type-II superconductors obtained by relaxing the symmetry constraints from $C_{4h}$ to $C_1$.}\label{fig:CheckingSymmetryConstraints}
\end{figure}

In order to validate the reasonability of imposing constraints of $D_{4h}$ symmetry, the topology optimization model corresponding to Fig.~\ref{fig:LHTS_vf_H3=12}c is solved by relaxing the symmetry constraints from $C_{4h}$ to $C_1$, yielding the results shown in Fig.~\ref{fig:CheckingSymmetryConstraints}. The corresponding value of the optimization objective in Eq.~\ref{eq:ObjFunction} is listed in Tab.~\ref{tab:CheckingSymmetryConstraintsObjValues}. As indicated by the bold entry in Tab.~\ref{tab:CheckingSymmetryConstraintsObjValues}, the reasonability of imposing symmetry constraints can be confirmed. When the symmetry constraints are removed, the convergent histories of the iterative algorithm exhibit severe oscillations due to numerical errors. The numerical errors can cause the asymmetries in the resulting topologies. Therefore, imposing symmetry constraints can effectively mitigate the negative influence of numerical errors on convergence robustness.

\begin{table}[!htbp]
\centering
\begin{tabular}{l|ccccc}
  \toprule
        & Fig.~\ref{fig:LHTS_vf_H3=12}c 
        & Fig.~\ref{fig:CheckingSymmetryConstraints}a 
        & Fig.~\ref{fig:CheckingSymmetryConstraints}b 
        & Fig.~\ref{fig:CheckingSymmetryConstraints}c 
        & Fig.~\ref{fig:CheckingSymmetryConstraints}d \\
  \midrule
  $J$ & $\mathbf{1.5066}$ & $1.5105$ & $1.5648$ & $1.7261$ & $1.6249$ \\
  \bottomrule
\end{tabular}
\caption{Objective values for the optimized topologies in Figs. \ref{fig:LHTS_vf_H3=12}c and \ref{fig:CheckingSymmetryConstraints}, where the objective value corresponding to the result with $D_{4h}$ symmetry is noted in bold.}\label{tab:CheckingSymmetryConstraintsObjValues}
\end{table}

\begin{figure}[!htbp]
  \centering
  \subfigure[$\mathbf{H}=1.6\mathbf{k}$]
  {\includegraphics[width=0.18\textwidth]{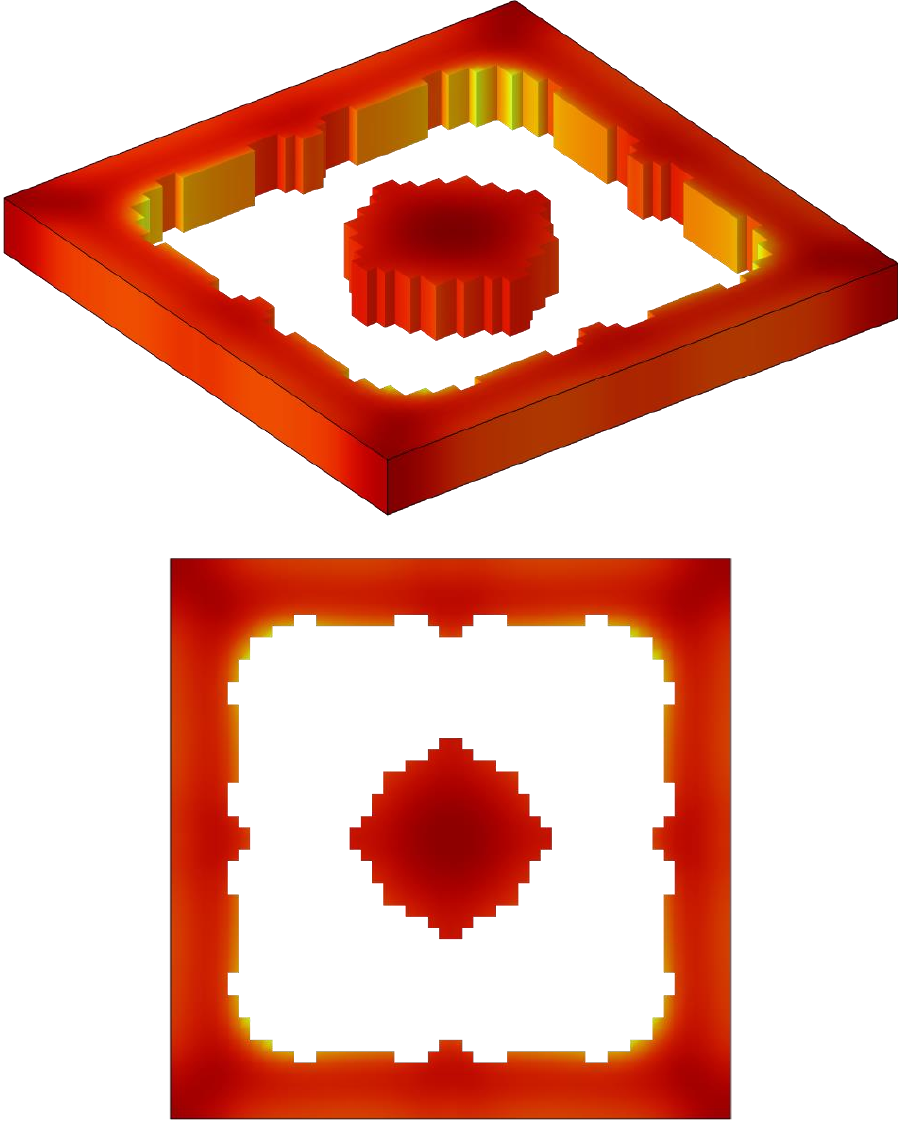}}
  \subfigure[$\mathbf{H}=2.0\mathbf{k}$]
  {\includegraphics[width=0.18\textwidth]{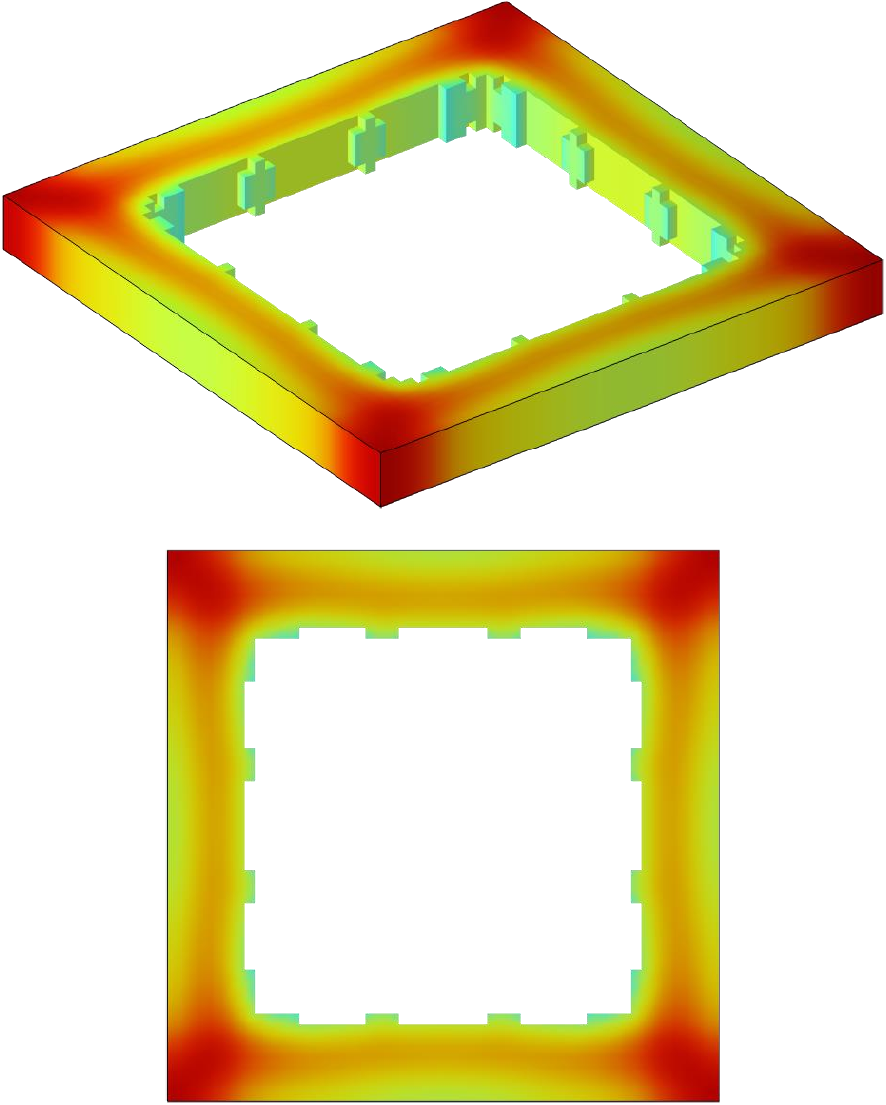}}
  {\includegraphics[height=0.22\textwidth]{Figures/HTS_VF_Leg.pdf}}
  \caption{Stereo and top views of the optimized topologies together with the modular distribution of the order parameter at the terminal time for low-temperature type-II superconductors optimized by using the recast objective function in Eq.~\ref{eq:ModifiedObjEq} under the applied magnetic fields of $1.6\mathbf{k}$ and $2.0\mathbf{k}$.}\label{fig:LHTS_ModifiedObj}
\end{figure}

The optimization objective in Eq.~\ref{eq:TOOPModelLowTempTypeIISuperconductor} can be recast to minimize the least-square difference between the order parameters of the mixed and Meissner states under an applied magnetic field. Since the order parameter in the Meissner state has a unitary modulus, the recast optimization objective takes the form expressed as
\begin{equation}\label{eq:ModifiedObjEq}
\begin{split}
  J & = \int_0^{T_t} \int_\Omega \left( \psi \psi^* - \psi_M \psi_M^* \right)^2 I_d \left(\rho_p\right) H \left( t-T_m \right) \,\mathrm{d}\Omega \mathrm{d}t \\
    & = \int_0^{T_t} \int_\Omega \left( \psi_r^2 + \psi_i^2 - 1 \right)^2 I_d \left(\rho_p\right) H \left( t-T_m \right) \,\mathrm{d}\Omega \mathrm{d}t,
\end{split}
\end{equation}
where $\psi_M$ with $\left|\psi_M\right|=1$ is the order parameter of the Meissner state. Topology optimization with this recast objective can be used to delay the appearance of regions in the normal state before the second-order phase transition at the upper critical magnetic field. This is equivalent to enhancing the robustness of the mixed state close to that second-order phase transition. The adjoint analysis of the recast optimization objective can be carried out by replacing $\left|\left[ \mathbf{j}_s \right]_{\forall\mathbf{x}\in\Omega}\right|^2$ as $\left( \psi_r^2 + \psi_i^2 - 1 \right)^2 I_d \left(\rho_p\right)$ in the related equations in Sections \ref{sec:TOOPAdjAnalysisLowTempTypeIISupercond}, \ref{sec:NumericalImplementationLowTempTypeIISupercond} and \ref{sec:AppendixTOOPSuperCond}. As shown in Fig.~\ref{fig:LHTS_H3_vf=05}d\&e, the regions in the intermediate and normal states appear under the applied magnetic fields of $1.6\mathbf{k}$ and $2.0\mathbf{k}$, respectively; and the second-order phase transition at the upper critical magnetic field has started to occur under the applied magnetic field of $2.0\mathbf{k}$. By recasting the optimization objective as that in Eq.~\ref{eq:ModifiedObjEq}, the optimized topologies are derived as shown in Fig.~\ref{fig:LHTS_ModifiedObj} for the applied magnetic fields of $1.6\mathbf{k}$ and $2.0\mathbf{k}$. Compared to the results in Fig.~\ref{fig:LHTS_H3_vf=05}d\&e, the regions in the intermediate and normal states are shrunk or removed. Therefore, it can be concluded that topology optimization of low-temperature type-II superconductors with the recast optimization objective in Eq.~\ref{eq:ModifiedObjEq} can delay the appearance of regions in the intermediate and normal states before the occurrence of second-order phase transitions at the upper critical magnetic fields.


The fabrication of the optimized superconductors can be addressed by using the advanced 3D focused electron beam induced deposition (3D FEBID) technique, which can be harnessed to fabricate nanoscale superconductors with complex geometries \cite{ZhakinaDonnellyAFM2025}. By incorporating constraints for extruded geometries, the topology optimization model can ensure that the resulting designs are feasible for electron beam lithography-based nanofabrication techniques.

\section{Conclusions}\label{sec:ConclusionsTOOPSuperCond}

Topology optimization of low-temperature type-II superconductors has been presented based on the material distribution method, where the temporal evolution of the order parameter and vector potential is described by the time-dependent Ginzburg-Landau equations under the Weyl gauge. The geometric configuration of a low-temperature type-II superconductor is expressed by a material density, derived from a design variable via a PDE filter, piecewise homogenization and threshold projection. The piecewise homogenization can remove gradients of the filtered variable to ensure the robustness of the topology optimization procedure. Using the material density, material interpolation is applied to the Ginzburg-Landau parameter, the Landau free energy, and the magnetic energy via a $q$-parameter scheme.
To enhance the robustness of the mixed state, the objective function is set to minimize the supercurrent density under an applied magnetic field. An essential volume-fraction constraint of the superconducting material is imposed to prevent trivial solutions with the design domains filled entirely with dielectric/vacuum. To circumvent the complexity introduced by numerical discretization of time derivatives in the relevant equations, the continuous adjoint analysis is employed. Furthermore, because the order parameter is complex, the time-dependent Ginzburg-Landau equations are split into real and imaginary parts. This ensures consistency between the real-valued properties of the design variable and the derived adjoint sensitivity of the optimization objective.

The effects of volume fraction, applied magnetic field and Ginzburg-Landau parameter have been investigated by numerically solving the topology optimization model. In the optimized topologies, the penetration of flux lines is significantly delayed from structural edges and corners; instead, the flux lines are effectively pinned and confined by the geometric features, including microcavities, edges and corners.
The arrangement of flux lines is governed by the interplay among the circulating supercurrents and material interfaces. At relatively low volume fractions, the common interaction of vortex-vortex repulsions and Bean-Livingston barriers is the dominant factor. As the volume fraction increases sufficiently, repulsive and attractive forces among the circulating supercurrents become dominant, confining the flux lines around vortices of Meissner currents. Concurrently, a higher volume fraction increases the density of flux lines within the mixed state.
When the applied magnetic field is low, a superheated Meissner state is achievable in the optimized topologies. As the field increases, flux lines penetrate the optimized topologies until the emergence of isolated regions in the normal state, signaling the onset of second-order phase transitions at the upper critical magnetic fields. 
Furthermore, because the design domain is nondimensionalized by the penetration depth of the superconducting material, the diameters of the flux lines within the optimized topologies decrease as the Ginzburg-Landau parameter increases. Across all investigated values of the Ginzburg-Landau parameter, the confinement of flux lines is consistently maintained through the interactions of circulating supercurrents and Meissner currents.
Additionally, by recasting the optimization objective as the least-square difference between the order parameters of the mixed and Meissner states, topology optimization can be adapted to delay the appearance of regions in the intermediate and normal states before the occurrence of second-order phase transitions at the upper critical magnetic fields.

When implementing topology optimization of a low-temperature type-II superconductor by using the presented approach, one needs to first specify the superconducting material and operating temperature. Then, the relevant material parameters can be determined, including the Ginzburg-Landau parameter, penetration depth, coherence length, dimensionless friction coefficient, conductivity, and values of the critical fields. Using these material parameters, topology optimization can be performed to yield an optimized geometry expressed in unit of the penetration depth for physical realization.

In the future, topology optimization of low-temperature type-II superconductors can be extended in several promising directions. First, extending it to high-temperature type-II superconductors can lead to optimized configurations that differ from those of low-temperature type-II superconductors, primarily due to the anisotropy inherent in their layered crystal structures \cite{LawrenceDoniach1970}. Second, extending it to the cases under the Coulomb and Lorentz gauges would allow for a comprehensive investigation of type-II superconductors for both electric and magnetic responses under externally applied electric currents and homogeneous magnetic fields. Third, it can be adapted to inversely design structures with superconductor-normal metal interfaces, incorporating the effects of surface tension. Finally, from an application perspective, it holds strong potential for the inverse design of key devices for nuclear magnetic resonance, and superconducting quantum interference devices leveraging Josephson effects in quantum computing.

\section{Acknowledgements}\label{sec:AcknowledgementsTOOPSuperCond}

Y.D. and J.G.K. acknowledge support from the European Research Council via ERC-SygG (HiSCORE, 951459), from the Deutsche Forschungsgemeinschaft (DFG) via CRC 1527 HyPERiON, and from the German Research Foundation (Deutsche
Forschungsgemeinschaft, DFG) under Germany's Excellence Strategy - 2082/2 - 390761711.

The authors would like to express their sincere gratitude to Prof.\ Ole Sigmund (Technical University of Denmark) for his valuable discussions and insightful suggestions. They also thank Prof.\ Krister Svanberg (KTH Royal Institute of Technology) for kindly providing the MATLAB codes for the Method of Moving Asymptotes. 

\section{Appendix} \label{sec:AppendixTOOPSuperCond}

In this section, details are provided for the derivation of the variational formulations for the split and gauged time-dependent Ginzburg-Landau equations and adjoint analysis of the topology optimization model.

\subsection{Variational formulations of split and gauged time-dependent Ginzburg-Landau equations} \label{sec:PartialIntegrationTDGLEqus}

The variational formulations for the split and gauged time-dependent Ginzburg-Landau equations in Eqs. \ref{eq:SplitGaugeTransformedTDGLEqs} can be derived based on Galerkin's variational method by implementing the following integrals by parts:
\begin{equation}\label{equ:PartialIntegrationPsir}
\begin{split}
  & \int_\Omega \eta {\partial \psi_r \over \partial t} \tilde{\psi}_r - \kappa^{-1} \nabla \cdot \left( \kappa^{-1} \nabla \psi_r\right) \tilde{\psi}_r - \kappa^{-1} \nabla \psi_i \cdot \mathbf{A} \tilde{\psi}_r - \kappa^{-1} \left( \mathbf{A} \cdot \nabla \right) \psi_i \tilde{\psi}_r \\
  & + \mathbf{A}^2 \psi_r \tilde{\psi}_r - I_d \psi_r \tilde{\psi}_r + I_d \left(\psi_r^2 + \psi_i^2\right) \psi_r \tilde{\psi}_r \, \mathrm{d}\Omega \\
= & \int_\Omega \eta {\partial \psi_r \over \partial t} \tilde{\psi}_r + \kappa^{-2} \nabla \psi_r \cdot \nabla 
  \tilde{\psi}_r + \left[ \kappa^{-1} \psi_i \left( \nabla \cdot \mathbf{A} \right) \tilde{\psi}_r + \psi_i \mathbf{A} \cdot \nabla \left( \kappa^{-1} \tilde{\psi}_r \right) \right] \\
  & - \kappa^{-1} \left( \mathbf{A} \cdot \nabla \right) \psi_i \tilde{\psi}_r + \mathbf{A}^2 \psi_r \tilde{\psi}_r - I_d \psi_r \tilde{\psi}_r + I_d \left(\psi_r^2 + \psi_i^2\right) \psi_r \tilde{\psi}_r \, \mathrm{d}\Omega \\
  & + \int_{\partial\Omega} - \kappa^{-2} \nabla \psi_r \cdot \mathbf{n} \tilde{\psi}_r - \kappa^{-1} \psi_i \left( \mathbf{A} \cdot \mathbf{n} \right) \tilde{\psi}_r \, \mathrm{d}\partial\Omega \\
= & \int_\Omega \eta {\partial \psi_r \over \partial t} \tilde{\psi}_r + \kappa^{-2} \nabla \psi_r \cdot \nabla 
  \tilde{\psi}_r + \psi_i \mathbf{A} \cdot \nabla \left( \kappa^{-1} \tilde{\psi}_r \right) - \kappa^{-1} \left( \mathbf{A} \cdot \nabla \right) \psi_i \tilde{\psi}_r \\
  & + \mathbf{A}^2 \psi_r \tilde{\psi}_r - I_d \psi_r \tilde{\psi}_r + I_d \left(\psi_r^2 + \psi_i^2\right) \psi_r \tilde{\psi}_r \, \mathrm{d}\Omega + \int_{\partial\Omega} \kappa^{-2} \gamma \psi_r \tilde{\psi}_r \, \mathrm{d}\partial\Omega,
\end{split}
\end{equation}
\begin{equation}\label{equ:PartialIntegrationPsii}
\begin{split}
  & \int_\Omega \eta {\partial \psi_i \over \partial t} \tilde{\psi}_i - \kappa^{-1} \nabla \cdot \left( \kappa^{-1} \nabla \psi_i \right) \tilde{\psi}_i + \kappa^{-1} \nabla \psi_r \cdot \mathbf{A} \tilde{\psi}_i + \kappa^{-1} \left( \mathbf{A} \cdot \nabla \right) \psi_r \tilde{\psi}_i \\
  & + \mathbf{A}^2 \psi_i \tilde{\psi}_i - I_d \psi_i \tilde{\psi}_i + I_d \left(\psi_r^2 + \psi_i^2\right) \psi_i \tilde{\psi}_i \, \mathrm{d}\Omega \\
= & \int_\Omega \eta {\partial \psi_i \over \partial t} \tilde{\psi}_i + \kappa^{-2} \nabla \psi_i \cdot \nabla 
  \tilde{\psi}_i - \left[ \kappa^{-1} \psi_r \left(\nabla \cdot \mathbf{A} \right) \tilde{\psi}_i + \psi_r \mathbf{A} \cdot \nabla \left( \kappa^{-1} \tilde{\psi}_i \right) \right] \\
  & + \kappa^{-1} \left( \mathbf{A} \cdot \nabla \right) \psi_r \tilde{\psi}_i + \mathbf{A}^2 \psi_i \tilde{\psi}_i - I_d \psi_i \tilde{\psi}_i + I_d \left(\psi_r^2 + \psi_i^2\right) \psi_i \tilde{\psi}_i \, \mathrm{d}\Omega \\
  & + \int_{\partial\Omega} - \kappa^{-2} \nabla \psi_i \cdot \mathbf{n} \tilde{\psi}_i + \kappa^{-1} \psi_r \left( \mathbf{A} \cdot \mathbf{n} \right) \tilde{\psi}_i \, \mathrm{d}\partial\Omega \\
= & \int_\Omega \eta {\partial \psi_i \over \partial t} \tilde{\psi}_i + \kappa^{-2} \nabla \psi_i \cdot \nabla 
  \tilde{\psi}_i - \psi_r \mathbf{A} \cdot \nabla \left( \kappa^{-1} \tilde{\psi}_i \right) + \kappa^{-1} \left( \mathbf{A} \cdot \nabla \right) \psi_r \tilde{\psi}_i \\
  & + \mathbf{A}^2 \psi_i \tilde{\psi}_i - I_d \psi_i \tilde{\psi}_i + I_d \left(\psi_r^2 + \psi_i^2\right) \psi_i \tilde{\psi}_i \, \mathrm{d}\Omega + \int_{\partial\Omega} \kappa^{-2} \gamma \psi_i \tilde{\psi}_i \, \mathrm{d}\partial\Omega,
\end{split}
\end{equation}
and
\begin{equation}\label{equ:PartialIntegrationA}
\begin{split}
  & \int_\Omega \sigma {\partial \mathbf{A} \over \partial t} \cdot \tilde{\mathbf{A}} + \left\{ \nabla \times \left[ w_p \left( \nabla \times \mathbf{A} - \mathbf{H} \right) \right] \right\} \cdot \tilde{\mathbf{A}} - \kappa^{-1} \left( \psi_r \nabla \psi_i - \psi_i \nabla \psi_r \right) \cdot \tilde{\mathbf{A}} \\
  & + \left(\psi_r^2 + \psi_i^2\right) \mathbf{A} \cdot \tilde{\mathbf{A}} \,\mathrm{d}\Omega \\
= & \int_\Omega \sigma {\partial \mathbf{A} \over \partial t} \cdot \tilde{\mathbf{A}} + w_p \left( \nabla \times \mathbf{A} 
  - \mathbf{H} \right) \cdot \left( \nabla \times \tilde{\mathbf{A}} \right) - \kappa^{-1} \left( \psi_r \nabla \psi_i - \psi_i \nabla \psi_r \right) \cdot \tilde{\mathbf{A}} \\
  & + \left(\psi_r^2 + \psi_i^2\right) \mathbf{A} \cdot \tilde{\mathbf{A}} \,\mathrm{d}\Omega + \int_{\partial\Omega} w_p \tilde{\mathbf{A}} \cdot \left[ \mathbf{n} \times \left( \nabla \times \mathbf{A} - \mathbf{H} \right) \right] \, \mathrm{d}\partial\Omega \\
= & \int_\Omega \sigma {\partial \mathbf{A} \over \partial t} \cdot \tilde{\mathbf{A}} + w_p \left( \nabla \times \mathbf{A} 
  - \mathbf{H} \right) \cdot \left( \nabla \times \tilde{\mathbf{A}} \right) - \kappa^{-1} \left( \psi_r \nabla \psi_i - \psi_i \nabla \psi_r \right) \cdot \tilde{\mathbf{A}} \\
  & + \left(\psi_r^2 + \psi_i^2\right) \mathbf{A} \cdot \tilde{\mathbf{A}} \,\mathrm{d}\Omega.
\end{split}
\end{equation}

Based on Eqs. \ref{equ:PartialIntegrationPsir}, \ref{equ:PartialIntegrationPsii} and \ref{equ:PartialIntegrationA}, the variational formulations for the split and gauged time-dependent Ginzburg-Landau equations can be derived as
\begin{equation}\label{equ:VariationalFormsSplitTDGLOrderParameter}
\left\{\begin{split}
& \left\{\begin{split}
& \left\{\begin{split}
& \text{Find} ~ \psi_r \in \mathcal{H}\left(\left(0,T_t\right);\mathcal{H}\left(\Omega\right)\right) ~ \text{with} ~ \left[\psi_r\right]_{t=0} = \psi_{r0} ~\mathrm{at}~ \forall\mathbf{x}\in\Omega \\
& \text{for} ~ \forall \tilde{\psi}_r \in \mathcal{H}\left(\left(0,T_t\right);\mathcal{H}\left(\Omega\right)\right), ~  \text{such~that} \\
& \int_0^{T_t} \int_\Omega \eta {\partial \psi_r \over \partial t} \tilde{\psi}_r + \kappa^{-2} \nabla \psi_r \cdot \nabla \tilde{\psi}_r + \psi_i \mathbf{A} \cdot {\nabla \left( \kappa^{-1} \tilde{\psi}_r \right)} - \kappa^{-1} \left( \mathbf{A} \cdot \nabla \right) \psi_i \tilde{\psi}_r \\
& + \mathbf{A}^2 \psi_r \tilde{\psi}_r - I_d \left[ 1 - \left(\psi_r^2 + \psi_i^2\right) \right] \psi_r \tilde{\psi}_r \, \mathrm{d}\Omega\mathrm{d}t + \int_0^{T_t} \int_{\partial\Omega} \kappa^{-2} \gamma \psi_r \tilde{\psi}_r \, \mathrm{d}\partial\Omega\mathrm{d}t = 0,
\end{split}\right. \\
& \left\{\begin{split}
& \text{Find} ~ \psi_i \in \mathcal{H}\left(\left(0,T_t\right);\mathcal{H}\left(\Omega\right)\right) ~ \text{with} ~ \left[\psi_i\right]_{t=0} = \psi_{i0} ~\mathrm{at}~ \forall\mathbf{x}\in\Omega \\
& \text{for} ~ \forall \tilde{\psi}_i \in \mathcal{H}\left(\left(0,T_t\right);\mathcal{H}\left(\Omega\right)\right), ~  \text{such~that} \\
& \int_0^{T_t} \int_\Omega \eta {\partial \psi_i \over \partial t} \tilde{\psi}_i + \kappa^{-2} \nabla \psi_i \cdot \nabla \tilde{\psi}_i - \psi_r \mathbf{A} \cdot {\nabla \left( \kappa^{-1} \tilde{\psi}_i \right)} + \kappa^{-1} \left( \mathbf{A} \cdot \nabla \right) \psi_r \tilde{\psi}_i \\
& + \mathbf{A}^2 \psi_i \tilde{\psi}_i - I_d \left[ 1 - \left(\psi_r^2 + \psi_i^2\right) \right] \psi_i 
  \tilde{\psi}_i \, \mathrm{d}\Omega\mathrm{d}t + \int_0^{T_t} \int_{\partial\Omega} \kappa^{-2} \gamma \psi_i \tilde{\psi}_i \, \mathrm{d}\partial\Omega\mathrm{d}t = 0,
\end{split}\right. \\
\end{split}\right. \\
& \left\{\begin{split}
& \text{Find} ~ \mathbf{A} \in \left(\mathcal{H}\left(\left(0,T_t\right);\mathcal{H}\left(\Omega\right)\right)\right)^3 ~ \text{with} ~ \left[\mathbf{A}\right]_{t=0} = \mathbf{A}_0 ~ \mathrm{at} ~ \forall\mathbf{x}\in\Omega \\ 
& \text{for} ~ \forall \tilde{\mathbf{A}} \in \left(\mathcal{H}\left(\left(0,T_t\right);\mathcal{H}\left(\Omega\right)\right)\right)^3, ~ \text{such~that} \\
& \int_0^{T_t} \int_\Omega \sigma {\partial \mathbf{A} \over \partial t} \cdot \tilde{\mathbf{A}} + w_p \left( \nabla \times \mathbf{A} - \mathbf{H} \right) \cdot \left( \nabla \times \tilde{\mathbf{A}} \right) \\
& - \kappa^{-1} \left( \psi_r \nabla \psi_i - \psi_i \nabla \psi_r \right) \cdot \tilde{\mathbf{A}} + \left(\psi_r^2 + \psi_i^2\right) \mathbf{A} \cdot \tilde{\mathbf{A}} \,\mathrm{d}\Omega\mathrm{d}t = 0.
\end{split}\right.
\end{split}\right.
\end{equation}
 
It is noted that the partial integrals are implemented on the terms $- \kappa^{-1} \nabla \psi_i \cdot \mathbf{A} \tilde{\psi}_r$ in Eq.~\ref{equ:PartialIntegrationPsir} and $\kappa^{-1} \nabla \psi_r \cdot \mathbf{A} \tilde{\psi}_i$ in Eq.~\ref{equ:PartialIntegrationPsii} to impose $\mathbf{n} \cdot \mathbf{A} = 0, ~ \forall \left(\mathbf{x},t\right) \in \partial\Omega \times \left(0, +\infty\right)$ in Eq.~\ref{eq:SplitGaugeTransformedTDGLEqsBndConds} and the Gaussian theorem is used for the combined term $\left\{ \nabla \times \left[ w_p \left( \nabla \times \mathbf{A} - \mathbf{H} \right) \right] \right\} \cdot \tilde{\mathbf{A}}$ in Eq.~\ref{equ:PartialIntegrationA} to impose the boundary condition of $\mathbf{n} \times \left( \nabla \times \mathbf{A} - \mathbf{H} \right) = \mathbf{0}, ~ \forall \left(\mathbf{x},t\right) \in \partial\Omega \times \left(0, +\infty\right)$ in Eq.~\ref{eq:SplitGaugeTransformedTDGLEqsBndConds}.

\subsection{Details for adjoint analysis}\label{sec:DetailsAdjointAnalysis}

Based on the Lagrangian multiplier based adjoint analysis method \cite{HinzeSpringer2009}, adjoint analysis can be implemented for the optimization objective in Eq.~\ref{eq:ObjFunction} and the volume fraction in Eq.~\ref{eq:VolumeFraction} to derive the adjoint sensitivities used to iteratively evolve the design variables for the geometric configurations of the low-temperature type-II superconductors.
 
For the optimization objective in Eq.~\ref{eq:ObjFunction}, the augmented Lagrangian can be derived as
\begin{equation}
\begin{split}
\hat {J} = 
& \int_0^{T_t} \int_\Omega \left|\left[ \mathbf{j}_s \right]_{\forall\mathbf{x}\in\Omega}\right|^2 H \left( t-T_m \right) \\
& + \eta {\partial \psi_r \over \partial t} \psi_{ra} + \kappa^{-2} \nabla \psi_r \cdot \nabla 
  \psi_{ra} + \psi_i \mathbf{A} \cdot \nabla \left( \kappa^{-1} \psi_{ra} \right) - \kappa^{-1} \left( \mathbf{A} \cdot \nabla \right) \psi_i \psi_{ra} \\
& + \mathbf{A}^2 \psi_r \psi_{ra} - I_d \psi_r \psi_{ra} + I_d \left(\psi_r^2 + \psi_i^2\right) \psi_r 
  \psi_{ra} \\
& + \eta {\partial \psi_i \over \partial t} \psi_{ia} + \kappa^{-2} \nabla \psi_i \cdot \nabla 
  \psi_{ia} - \psi_r \mathbf{A} \cdot \nabla \left( \kappa^{-1} \psi_{ia} \right) + \kappa^{-1} \left( \mathbf{A} \cdot \nabla \right) \psi_r \psi_{ia} \\
& + \mathbf{A}^2 \psi_i \psi_{ia} - I_d \psi_i \psi_{ia} + I_d \left(\psi_r^2 + \psi_i^2\right) \psi_i 
  \psi_{ia} \\
& + \sigma {\partial \mathbf{A} \over \partial t} \cdot \mathbf{A}_a + w_p \left( \nabla \times \mathbf{A} - \mathbf{H} \right) \cdot \left( \nabla \times \mathbf{A}_a \right) - \kappa^{-1} \left( \psi_r \nabla \psi_i - \psi_i \nabla \psi_r \right) \cdot \mathbf{A}_a \\
& + \left(\psi_r^2 + \psi_i^2\right) \mathbf{A} \cdot 
  \mathbf{A}_a \,\mathrm{d}\Omega\mathrm{d}t + \int_\Omega r_\rho^2 \nabla \rho_f \cdot \nabla \rho_{fa} + \rho_f \rho_{fa} - \rho \rho_{fa} \,\mathrm{d}\Omega,
\end{split}
\end{equation}
where $\hat{J}$ is the augmented Lagrangian of $J$. Based on the variational method, the first-order variational of $\hat{J}$ can be derived as
\begin{equation}
\begin{split}
\delta \hat {J} = 
& \int_\Omega \left[ \eta \left( \delta \psi_r \psi_{ra} + \delta \psi_i \psi_{ia} \right) + \sigma \delta \mathbf{A} \cdot \mathbf{A}_a \right]_{t=T_t} \,\mathrm{d}\Omega \\
& + \int_0^{T_t} \int_\Omega \Bigg[ {\partial \left|\left[ \mathbf{j}_s \right]_{\forall\mathbf{x}\in\Omega}\right|^2 \over \partial \psi_r} \delta \psi_r + {\partial \left|\left[ \mathbf{j}_s \right]_{\forall\mathbf{x}\in\Omega}\right|^2 \over \partial \nabla \psi_r} \cdot \nabla \delta \psi_r + {\partial \left|\left[ \mathbf{j}_s \right]_{\forall\mathbf{x}\in\Omega}\right|^2 \over \partial \psi_i} \delta \psi_i \\
& + {\partial \left|\left[ \mathbf{j}_s \right]_{\forall\mathbf{x}\in\Omega}\right|^2 \over \partial \nabla \psi_i} \cdot \nabla \delta \psi_i + {\partial \left|\left[ \mathbf{j}_s \right]_{\forall\mathbf{x}\in\Omega}\right|^2 \over \partial \mathbf{A}} \cdot \delta \mathbf{A} + {\partial \left|\left[ \mathbf{j}_s \right]_{\forall\mathbf{x}\in\Omega}\right|^2 \over \partial \nabla \times \mathbf{A}} \cdot \left( \nabla \times \delta \mathbf{A} \right) \Bigg] \\
& H \left( t-T_m \right) - \eta {\partial \psi_{ra} \over \partial t} \delta \psi_r + \kappa^{-2} \nabla \delta \psi_r \cdot \nabla \psi_{ra} + \kappa^{-1} \delta \psi_i \mathbf{A} \cdot \nabla \psi_{ra} + \kappa^{-1} \psi_i \\
& \delta \mathbf{A} \cdot \nabla \psi_{ra} - \kappa^{-1} \left( \delta \mathbf{A} \cdot \nabla \right) \psi_i \psi_{ra} - \kappa^{-1} \left( \mathbf{A} \cdot \nabla \right) \delta \psi_i \psi_{ra} + 2 \mathbf{A} \cdot \delta \mathbf{A} \psi_r \psi_{ra} \\
& + \mathbf{A}^2 \delta \psi_r \psi_{ra} - I_d \delta \psi_r \psi_{ra} + 2 I_d \left(\psi_r \delta \psi_r + \psi_i \delta \psi_i\right) \psi_r \psi_{ra} + I_d \left(\psi_r^2 + \psi_i^2\right) \\
& \delta \psi_r \psi_{ra} - \eta {\partial \psi_{ia} \over \partial t} \delta \psi_i + \kappa^{-2} \nabla \delta \psi_i \cdot \nabla \psi_{ia} - \kappa^{-1} \delta \psi_r \mathbf{A} \cdot \nabla \psi_{ia} - \kappa^{-1} \psi_r \delta \mathbf{A} \\
& \cdot \nabla \psi_{ia} + \kappa^{-1} \left( \delta \mathbf{A} \cdot \nabla \right) \psi_r \psi_{ia} + \kappa^{-1} \left( \mathbf{A} \cdot \nabla \right) \delta \psi_r \psi_{ia} + 2 \mathbf{A} \cdot \delta \mathbf{A} \psi_i \psi_{ia} \\
& + \mathbf{A}^2 \delta \psi_i \psi_{ia} - I_d \delta \psi_i \psi_{ia} + 2 I_d \left(\psi_r \delta\psi_r + \psi_i \delta\psi_i\right) \psi_i \psi_{ia} + I_d \left(\psi_r^2 + \psi_i^2\right) \\
& \delta \psi_i \psi_{ia} - \sigma {\partial \mathbf{A}_a \over \partial t} \cdot \delta \mathbf{A} + w_p \left( \nabla \times \delta \mathbf{A} \right) \cdot \left( \nabla \times \mathbf{A}_a \right) \\
& - \kappa^{-1} \left( \delta \psi_r \nabla \psi_i + \psi_r \nabla \delta \psi_i - \delta \psi_i \nabla \psi_r - \psi_i \nabla \delta \psi_r \right) \cdot \mathbf{A}_a \\
& + 2 \left(\psi_r \delta \psi_r + \psi_i \delta \psi_i\right) \mathbf{A} \cdot 
  \mathbf{A}_a + \left(\psi_r^2 + \psi_i^2\right) \delta \mathbf{A} \cdot 
  \mathbf{A}_a \,\mathrm{d}\Omega\mathrm{d}t \\
& + \int_0^{T_t} \Bigg\{ \sum_{n=1}^N \int_{\Omega_n} {\partial \rho_p \over \partial \rho_e} {\partial \rho_e \over \partial \rho_f} \Bigg[ {\partial \left|\left[ \mathbf{j}_s \right]_{\forall\mathbf{x}\in\Omega}\right|^2 \over \partial \rho_p} H \left( t-T_m \right) + 2 \kappa^{-1} {\partial\kappa^{-1} \over \partial \rho_p} \nabla \psi_r \cdot \nabla \psi_{ra} \\
& + {\partial\kappa^{-1} \over \partial \rho_p} \psi_i \mathbf{A} \cdot \nabla \psi_{ra} - {\partial\kappa^{-1} \over \partial \rho_p} \left( \mathbf{A} \cdot \nabla \right) \psi_i \psi_{ra} - {\partial I_d \over \partial \rho_p} \psi_r \psi_{ra} + {\partial I_d \over \partial \rho_p} \left(\psi_r^2 + \psi_i^2\right) \\
& \psi_r \psi_{ra} + 2 \kappa^{-1} {\partial \kappa^{-1} \over \partial \rho_p} \nabla \psi_i \cdot \nabla \psi_{ia} - {\partial \kappa^{-1} \over \partial \rho_p} \psi_r \mathbf{A} \cdot \nabla \psi_{ia} + {\partial \kappa^{-1} \over \partial \rho_p} \left( \mathbf{A} \cdot \nabla \right) \psi_r \psi_{ia} \\
& - {\partial I_d \over \partial \rho_p} \psi_i \psi_{ia} + {\partial I_d \over \partial \rho_p} \left(\psi_r^2 + \psi_i^2\right) \psi_i \psi_{ia} + {\partial w_p \over \partial \rho_p} \left( \nabla \times \mathbf{A} - \mathbf{H} \right) \\
& \cdot \left( \nabla \times \mathbf{A}_a \right) - {\partial \kappa^{-1} \over \partial \rho_p} \left( \psi_r \nabla \psi_i - \psi_i \nabla \psi_r \right) \cdot \mathbf{A}_a \Bigg] \delta \rho_f \,\mathrm{d}\Omega \Bigg\} \mathrm{d}t \\
& + \int_\Omega r_\rho^2 \nabla \delta \rho_f \cdot \nabla \rho_{fa} + \delta \rho_f \rho_{fa} - \delta \rho \rho_{fa} \,\mathrm{d}\Omega
\end{split}
\end{equation}
with
\begin{equation}
\left\{\begin{split}
  & \forall \delta \psi_r \in \mathcal{H}\left(\left(0,T_t\right);\mathcal{H}\left(\Omega\right)\right) \\
  & \forall \delta \psi_i \in \mathcal{H}\left(\left(0,T_t\right);\mathcal{H}\left(\Omega\right)\right) \\
  & \forall \delta \mathbf{A} \in \left(\mathcal{H}\left(\left(0,T_t\right);\mathcal{H}\left(\Omega\right)\right)\right)^3 \\
  & \forall \delta \rho_f \in \mathcal{H} \left(\Omega\right) \\
  & \forall \delta \rho \in \mathcal{L}^2 \left(\Omega\right)
\end{split}\right..
\end{equation}
According to the Karush-Kuhn-Tucker conditions of the PDE constrained optimization problem \cite{HinzeSpringer2009}, the first-order variational of $\hat{J}$ to $\psi_r$ can be set to be zero as
\begin{equation}
\begin{split}
& \int_\Omega \left[ \eta \delta \psi_r \psi_{ra} \right]_{t=T_t} \,\mathrm{d}\Omega \\ 
& + \int_0^{T_t} \int_\Omega \left( {\partial \left|\left[ \mathbf{j}_s \right]_{\forall\mathbf{x}\in\Omega}\right|^2 \over \partial \psi_r} \delta \psi_r + {\partial \left|\left[ \mathbf{j}_s \right]_{\forall\mathbf{x}\in\Omega}\right|^2 \over \partial \nabla \psi_r} \cdot \nabla \delta \psi_r \right) H \left( t-T_m \right) \\
& - \eta {\partial \psi_{ra} \over \partial t} \delta \psi_r + \kappa^{-2} \nabla \delta \psi_r \cdot \nabla \psi_{ra} + \kappa^{-1} \left[ \left( \mathbf{A} \cdot \nabla \right) \delta \psi_r \right] \psi_{ia} + \kappa^{-1} \psi_i \nabla \delta \psi_r \\
& \cdot \mathbf{A}_a + \Big[ \mathbf{A}^2 \psi_{ra} - I_d \psi_{ra} + 2 I_d \psi_r \left( \psi_r \psi_{ra} + \psi_i \psi_{ia} \right) + I_d \left(\psi_r^2 + \psi_i^2\right) \psi_{ra} \\
& - \kappa^{-1} \mathbf{A} \cdot \nabla \psi_{ia} - \kappa^{-1} \nabla \psi_i \cdot \mathbf{A}_a + 2 \psi_r \mathbf{A} \cdot \mathbf{A}_a \Big] \delta \psi_r \,\mathrm{d}\Omega\mathrm{d}t = 0;
\end{split}
\end{equation}
the first-order variational of $\hat{J}$ to $\psi_i$ can be set to be zero as
\begin{equation}
\begin{split}
& \int_\Omega \left[ \eta \delta \psi_i \psi_{ia} \right]_{t=T_t} \,\mathrm{d}\Omega \\
& + \int_0^{T_t} \int_\Omega \left( {\partial \left|\left[ \mathbf{j}_s \right]_{\forall\mathbf{x}\in\Omega}\right|^2 \over \partial \psi_i} \delta \psi_i + {\partial \left|\left[ \mathbf{j}_s \right]_{\forall\mathbf{x}\in\Omega}\right|^2 \over \partial \nabla \psi_i} \cdot \nabla \delta \psi_i \right) H \left( t-T_m \right) \\
& - \eta {\partial \psi_{ia} \over \partial t} \delta \psi_i + \kappa^{-2} \nabla \delta \psi_i \cdot \nabla \psi_{ia} - \kappa^{-1} \left[ \left( \mathbf{A} \cdot \nabla \right) \delta \psi_i \right] \psi_{ra} - \kappa^{-1} \psi_r \nabla \delta \psi_i \\
& \cdot \mathbf{A}_a + \Big[ \mathbf{A}^2 \psi_{ia} - I_d \psi_{ia} + 2 I_d \psi_i \left( \psi_r \psi_{ra} + \psi_i \psi_{ia} \right) + I_d \left(\psi_r^2 + \psi_i^2\right) \psi_{ia} \\
& + \kappa^{-1} \mathbf{A} \cdot \nabla \psi_{ra} + \kappa^{-1} \nabla \psi_r \cdot \mathbf{A}_a + 2 \psi_i \mathbf{A} \cdot 
  \mathbf{A}_a \Big] \delta \psi_i \,\mathrm{d}\Omega\mathrm{d}t = 0;
\end{split}
\end{equation}
the first-order variational of $\hat{J}$ to $\mathbf{A}$ can be set to be zero as
\begin{equation}
\begin{split}
& \int_\Omega \left[ \sigma \delta \mathbf{A} \cdot \mathbf{A}_a \right]_{t=T_t} \,\mathrm{d}\Omega \\
& + \int_0^{T_t} \int_\Omega \left[ {\partial \left|\left[ \mathbf{j}_s \right]_{\forall\mathbf{x}\in\Omega}\right|^2 \over \partial \mathbf{A}} \cdot \delta \mathbf{A} + {\partial \left|\left[ \mathbf{j}_s \right]_{\forall\mathbf{x}\in\Omega}\right|^2 \over \partial \nabla \times \mathbf{A}} \cdot \left( \nabla \times \delta \mathbf{A} \right) \right] H \left( t-T_m \right) \\
& - \sigma {\partial \mathbf{A}_a \over \partial t} \cdot \delta \mathbf{A} + w_p \left( \nabla \times \delta \mathbf{A} \right) \cdot \left( \nabla \times \mathbf{A}_a \right) + \left(\psi_r^2 + \psi_i^2\right) \delta \mathbf{A} \cdot \mathbf{A}_a + \kappa^{-1} \delta \mathbf{A} \\
& \cdot \left( \psi_i \nabla \psi_{ra} - \psi_r \nabla \psi_{ia} \right) - \kappa^{-1} \left\{ \left[ \left( \delta \mathbf{A} \cdot \nabla \right) \psi_i \right] \psi_{ra} - \left[ \left( \delta \mathbf{A} \cdot \nabla \right) \psi_r \right] \psi_{ia} \right\} \\
& + 2 \mathbf{A} \cdot \delta \mathbf{A} \left( \psi_r \psi_{ra} + \psi_i \psi_{ia} \right) \, \mathrm{d}\Omega\mathrm{d}t = 0. \\
\end{split}
\end{equation}
and the first-order variational of $\hat{J}$ to $\rho_f$ can be set to be zero as
\begin{equation}
\begin{split}
& \int_0^{T_t} \Bigg\{ \sum_{n=1}^N \int_{\Omega_n} {\partial \rho_p \over \partial \rho_e} {\partial \rho_e \over \partial \rho_f} \Bigg[ {\partial \left|\left[ \mathbf{j}_s \right]_{\forall\mathbf{x}\in\Omega}\right|^2 \over \partial \rho_p} H \left( t-T_m \right) + 2 \kappa^{-1} {\partial\kappa^{-1} \over \partial \rho_p} \nabla \psi_r \cdot \nabla \psi_{ra} \\
& + {\partial\kappa^{-1} \over \partial \rho_p} \psi_i \mathbf{A} \cdot \nabla \psi_{ra} - {\partial\kappa^{-1} \over \partial \rho_p} \left[ \left( \mathbf{A} \cdot \nabla \right) \psi_i \right] \psi_{ra} - {\partial I_d \over \partial \rho_p} \psi_r \psi_{ra} + {\partial I_d \over \partial \rho_p} \left(\psi_r^2 + \psi_i^2\right) \\
& \psi_r \psi_{ra} + 2 \kappa^{-1} {\partial \kappa^{-1} \over \partial \rho_p} \nabla \psi_i \cdot \nabla \psi_{ia} - {\partial \kappa^{-1} \over \partial \rho_p} \psi_r \mathbf{A} \cdot \nabla \psi_{ia} + {\partial \kappa^{-1} \over \partial \rho_p} \left[ \left( \mathbf{A} \cdot \nabla \right) \psi_r \right] \psi_{ia} \\
& - {\partial I_d \over \partial \rho_p} \psi_i \psi_{ia} + {\partial I_d \over \partial \rho_p} \left(\psi_r^2 + \psi_i^2\right) \psi_i \psi_{ia} + {\partial w_p \over \partial \rho_p} \left( \nabla \times \mathbf{A} - \mathbf{H} \right) \\
& \cdot \left( \nabla \times \mathbf{A}_a \right) - {\partial \kappa^{-1} \over \partial \rho_p} \left( \psi_r \nabla \psi_i - \psi_i \nabla \psi_r \right) \cdot \mathbf{A}_a \Bigg] \delta \rho_f \,\mathrm{d}\Omega \Bigg\} \mathrm{d}t \\
& + \int_\Omega r_\rho^2 \nabla \delta \rho_f \cdot \nabla \rho_{fa} + \delta \rho_f \rho_{fa} \,\mathrm{d}\Omega = 0.
\end{split}
\end{equation}
Then, the adjoint sensitivity can be derived as
\begin{equation}
\begin{split}
\delta \hat {J} = \int_\Omega - \rho_{fa} \delta \rho \,\mathrm{d}\Omega.
\end{split}
\end{equation}
Without losing arbitrariness, the variational formulations of the adjoint system composed of Eqs. \ref{eq:AdjSensDesignObj}, \ref{equ:VariationalFormsAdjEquForOrderParameter} and \ref{equ:VariationalFormsAdjEquForPDEFilter} can be derived by setting 
\begin{equation}
\left.\begin{split}
  \tilde{\psi}_{ra} = \delta \psi_r & \\
  \tilde{\psi}_{ia} = \delta \psi_i & \\
  \tilde{\mathbf{A}}_a = \delta \mathbf{A} & \\
  \tilde{\rho}_{fa} = \delta \rho_f & 
\end{split}\right\} \text{with} 
\left\{\begin{split}
  & \forall \tilde{\psi}_{ra} \in \mathcal{H}\left(\left(0,T_t\right);\mathcal{H}\left(\Omega\right)\right) \\
  & \forall \tilde{\psi}_{ia} \in \mathcal{H}\left(\left(0,T_t\right);\mathcal{H}\left(\Omega\right)\right) \\
  & \forall \tilde{\mathbf{A}}_a \in \left(\mathcal{H}\left(\left(0,T_t\right);\mathcal{H}\left(\Omega\right)\right)\right)^3 \\
  & \forall \tilde{\rho}_{fa} \in \mathcal{H} \left(\Omega\right)
\end{split}\right..
\end{equation}

For the volume fraction in Eq.~\ref{eq:VolumeFraction}, the augmented Lagrangian can be derived as
\begin{equation}
\begin{split}
\hat{v} = 
& \int_\Omega \rho_p + r_\rho^2 \nabla \rho_f \cdot \nabla \rho_{fa} + \rho_f \rho_{fa} - \rho \rho_{fa} \,\mathrm{d}\Omega,
\end{split}
\end{equation}
based on the variational formulation of the PDE filter in Eq.~\ref{eq:PDEFilter} expressed as
\begin{equation}
\left\{\begin{split}
& \text{Find} ~ \rho_f \in \mathcal{H}\left(\Omega\right) ~ \text{for} ~ \forall \tilde{\rho}_f \in \mathcal{H}\left(\Omega\right) ~ \text{and} ~ \rho \in \mathcal{L}^2\left(\Omega\right), \\
& \text{such~that} \int_\Omega r_\rho^2 \nabla \rho_f \cdot \nabla \tilde{\rho}_f + \rho_f \tilde{\rho}_f - \rho \tilde{\rho}_f \,\mathrm{d}\Omega = 0,
\end{split}\right.
\end{equation}
where $\tilde{\rho}_f$ is the test function of $\rho_f$; and $\hat{v}$ is the augmented Lagrangian of $v$. The first-order variational of the augmented Lagrangian is
\begin{equation}
\begin{split}
\delta \hat {v} = \sum_{n=1}^N \int_{\Omega_n} {\partial \rho_p \over \partial \rho_e} {\partial \rho_e \over \partial \rho_f} \delta \rho_f \,\mathrm{d}\Omega + \int_\Omega r_\rho^2 \nabla \delta \rho_f \cdot \nabla \rho_{fa} + \delta \rho_f \rho_{fa} - \delta \rho \rho_{fa} \,\mathrm{d}\Omega.
\end{split}
\end{equation}
According to the Karush-Kuhn-Tucker conditions of the PDE constrained optimization problem, the first variational of $\delta \hat {v}$ to $\rho_f$ can be set to be zero as
\begin{equation}
\begin{split}
\sum_{n=1}^N \int_{\Omega_n} {\partial \rho_p \over \partial \rho_e} {\partial \rho_e \over \partial \rho_f} \delta \rho_f \,\mathrm{d}\Omega + \int_\Omega r_\rho^2 \nabla \delta \rho_f \cdot \nabla \rho_{fa} + \delta \rho_f \rho_{fa} \,\mathrm{d}\Omega = 0;
\end{split}
\end{equation}
then, the adjoint sensitivity of the volume fraction can be derived as
\begin{equation}
\begin{split}
\delta \hat {v} = \int_\Omega - \rho_{fa} \delta \rho \,\mathrm{d}\Omega.
\end{split}
\end{equation}
Without losing arbitrariness, the variational formulations of the adjoint system composed of Eqs. \ref{eq:AdjSensVolumeFraction} and \ref{eq:VariationalFormAdjEquVolumeFraction} can be derived by setting 
\begin{equation}
  \tilde{\rho}_{fa} = \delta \rho_f ~ \text{with} ~ \forall \tilde{\rho}_{fa} \in \mathcal{H} \left(\Omega\right).
\end{equation}

\end{document}